\RequirePackage{ifpdf}
\RequirePackage{ifpdf}
\documentclass[11pt,letterpaper]{article}
\pdfoutput=1
\usepackage{jheppub}
\usepackage{amsmath,amssymb,amsfonts,bm,amscd, yfonts,dsfont,bbm}
\usepackage{graphicx}
\usepackage{subcaption}
\usepackage{centernot}
\usepackage{stmaryrd}
\usepackage{verbatim}
\usepackage{fancybox}
\usepackage{changepage}
\usepackage{slashed}
\usepackage{url}
\usepackage{array}
\usepackage{hhline}
\usepackage{tabularx}
\usepackage{tikz}
\usepackage[utf8]{inputenc}

\usepackage{rotating,booktabs}
\usepackage[verbose]{placeins}
\usepackage{blindtext}

\graphicspath {{figures/}}

\newcommand{\bea}{\begin{eqnarray}}
\newcommand{\eea}{\end{eqnarray}}
\newcommand{\be}{\begin{equation}\begin{aligned}}
\newcommand{\ee}{\end{aligned}\end{equation}}

\newcommand{\del}{\partial}
\newcommand{\delbar}{\overline{\partial}}
\newcommand{\zbar}{\overline{z}}

\newcommand{\Z}{{\mathbb Z}}
\newcommand{\R}{{\mathbb R}}
\newcommand{\C}{{\mathbb C}}

\newcommand{\cF}{{\mathcal{F}}}
\newcommand{\cN}{{\mathcal{N}}}

\newcommand{\fr}{{\frak r}}

\newcommand{\fq}{{\frak q}}
\newcommand{\fv}{{\frak v}}
\newcommand{\pbra}[1]{\left(#1\right)}
\newcommand{\bbra}[1]{\left[#1\right]}

\newcommand{\dbra}[1]{[\![ #1 ]\!] }
\newcommand{\rnum}[1]{\lowercase\expandafter{\romannumeral #1\relax}}
\newcommand{\Rnum}[1]{\uppercase\expandafter{\romannumeral #1\relax}}
\newcolumntype{x}[1]{>{\centering\arraybackslash}p{#1}}

\def\Tr{{\rm Tr \,}}

\def\eg{{\textit{e.g.}}}
\def\ie{{\textit{i.e.}}}

\def\ft{\frak{t}} 

\def\frak{\mathfrak}

\def\tilde{\widetilde}
\def\hat{\widehat}
\def\bar{\overline}

\def\CA{{\mathcal A}}
\def\CB{{\mathcal B}}
\def\CC{{\mathcal C}}
\def\CE{{\mathcal E}}
\def\CF{{\mathcal F}}

\def\CH{{\mathcal H}}
\def\CI{{\mathcal I}}

\def\CL{{\mathcal L}}
\def\CM{{\mathcal M}}
\def\CN{{\mathcal N}}
\def\CO{{\mathcal O}}
\def\CP{{\mathcal P}}

\def\CS{{\mathcal S}}
\def\CT{{\mathcal T}}

\def\CW{{\mathcal W}}

\renewcommand{\bar}{\overline}
\renewcommand{\hat}{\widehat}

\def\^{{\wedge}}
\def\*{{\star}}

\newcommand{\beq}{\begin{equation}\begin{aligned}}
\newcommand{\eeq}{\end{aligned}\end{equation}}

\def\CD{{\mathcal D}}

\title{Argyres-Douglas Theories, Chiral Algebras and Wild Hitchin Characters}

\author[a]{Laura Fredrickson,}
\author[b,c]{Du Pei,}
\author[d,e,f]{Wenbin Yan\footnote{Primary affiliation: Yau Mathematical Sciences Center, Tsinghua University, Beijing, China},}
\author[b]{and Ke Ye}

\affiliation[a]{Department of Mathematics, Stanford University, Stanford, CA 94305, USA}

\affiliation[b]{Walter Burke Institute for Theoretical Physics,
California Institute of Technology, \\Pasadena, CA 91125, USA}

\affiliation[c]{Center for Quantum Geometry of Moduli Spaces, Department of Mathematics, University of Aarhus, \\DK-8000, Denmark}

\affiliation[d]{Yau Mathematical Sciences Center, Tsinghua University, Haidian District, Beijing 100094, China}

\affiliation[e]{Center for Mathematical Sciences and Applications, Harvard University, \\Cambridge, MA 02138, USA}

\affiliation[f]{Jefferson Physical Laboratory, Harvard University, Cambridge, MA 02138, USA}

\emailAdd{lfredrickson@stanford.edu}
\emailAdd{pei@caltech.edu}
\emailAdd{wbyan@cmsa.fas.harvard.edu}
\emailAdd{kye@caltech.edu}

\abstract{We use Coulomb branch indices of Argyres-Douglas theories on $S^1 \times L(k,1)$ to quantize moduli spaces $\CM_H$ of wild/irregular Hitchin systems. In particular, we obtain formulae for the ``wild Hitchin characters'' --- the graded dimensions of the Hilbert spaces from quantization --- for four infinite families of $\CM_H$, giving access to many interesting geometric and topological data of these moduli spaces. We observe that the wild Hitchin characters can always be written as a sum over fixed points in ${\cal M}_H$ under the $U(1)$ Hitchin action, and a limit of them can be identified with matrix elements of the modular transform $ST^kS$ in certain two-dimensional chiral algebras. 
 Although naturally fitting into the geometric Langlands program, the appearance of chiral algebras, which was known previously to be associated with Schur operators but not Coulomb branch operators, is somewhat surprising. 
\\
\\
\\
\\
\\
{\tt CALT-TH-2016-038}\\

}

\begin{document}

\maketitle

\section{Introduction}\label{Sec 1: intro}

Since their invention, string theory and supersymmetric quantum field theory have been ``unreasonably effective'' in predicting new structures in mathematics. In \cite{Gukov:2016lki}, another relation was proposed, linking the quantization of a large class of hyper-K\"ahler manifolds and BPS spectra of superconformal theories,\footnote{See (2.26) of \cite{Gukov:2016lki}. We have stated the proposal here at the categorified level.}
\be\label{CBIQuant}
\boxed{\genfrac{}{}{0pt}{}{\text{Space of Coulomb BPS states of}}{\text{4d $\CN=2$ SCFT $\CT$ on $L(k,1)$}}} \quad {=} \quad \boxed{\genfrac{}{}{0pt}{}{\text{Hilbert space from }}{\text{quantization of $({^L}\!{\CM}_{\CT},k\omega_I)$}}}\ .
\ee
Here, the hyper-K\"ahler space ${^L}{\CM}_{\CT}$ is the mirror of the Coulomb branch $\CM_{\CT}$ of $\CT$ on $\R^3\times S^1$, with $\omega_{I}$ being one of the three real symplectic structures, and ``Coulomb BPS states'' refer to those which contribute to the superconformal index in the Coulomb branch limit \cite{Gadde:2011uv}. Each side of \eqref{CBIQuant} admits a natural grading, coming from the $U(1)_r\subset SU(2)_R\times U(1)_r$ R-symmetry of the 4d $\CN=2$ SCFT, and the proposal \eqref{CBIQuant} is a highly non-trivial isomorphism between two graded vector spaces.

This relation was studied in \cite{Gukov:2016lki} for theories of class $\CS$ \cite{Gaiotto:2009we,Gaiotto:2009hg}. For a given Riemann surface $\Sigma$, possibly with regular singularities (or ``tame ramifications''), and a compact simple Lie group $G$, the Coulomb branch $\CM_{\CT}$ of the theory $T[\Sigma,G]$ compactified on $S^1$ is the Hitchin moduli spaces $\CM_H(\Sigma,G)$ \cite{Bershadsky:1995vm, Harvey:1995tg, Seiberg:1996nz}, whose mirror ${^L}{\CM}_{\CT}$ is given by $\CM_H(\Sigma,^L\!G)$ associated with the Langlands dual group $^L\!G$ via the geometric Langlands correspondence \cite{beilinson1991quantization, hausel2003mirror, Kapustin:2006pk, Gukov:2006jk}, and the $U(1)_r$ action on it becomes the so-called Hitchin action \cite{hitchin1987self}. Quantizing the Hitchin moduli space gives the Hilbert space of complex Chern-Simons theory $\CH(\Sigma,^L\!G_\C;k)$, whose graded dimension --- the \emph{Hitchin character}\footnote{The graded dimension (see \eqref{graded}) is the same as the character of the $U(1)$ Hitchin action, lifted from $\CM_H$ to acting on $\CH$, and hence the name ``Hitchin character.''} --- is given by the ``equivariant Verlinde formula'' proposed in \cite{Gukov:2015sna} and later proved in \cite{Andersen:2016hoj, 2016arXiv160801754H}. The authors of \cite{Gukov:2016lki} have verified relation \eqref{CBIQuant} by matching the lens space Coulomb index of class $\CS$ theories and the Hitchin characters, 
\be
{\cal I}_{\rm Coulomb} (T[\Sigma, G]; L(k,1)\times S^1)= \dim_{\ft} \CH(\Sigma, {^L}G_{\mathbb{C}}; k).
\label{EVF=CBI}
\ee

In the present paper, we further explore the connection in \eqref{CBIQuant} for a wider class of 4d $\CN=2$ theories including the $A_1$ Argyres-Douglas (AD) theories. In the process, we introduce another player into the story, making \eqref{CBIQuant} a triangle,
\be\label{Triangle}
\begin{array}{rcl}
\text{Coulomb index of $\CT$} & \longleftrightarrow & \text{quantization of ${^L}{\CM}_{\CT}$}
\\
\\
$\rotatebox[origin=c]{-45}{$\longleftrightarrow$}$   &  & $\rotatebox[origin=c]{45}{$\longleftrightarrow$} $ \\
\\
& \text{chiral algebra $\chi_{\CT}$}
&
\end{array}
\ee
where the chiral algebra $\chi_{\CT}$ is associated with the 4d $\CN=2$ theory $\CT$ \`a la \cite{Beem:2013sza}. We observe that fixed points on ${\CM}_{\CT}$ under $U(1)_r$ are in bijection with highest-weight representations of $\chi_{\CT}$,\footnote{In the physics literature --- and also in this paper --- ``chiral algebra'' and ``vertex operator algebra'' (VOA) are often used interchangeably, while in the math literature, the two have different emphasis on, respectively, geometry and representation theory. The ``highest-weight representations of $\chi_\CT$'' here denotes a suitable subcategory, closed under modular transform, of the full category of modules of vertex operator algebra. The precise statement will be clear in Section \ref{Sec: CA}.} and in addition the $\ft \rightarrow \exp(2\pi i)$ limit of the Hitchin character can be expressed in terms modular transformation matrix of those representations. The appearance of the chiral algebra is anticipated from the geometric Langlands program, as the triangle above can be understood as an analogue of the ``geometric Langlands triangle'' formed by A-model, B-model and $\CD$-modules for general $\CM_\CT$. However, the role of 
the chiral algebra $\chi_\CT$  in the counting of \emph{Coulomb} branch BPS states is somewhat unexpected, since the chiral algebra is related to the Schur operators of $\CT$ \cite{Beem:2013sza, Beem:2014rza, Lemos:2014lua, Cordova:2015nma}, which contains the Higgs branch operators but not the Coulomb branch operators at all! The current paper shows that, the Coulomb branch index is related to $\chi_{\CT}$ through modular transformations.

Argyres-Douglas theories form a class of very interesting 4d $\CN = 2$ strongly-interacting, ``non-Lagrangian'' SCFTs. They were originally discovered by studying singular loci in the Coulomb branch of $\CN = 2$ gauge theories \cite{Argyres:1995jj, Argyres:1995xn, Eguchi:1996vu}, where mutually non-local dyons become simultaneously massless. The hallmarks of this class of theories are the fixed values of coupling constants and the fractional scaling dimensions of their Coulomb branch operators. Like the class $\CS$ theories, Argyres-Douglas theories can also be engineered by compactifying M5-branes on a \textit{Riemann sphere} $\Sigma = \mathbb{C}\mathbf{P}^1$, but now with irregular singularities --- or ``wild ramifications'' \cite{Bonelli:2011aa, Xie:2012hs, Wang:2015mra}.  Their Coulomb branch $\CM_H(\Sigma,G)$ on $\R^3\times S^1$ and their mirrors $\CM_H(\Sigma,^LG)$ are sometimes called \textit{wild Hitchin moduli spaces}. The study of these spaces and their role in the geometric Langlands correspondence (see \eg~\cite{Witten:2007td} and references therein) is a very interesting subject and under active development. Over the past few years, much effort has been made to give a precise definition of the moduli space, and analogues for many well-known theorems in the unramified or tamely ramified cases were only established recently (see \cite{biquard2004wild, boalch2015twisted, boalch2011geometry}, as well as the short survey \cite{boalch2012hyperkahler} and references therein). In this paper, relation \eqref{CBIQuant} enables us to obtain the \emph{wild Hitchin characters} for many moduli spaces. Just like their cousins in the unramified or tamely ramified cases \cite{Gukov:2015sna}, wild Hitchin characters encode rich algebraic and geometric information about $\CM_H$, with some of the invariants $\CM_H$ being able to be directly read off from the formulae. This enables us to make concrete predictions about the moduli space. 

For instance, the $L(k,1)$ Coulomb index of the original Argyres-Douglas theory \cite{Argyres:1995jj}, which in the notation of \cite{Xie:2012hs} will be called the $(A_1,A_2)$ theory, is given by
\be
\CI_{(A_1,A_2)} = \frac{1-\ft^{-\frac{1}{5}} - \ft^{\frac{1}{5}}+ \ft^{\frac{k}{5}}}{(1-\ft^{\frac{6}{5}})(1-\ft^{-\frac{1}{5}})},
\ee
and it is easy to verify that it agrees with the wild Hitchin character of the mirror of the Coulomb branch ${^L}\CM_{(A_1,A_2)}={^L}\CM_{2,3}$ (the precise meaning of this notation will be clarified shortly), 
\be
\dim_{\ft}\CH({^L} \CM_{2,3}) & = \frac{1}{(1-\ft^{\frac{2}{5}})(1-\ft^{\frac{3}{5}})}+\frac{\ft^{\frac{k}{5}}}{(1-\ft^{\frac{6}{5}})(1-\ft^{-\frac{1}{5}})},
\ee
with the two terms coming from the two $U(1)$ fixed points. And the two fixed points correspond to the two highest weight representations of the non-unitary $(2,5)$ Virasoro minimal model --- famously known as the Lee-Yang model --- via a detailed dictionary which will be provided in later sections.

This paper is organized as follows: In Section~\ref{Sec 2: review}, we first briefly recall how the wild Hitchin moduli space $\CM_H$ arises from brane geometry and how it is related to general Argyres-Douglas theories. We then proceed to describe $\CM_H$, introduce the $U(1)$ Hitchin action on it and discuss its geometric quantization.

In Section~\ref{Sec 3: CBI}, we obtain the Coulomb branch indices of Argyres-Douglas theories, expressed as integral formulae. We follow the prescription in \cite{Maruyoshi:2016tqk, Maruyoshi:2016aim, Agarwal:2016pjo} by starting with $\CN = 1$ Lagrangian theories that flow to Argyres-Douglas theories in the IR. The TQFT structure for the index is presented in Appendix~\ref{sec:TQFT}.

In Section~\ref{Sec 4: Verlinde}, we present the wild Hitchin characters, decomposed into summations over the fixed points. Using the character formulae we explore the geometric properties of the moduli space. Confirmation from direct mathematical computation is given in Appendix~\ref{sec: fixed}. We then study the large-$k$ limits of the wild Hitchin characters, giving a physical interpretation of some fixed points in $\CM_H$ as massive vacua on the Higgs branch of the 3d mirror theory. We also study the symmetry mixing upon dimensional reduction, following \cite{Buican:2015hsa}. Further details are given in Appendix~\ref{sec: 3d red} and~\ref{app: Massive}.  

In Section~\ref{Sec: CA}, we study the relation between Hitchin characters and chiral algebras, and demonstrate that a limit of wild Hitchin characters can be identified with matrix elements of the modular transformation $ST^kS$. Further, we check the correspondence between the fixed points on $\CM_H$ and the highest-weight modules for various examples.

\section{Wild Hitchin moduli space and Argyres-Douglas theories}\label{Sec 2: review}

Embedding a given physical or mathematical problem into string theory usually leads to new insights and generalizations. In \cite{Gukov:2015sna, Gukov:2016lki}, the problem of quantizing the Hitchin moduli space was studied using the following brane set-up
\beq
\begin{matrix}
{\mbox {\textrm{fivebranes:}}} & \qquad & L(k,1)\times S^1& \times & \Sigma &  & \\
& \qquad &   &  & \cap \\
{\mbox{\rm space-time:}} & \qquad & L(k,1)\times S^1 & \times & T^* \Sigma &  &  & \times & \R^3\\
& \qquad &  \!\!\!\!\!\!\!\!\!\!\!\!\circlearrowright &  &\!\!\!\!\!\!\! \circlearrowright &  &  &  &  \circlearrowright \\
{\mbox{\rm symmetries:}} & \qquad & \!\!\!\!\!\! SO(4)_E &   & \!\!\!\! U(1)_N &   &   &   & SU(2)_R
\end{matrix}
\label{IndGeo}
\eeq
We will first review how the Hitchin moduli space arises from this geometry, and how adding irregular singularities to $\Sigma$ leads to a relation between the general Argyres-Douglas theories and wild Hitchin systems.  

\subsection{Hitchin equations from six dimensions}
\label{sec:2.1}
Hitchin moduli spaces were first introduced to physics in the context of string theory and its dimensional reduction in the pioneering work of \cite{Bershadsky:1995vm, Harvey:1995tg, Seiberg:1996nz} in the past century, and were highlighted in the gauge theory approach to geometric Langlands program \cite{Kapustin:2006pk, Gukov:2006jk, Witten:2007td}. In our brane setting \eqref{IndGeo}, which is closely related to the system studied in detail in \cite{Gaiotto:2009hg}, one can first reduce the M5-branes on the $S^1$ to obtain D4-branes, whose world-volume theory is given by the 5d $\CN=2$ super-Yang-Mills theory. We consider theories with gauge group $G$ of type ADE. In addition to the gauge fields, this theory also contains five real scalars $Y^I$ with $I = 1, 2, \dots, 5$, corresponding to the motion of the branes in the five transverse directions. Further topological twisting along $\Sigma$ enables us to identify $\varphi(z) = Y^1 + i Y^2$ as a $(1,0)$-form on $\Sigma$ with respect to the complex structure of $\Sigma$. As a consequence, the BPS equations in the remaining three space-time dimensions are precisely the Hitchin equations \cite{hitchin1987self},
\be
& F_A + \left[\varphi, \varphi^{\dagger}\right] = 0,\\[0.5em]
& {\bar {\partial}}_A \varphi = 0.
\label{HitchinEqs}
\ee
Here $F_A$ is the curvature two-form of $A = A_z dz + A_{\bar z} d{\bar z}$ valued in the adjoint bundle of the principle $G$-bundle $P$, and ${\bar {\partial}}_A$ is the $(0,1)$ part of the covariant derivative $d_A$. We will call $\varphi \in \Gamma(\Sigma, {\rm ad}(P) \otimes_\C K)$ the \textit{Higgs field} following standard mathematical nomenclature. Regarded as a sigma model, the target space of the three-dimensional theory is identified with the Hitchin moduli space $\CM_H(\Sigma, G)$ --- solutions to the Hitchin equations modulo gauge transformations.  

One can allow the Riemann surface $\Sigma$ to have a finite number of marked points $\{ p_1, p_2, \dots, p_s \}$ for $s \geq 0$. In the neighborhood of each marked point $p_i$, the gauge connection and the Higgs field take the asymptotic form:
\be
& A \sim \alpha d \theta,\\[0.5em]
& \varphi \sim \left(\frac{u_n}{z^n} + \frac{u_{n-1}}{z^{n-1}} + \dots \frac{u_1}{z} + \text{regular}\right)dz.
\label{HiggsSing}
\ee
Here $\alpha\in\frak{g}$ and $u_i\in \frak{g}_\C$ are collectively called the \textit{ramification data},\footnote{We use the convention that elements in $\frak{g}=\mathrm{Lie}\,G$ are anti-Hermitian.} and they are fixed in defining $\CM_H$ to ensure that the moduli space is symplectic (more precisely, gauge-invariant combinations of them are fixed). When the order of the pole is $n = 1$, we call the puncture \textit{tame} or \textit{regular}. From the M-theory geometry, adding a regular puncture corresponds to the insertion of a set of defect M5-branes placed at the point $p_i$ of $\Sigma$, occupying the four spacetime dimensions as well as the cotangent space at $p_i\in \Sigma$. Set-up \eqref{IndGeo} becomes
\beq
\begin{matrix}
{\mbox{\textrm{fivebranes:}}}& L(k,1)_b & \times & \Sigma & \times &S^1 \\
&   &  & \cap \\
{\mbox{\rm space-time:}}& L(k,1)_b & \times & T^* \Sigma & \times & S^1 & \times & \R^3\\
&   &  & \cup \\
{\mbox{\textrm{``defect'' fivebranes:}}} & L(k,1)_b &\times& T^*|_{p_i}\Sigma & \times & S^1 &.
\end{matrix}
\label{3d3dDefects}
\eeq
The defect fivebranes give rise to a codimension-two singularity in the 6d (2,0) theory and introduce a flavor symmetry of the effective 4d theory $T[\Sigma, G]$ \cite{Gaiotto:2009we, Chacaltana:2012zy}. If $u_1$ is nilpotent, then the flavor symmetry is given by the commutant subgroup of the nilpotent embedding $\mathfrak{su}(2) \rightarrow \mathfrak{g}$; if $u_1$ is semi-simple, the flavor symmetry is explicitly broken by mass deformations \cite{Gaiotto:2009we, Nanopoulos:2009uw}. The ramification data $\alpha$ and $u_1$ is acted upon by the affine Weyl group of $G$, and the conjugacy class of the monodromy in the complexified gauge connection $\CA_z = A_z +i \varphi$ is an invariant of the ramification data.

When $n > 1$ the puncture will be called \textit{wild} or \textit{irregular}, which will play a central role in the present paper. The leading coefficient matrix $u_n$ is allowed to be either semi-simple or nilpotent as in the tame case. However, now the monodromy of $\CA_z$ around $p_i$ needs to be supplemented by more sophisticated data --- the Stokes matrices --- to fully characterize the irregular puncture \cite{wasow2002asymptotic} (see \eg~\cite{Witten:2007td} for more detail and explicit examples).

The Hitchin moduli space $\CM_H(\Sigma, G)$ with fixed local ramification data is hyper-K\"ahler, admitting a family of complex structures parametrized by an entire $\mathbb{C}\mathbf{P}^1$. There are three distinguished ones $(I, J, K)$, and the corresponding symplectic forms are denoted as $\omega_I, \omega_J, \omega_K$. The complex structure $I$ is inherited from the complex structure of the Riemann surface $\Sigma$, over which ${\bar \partial}_A$ defines a holomorphic structure on $E$, and the triple $(E, {\bar \partial}_A, \varphi)$ parametrizes a \textit{Higgs bundle} on $\Sigma$. This is usually referred to as the \emph{holomorphic} or \emph{algebraic} perspective. Alternatively, one can also employ the \emph{differential geometric} point of view, identifying $\CM_H$ as the moduli space of flat $G_\C$-connections on $\Sigma$ with the prescribed singularity near the puncture, and the complex structure $J$ comes from the complex structure of $G_\C$. There is also the \emph{topological} perspective, viewing $\CM_H$ as the character variety $\mathrm{Hom}(\pi_1\Sigma,G_\C)$, with  boundary holonomies in given conjugacy classes (and with inclusion of Stokes matrices in the wildly ramified case). Non-abelian Hodge theory states that the three constructions give canonically isomorphic moduli spaces \cite{hitchin1987self,simpson1988constructing, donaldson1987twisted, corlette1988flat}. In the wild case, the isomorphism between the Hitchin moduli space $\CM_H$ and moduli space of flat $G_\C$-connections was proved in \cite{Sabbah99, biquard2004wild}, while \cite{biquard2004wild} proved the isomorphism between $\CM_H$ and moduli space of Higgs bundles, thus establishing the equivalence of first two perspectives. The wild character variety was later constructed and studied in \cite{Boalch:2002cn, Boalch:2013lqa, boalch2011geometry, boalch2015twisted}. In this paper, we will mainly adopt the holomorphic perspective but will occasionally switch between the three viewpoints as each offers unique insights into $\CM_H$.\footnote{In general, physical quantities know about the full moduli \emph{stack}, where all Higgs bundles including the \emph{unstable} ones are taken into account, as the path integral sums over all configurations. However, for co-dimension reasons, all wild Hitchin characters we will consider are the same for stacks and for spaces. In the tame or unramified cases, there can be differences, and working over the stack is usually preferable. See \cite[Sec.~5]{Andersen:2016hoj} for more details. }

For later convenience, we shall use below a different but equivalent formulation of Hitchin equations \eqref{HitchinEqs}. Fix a Riemann surface $\Sigma$ and a complex vector bundle $E$.
Given a  Higgs bundle $(\delbar_E, \varphi)$, \ie~a holomorphic structure on $E$ and a Higgs field,
we additionally equip $E$ with a Hermitian metric $h$.
Then there exists a unique \textit{Chern connection} $D$ compatible with the Hermitian metric whose $(0,1)$ part coincides with ${\bar \partial}_E$. The Hitchin equations are then equations
for the Hermitian metric $h$:
\be
& F_D + \left[ \varphi, \varphi^{\dagger_h} \right] = 0,\\[0.5em]
& {\bar \partial}_E\, \varphi = 0
\label{Hitchin2}
\ee
where $\varphi^{\dagger_h} = h^{-1} \varphi^{\dagger} h$ is the Hermitian conjugation of the Higgs field. The previous version of Hitchin equations, \eqref{HitchinEqs}, is in the ``unitary gauge" where the Hermitian metric is identity. The two conventions are related by a gauge transformation $g \in G_{\mathbb{C}}$ such that
\be
g^{-1} \circ {\bar \partial}_E \circ g = {\bar \partial}_{A_u}, \ \ \ \ g^{-1} \cdot \varphi \cdot g = \varphi_{u}, \ \ \ \ g^{\dagger} \cdot h \cdot g = {\rm Id}_u
\ee
where the subscript $u$ indicates unitary gauge.

The moduli space $\CM_H$ admits a natural map known as the Hitchin fibration \cite{hitchin1987stable},
\be
\CM_H & \rightarrow \CB,\\[0.5em]
(E, \varphi) & \mapsto \det(x dz - \varphi),
\ee
where $\CB$ is commonly referred to as the \emph{Hitchin base} and generic fibers are abelian varieties. As explained in \cite{Gaiotto:2009hg}, $\CB$ can be identified with the Coulomb branch of the theory $T[\Sigma, G]$ on $\mathbb{R}^4$, and the curve $ \det(x dz - \varphi) = 0$ with the Seiberg-Witten curve
of $T[\Sigma, G]$. 

\subsubsection*{The Hitchin action}
There is a $U(1)$ action on the Hitchin moduli space $\CM_H$.
As emphasized in \cite{Gukov:2015sna, Gukov:2016lki}, the existence of the $U(1)$ Hitchin action gives us control over the infinite-dimensional Hilbert space arising from quantizing $\CM_H$ in both the unramified or tamely ramified case,\footnote{Occasionally, it is also useful to talk about the complexified $\mathbb{C}^*$-action, and we will refer to both as the ``Hitchin action.''} and we will also focus in this paper on the wild Hitchin moduli spaces $\CM_H$ that admit similar $U(1)$ actions. 

We first recall that in the unramified case, the Hitchin action on the moduli space is given by
\be
(A, \varphi) \mapsto (A, e^{i \theta} \varphi).
\label{circleAction}
\ee
On the physics side, it coincides with the $U(1)_r$ symmetry of the 4d $\CN=2$ SCFT $T[\Sigma,G]$. A similar action   also exists for $\Sigma$ with tame ramifications, provided the singularities are given by
\be
& A \sim \alpha d \theta,\\[0.5em]
& \varphi \sim \text{nilpotent}.
\ee
However, near an irregular singularity, $\varphi$ acquires an higher order pole \eqref{HiggsSing} and the action \eqref{circleAction} has to rotate the $u_i$'s. As the definition of the $\CM_H$ depends on ramification data, this $U(1)$ action does not act on the moduli space --- it will transform it into different ones. One can attempt to partially avoid this problem by setting $u_1,u_2,\ldots,u_{n-1}$ to be zero\footnote{More generally, we should choose their values such that the $U(1)$-action on them can be cancelled by gauge tranformations.} --- similar to the case with tame ramifications --- but $u_n$ has to be non-zero in order for the singularity to be irregular. 

The way out is to modify \eqref{circleAction} such that it also rotates the $z$ coordinate by, \eg,
\be
z \mapsto e^{\frac{i\theta}{n-1}}z.
\ee
To have this action well-defined globally on $\Sigma$ highly constrains the topology of the Riemann surface, only allowing $\mathbb{C}\mathbf{P}^1$ with one wild singularity, or one wild and one tame singularities.\footnote{We will focus on such $\Sigma$ and the moduli spaces $\CM_H$ associated with them. Henceforth, by ``wild Hitchin moduli space'', we will be usually referring to these particular $\CM_H$, where the $U(1)$ action exists.} Interestingly, the $U(1)$ Hitchin action on $\CM_H$ exists whenever $T[\Sigma,G]$ is superconformal,
\be
\boxed{\genfrac{}{}{0pt}{}{\text{$\CM_H(\Sigma,G)$ admits}}{\text{$U(1)$ Hitchin action}} } \longleftrightarrow \boxed{\genfrac{}{}{0pt}{}{\text{$T[\Sigma,G]$ is a}}{\text{4d $\CN=2$ SCFT}}}\ .
\ee
This is because superconformal invariance for $T[\Sigma,G]$ implies the existence of $U(1)_r$ symmetry which define a $U(1)$ action on $\CM_H$. All possible choices for wild punctures of ADE type on the Riemann sphere are classified in \cite{Xie:2012hs, Wang:2015mra}, and the resulting theories $T[\Sigma, G]$ are called ``general Argyres-Douglas theories'', which we will review in the next subsection. In Section~\ref{sec: Wild moduli}, we will get back to geometry again to give a definition of the wild Hitchin moduli space and describe more precisely the $U(1)$ action on it.

\subsection{General Argyres-Douglas theories}\label{sec: AD}

In this section we take $G = SU(2)$, and moreover assume that the irregular singularity lies at $z = \infty$ (the north pole) on the Riemann sphere. Another regular puncture can also be added at $z = 0$ (the south pole). 

Near $z=\infty$, there can be two types of singular behaviors for the Higgs field $\varphi$; the leading coefficient can be either semisimple or nilpotent.\footnote{If the leading coefficient is not nilpotent, it can always be made semisimple by a gauge transformation. Also, notice that an semisimple element of $\frak{sl}(2,\C)$ is automatically regular. } A semisimple pole looks like
\be
\varphi(z) \sim z^{n-2}dz \left( \begin{array}{cc} a & 0\\ 0 & -a \end{array} \right) + \cdots
\label{SingularPhiInf}
\ee
with $n > 1$ an integer. For a nilpotent pole, it cannot be cast into this form by usual gauge transformations. But if we are allowed to use a local gauge transformation that has a branch cut on $\Sigma$, we can still diagonalize it into \eqref{SingularPhiInf}, but now with $n\in \mathbb{Z}+1/2$. We will not allow such gauge transformation globally in the definition of the moduli space $\CM_H$ since it creates extra poles at $z = 0$, but \eqref{SingularPhiInf} is still useful conceptually in local classifications. For example, one can read off the correct $U(1)$ action on $z$, 
\be
z \mapsto e^{-\frac{i\theta}{n-1}}.
\ee

In \cite{Xie:2012hs}, a puncture is called type \Rnum{1} if $n$ is integral, and  type \Rnum{2} if $n$ half-odd. We will use the notation $I_{2, K}$ for the singularity with $K = 2(n-2)$ and the subscript ``2" is referring to the $SL(2,\C)$ gauge group.

\subsubsection*{The $(A_1,A_{K-1})$ series}

If there is only one irregular singularity $I_{2, K}$ at the north pole, \eqref{SingularPhiInf} will only have non-negative powers of $z$. This kind of solution describes the $(A_1, A_{K-1})$ Argyres-Douglas theory in the notation of \cite{Xie:2012hs}. Historically, this class of theories was discovered from the maximally singular point on the Coulomb branch of $\CN = 2$ $SU(K)$ pure Yang-Mills theory \cite{Argyres:1995jj, Eguchi:1996vu}. The Seiberg-Witten curve (or the spectral curve from the Higgs bundle point of view) takes the form
\be
x^2 = z^K + v_2 z^{K-2} + \dots + v_{K-1} z + v_K.
\label{AN-1SW}
\ee

The Seiberg-Witten differential $\lambda = x dz$ has scaling dimension $1$, from which we can derive the scaling dimensions for $v_i$,
\be
\left[ v_i \right] = \frac{2i}{K+2}.
\label{uScaling}
\ee
For $i >  (K+2) / 2$, the scaling dimensions of the $v_i$'s are greater than $1$, and they are the expectation values of Coulomb branch operators. When $K$ is even, there is a mass parameter at $i = (K+2)/2$. The rest with $i <  (K+2) / 2$ are the coupling constants that give rise to $\CN = 2$ preserving deformations
\be
\Delta W \sim v_i \int d^4 x\, {\tilde Q}^4 \CO_i
\ee
for Coulomb branch operator $\CO_i$ associated to $v_{K+2-i}$, where ${\tilde Q}^4$ denotes the product of the four supercharges that do not annihilate $\CO_i$. Such deformation terms are also consistent with the pairing $\left[ v_i \right] + \left[ v_{K+2-i} \right] =2$.  If we promote all the couplings to the background chiral superfields, one can assign a $U(1)_r$ charge to them, which is equal to their scaling dimensions.\footnote{Our convention here for the $U(1)_r$ charge differs from the usual one as $r_{\rm usual} = - r$. In our convention, $U(1)_r$ charge  for chiral BPS operators will be the same as scaling dimensions. Notice that one can formally assign $U(1)_r$ charge to $z$ as well; the value will turn out to be minus the scaling dimension $-\bbra{z}$.}

The coupling constants and mass term parametrize deformations of $\CM_H$, thus not all $v_i$'s are part of the moduli. Moreover, to have a genuine $U(1)$ action on $\CM_H$ itself, the $v_i$'s with $i  \leq (K+2)/2$ ought to be set zero in the spectral curve in \eqref{AN-1SW}. On the other hand, those $v_i$'s with $i > (K+2) / 2$ are allowed to be non-zero, and in fact they parametrize the Hitchin base $\CB$. In what follows we denote this wild Hitchin moduli space as $\CM_{2, K}$, and its Langlands dual as ${^L}\CM_{2, K}$. The parameter $a$ in \eqref{SingularPhiInf} can be scaled away but the parameter $\alpha\in {\rm Lie} (\mathbb{T})$ corresponding to the monodromy of the gauge connection at the singularity enters as part of the definition of the moduli space $\CM_{2,K}(\alpha)$. As argued in \cite[Sec.~6]{Witten:2007td}, this monodromy has to vanish for odd $K$, but can be non-zero when $K$ is even.\footnote{Had the puncture been tame, such monodromy would be required to the zero to have a non-empty moduli space. However, in the wild case, due to Stokes phenomenon, $\alpha$ can take non-zero values. Now, $e^\alpha$ is a ``formal monodromy,'' and the real monodromy, which is required to be the identity, is a product of $e^\alpha$ with Stokes matrices.} On the physics side, this agrees with the fact that the $(A_1, A_{K-1})$ theory has no flavor symmetry when $K$ is odd, and generically a $U(1)$ symmetry when $K$ is even \cite{Argyres:2012fu}. This phenomenon is quite general, and works in the case with tame ramifications as well,
\be
\boxed{\genfrac{}{}{0pt}{}{\text{Monodromy parameters}}{\text{for the moduli space $\CM_H(\Sigma)$} }} \longleftrightarrow \boxed{\genfrac{}{}{0pt}{}{\text{flavor symmetries}}{\text{for the theory $T[\Sigma]$}}}.
\ee

\subsubsection*{The $(A_1,D_{K+2})$ series}

If $\Sigma$ also has a regular puncture on the south pole in addition to the irregular $I_{2,K}$ at the north pole, we will get the $(A_1, D_{K+2})$ Argyres-Douglas theory in the notation of \cite{Xie:2012hs}. Originally, this class of theories was discovered at the ``maximal singular point'' on the Coulomb branch of the $SO(2K+4)$ super-Yang-Mills theory \cite{Eguchi:1996vu}. 

To accommodate the regular puncture, the Higgs field should behave as
\be
\varphi(z) \sim z^{n-2}dz \left( \begin{array}{cc} a & 0\\ 0 & -a \end{array} \right) + \cdots + \frac{dz}{z} \left( \begin{array}{cc} m & 0\\ 0 & -m \end{array} \right).
\label{DSingularPhiInf}
\ee
Consequently, the Seiberg-Witten curve is
\be
x^2 = z^K + v_1 z^{K-1} + \dots + v_{K-1} z + v_K + \frac{v_{K+1}}{z} + \frac{m^2}{z^2}
\label{DN+2SW}
\ee
with the same expression for the scaling dimensions in \eqref{uScaling} except that $i$ now takes value from $1$ up to $K+1$. The parameter $m$ has the scaling dimension of mass, and it is identified as a mass parameter for the $SU(2)$ flavor symmetry associated with the regular puncture. Once again, we will turn off all the coupling constants and masses in the spectral curve since they describe deformations of the Coulomb branch moduli. Around the irregular puncture, the monodromy parameter $\alpha_1\in {\rm Lie}(\mathbb{T})$ of the gauge connection $A$ can be non-trivial. Moreover, it may not agree with the monodromy $\alpha_2$ around the regular puncture. Similar to the $(A_1,A_{K-1})$ case, $\alpha_1 = 0$ when $K$ is odd, and can be turned on when $K$ is even.  The corresponding moduli spaces, denoted as $\tilde{\CM}_{2,K}(\alpha_1,\alpha_2)$, and their Langlands dual ${^L}{\widetilde{\cal M}}_{2, K}(\alpha_1,\alpha_2)$ depend on those $\alpha$'s. 
\subsection{Geometry of the wild Hitchin moduli space}\label{sec: Wild moduli}

\begin{table}
\centering
\begin{tabular}{|c|c|c|c|}
\hline
AD theory & order of pole of $\varphi$ at $z = \infty, 0$ & moduli space $\CM_H$ & $\dim_{\mathbb{C}} \CM_H$ \\
\hline
$(A_1, A_{2N})$ & $(2N+1)/2, 0$ & $\CM_{2, 2N+1}$ & $2N$\\
$(A_1, A_{2N-1})$ & $N, 0$ & $\CM_{2, 2N}$ & $2N-2$\\
$(A_1, D_{2N+1})$ & $(2N-1)/2, 1$ & ${\tilde \CM}_{2, 2N-1}$ & $2N$\\
$(A_1, D_{2N})$ & $N-1, 1$ & ${\tilde \CM}_{2, 2N-2}$ & $2N-2$\\ \hline
\end{tabular}
\caption{\label{table:moduli summary}Summary of $A_1$ Argyres-Douglas theories, the order of singularities of the Higgs fields, the corresponding wild Hitchin moduli spaces and their dimensions.}
\end{table}

We have argued that the wild Hitchin moduli space can be realized as the Coulomb branch vacua of certain Argyres-Douglas theories compactified on a circle. They are summarized in Table~\ref{table:moduli summary}. In accordance with the physics construction, we will now turn to a pure mathematical description of the moduli space.

A mathematical definition of these moduli spaces depends on the singular behavior of the Higgs field $\varphi$ near irregular singularities, as in \cite{biquard2004wild, Witten:2007td}. When $K$ is even, the moduli spaces $\mathcal{M}_{2,K}$ and $\widetilde{\CM}_{2,K}$ are described in \cite{biquard2004wild}. Consequently, we turn to the case where $K = 2N+1$ is odd. The corresponding Higgs bundle moduli space is described in \cite{Fredrickson-Neitzke}, and we here describe the corresponding Hitchin moduli space.  To motivate the definition of $\CM_{2,2N+1}$, note that in this case, the leading coefficient matrix \eqref{HiggsSing} is nilpotent, which slightly differs from that of \cite{biquard2004wild}. However, one can diagonalize the Higgs field near the irregular singularity by going to the double cover of the disk centered at infinity (a ``lift"), so that locally the Higgs field looks like
\be
\varphi \sim u'_N z^{N + \frac{1}{2} } + \dots
\ee
with $u'_N$ regular semi-simple. This polar part of the Higgs field is not single-valued, so we futher impose a gauge transformation across the branch cut \cite{Witten:2007td}
\be
g_e = \left( \begin{array}{cc} 0 & 1\\ -1 & 0 \end{array} \right)
\ee

In our definition of $\CM_{2,2N+1}$, the local picture at the infinity follows from an equivariant version of the local picture of \cite{biquard2004wild} on the ramified disk with respect to the $\mathbb{Z}_2$-change of coordinate $w \rightarrow - w$ for $w^2 = z$. The ramification ``untwists'' the twisted Cartan so the local model is still diagonal, as in \cite{biquard2004wild}.  

Two perspectives on solutions of Hitchin's equations appear  in Section~\ref{sec:2.1}, and we use both in the following definition. A solution of Hitchin equations is a triple of $(\delbar_E, \varphi, h)$ consisting of a holomorphic structure, Higgs field, and Hermitian metric satisfying \eqref{Hitchin2}. Alternatively, a solution of Hitchin equations
in unitary gauge (\ie~$h=\mathrm{Id}$) is a pair $(A, \varphi)$ consisting of a unitary connection $d_A$ and  Higgs field $\varphi$ satisfying \eqref{HitchinEqs}. We use the notation $\varphi$ for the Higgs field in both perspectives for simplicity.

Next we describe the relevant data needed to specify the moduli space $\CM_{2, 2N+1}$.

\bigskip

\noindent \emph{Fixed Data:} Take $\mathbb{C}\mathbf{P}^1$ with a marked point at $\infty$. Fix a complex vector bundle $E \rightarrow \mathbb{C}\mathbf{P}^1$ of degree $0$ with a trivialization of $\mathrm{Det} E$, the determinant bundle. Let $\delbar_E$ be a holomorphic structure on $E$ which induces a fixed holomorphic structure on $\mathrm{Det} E$. Let $h$ be a Hermitian metric on $E$ which induces a fixed Hermitian structure on $\mathrm{Det} E$.

At $\infty$, we allow an irregular singularity, and fix the following data:
\begin{equation}\label{eq:localmodel}
 D_{\mathrm{model}}=d + \varphi_{\mathrm{model}} + \varphi^\dagger_{\mathrm{model}}
\end{equation}
where 
\begin{equation}
 \varphi_{\mathrm{model}}=\begin{pmatrix} -2 & 0\\ 0 & 2 \end{pmatrix}
 \frac{\mathrm{d}u}{u^{K+3}}. 
\end{equation}
(To explain the power appearing, note that if $u$ is the holomorphic coordinate on the ramified double cover of the disk at $0$, \ie~$u^{-2}=z$, then $u^{-(2N+4)}\mathrm{d}u= z^{N+ \frac{1}{2}} \mathrm{d} z$.)

\bigskip

\bigskip

\noindent \emph{Definition of the moduli space, $\mathcal{M}_{2,2N+1}$:}
Given a triple $(\delbar_E, \varphi, h)$, denote the lift of the unitary pair $(A, \varphi)$ by
\be(\tilde{A}, \tilde{\varphi}) = l \cdot (A, \varphi). \ee
A triple $(\delbar_E, \varphi, h)$ is in $\mathcal{M}_{2,2N+1}$ if it is a solution of Hitchin equations on $\mathbb{C}\mathbf{P}^1$ and on a neighborhood of $\infty$ the associated flat connection $\tilde{D}= \tilde{A} + \tilde{\varphi} + \tilde{\varphi}^\dagger$ differs from the local model in \eqref{eq:localmodel} by a deformation allowed by \cite{biquard2004wild}. Moreover, we say that $(\delbar_E, \varphi, h)$ and $(\delbar_E', \varphi', h')$ are gauge equivalent if there is some unitary gauge transformation $g$ by which $(A, \varphi)$ and $(A', \varphi')$ are gauge equivalent, and $g$ lifts to an allowed gauge transformation on the ramified disk around $\infty$.  More precisely, the lift $\tilde{g}=l' \circ g \circ l^{-1}$ must be an allowed unitary gauge transformation,
in the perspective of \cite{biquard2004wild},
from $l \cdot(A, \varphi)$ to $l' \cdot(A',\varphi')$ on the ramified disk around $\infty$. 

The moduli space $\widetilde{\CM}_{2,2N-1}$ can be defined similarly.

\bigskip

With the above definitions, it is expected that the symplectic form $\omega_I$ on $\mathcal{M}_{2,K}$ and ${\widetilde \CM}_{2, K}$ can be expressed just as that in \cite{hitchin1987self}:
\be
\omega_I = \frac{i}{\pi} \int \Tr \pbra{\delta A_z \wedge \delta A_{\bar z} - \delta \varphi \wedge \delta \varphi^{\dagger}}.
\label{symplectic I}
\ee
There is a $U(1)$ action on the moduli space $\mathcal{M}_{2,K}$ and ${\widetilde \CM}_{2, K}$, by composing the rotation of 
Higgs field with a rotation of the Riemann sphere. It is defined as: 
\be
 & z\ \  {\xrightarrow{\rho_{\theta}}}\ \  e^{-i \frac{2}{2+K} \theta} z,\\[0.5em]
 & \varphi\ \rightarrow\ \ e^{i \theta} \rho_\theta^* \varphi,\\[0.5em]
 & A\ \rightarrow\ \  \rho_\theta^* A.
 \label{circleAction'}
\ee
We say $(A, \varphi)$ is fixed by the $U(1)$ action if for all $\theta$, the rotated solution is gauge equivalent to the unrotated one. This $U(1)$ action is expected to be Hamiltonian with moment map $\mu$ such that
\be
d \mu = \iota_V \omega_I
\ee
where $V$ is the vector field generated by the $U(1)$ action. At the fixed points of the $U(1)$ action, there is evidence that this moment map agrees with the following quantity \cite{Fredrickson-Neitzke}:
\be
\mu = \frac{i}{2\pi} \int \mathrm{Tr} \left( \varphi \wedge \varphi^{\dagger}
-\mathrm{Id}\cdot |z|^{K} dz \wedge d\zbar \right).
\label{moment Mu}
\ee

\bigskip
In Appendix \ref{sec: fixed}, we compute the weights of the $U(1)$ action at the fixed points. Practically, rather than working with the Hitchin moduli space, we may instead work with the Higgs bundle moduli space diffeomorphic to $\mathcal{M}_{2,K}$ or ${\widetilde \CM}_{2, K}$. In the case $\CM_{2,2N+1}$, the corresponding Higgs bundle moduli space $\CM_{2,2N+1}^{\mathrm{Higgs}}$ is rigorously described in \cite{Fredrickson-Neitzke}.  For the other moduli spaces, we provide a general set-up of the definition for the Higgs bundle moduli space, and leave a rigorous treatment to future work. Unsurprisingly, the fixed data for the Higgs bundle moduli space is the same as the fixed data for the Hitchin moduli space. On the ramified double cover of the disk at $\infty$ with coordinate $u=z^{-1/2}$, the local model for the Higgs field is
\begin{equation}
 \varphi_{\mathrm{model}}=\begin{pmatrix} -2 & 0\\ 0 & 2 \end{pmatrix}
\frac{\mathrm{d}u}{u^{K+3}}=\begin{pmatrix} 1 & 0\\ 0 & -1 \end{pmatrix}
 z^{K/2}\mathrm{d}z, 
\end{equation}
as in \eqref{eq:localmodel}.
Additionally, the monodromy at $\infty$ on the ramified double cover at $\infty$ is trivial when $K$ is odd, but otherwise a free parameter. The monodromy is algebraically encoded in the data of a filtration structure of the holomorphic vector bundle $\mathcal{E}=(E, \delbar_E)$ at $\infty$.
The filtered vector bundle of $\mathcal{E}$ and the filtration structure at $\infty$ are denoted as $\mathcal{P}_\bullet \mathcal{E}$.

A  pair $(\mathcal{P}_\bullet \mathcal{E}, \varphi)$ consisting a filtered bundle $\mathcal{P}_\bullet \mathcal{E}$ and meromorphic Higgs field $\varphi$ with pole at $\infty$ (with no additional compatibility conditions) is in the Higgs bundle moduli space $\mathcal{M}_{2,K}^{\mathrm{Higgs}}$ if there is a holomorphic lift to the ramified disk in which
$\psi^*(\mathcal{P}_\bullet \mathcal{E}, \varphi)$ is ``unramifiedly good'' (in the sense of  \cite{Mochizuki-wild}), 
\ie 
\begin{equation}
 \psi^*\varphi = \varphi_{\mathrm{model}} + \mbox{holomorphic terms}
\end{equation}
and $\psi^*(\mathcal{P}_\bullet \mathcal{E})$ is the trivial filtration. In $\mathcal{M}^{\mathrm{Higgs}}_{2,K}$, $(\mathcal{P}_\bullet \mathcal{E}, \varphi)$ and $(\mathcal{P}_\bullet \mathcal{E}', \varphi')$ are identified if there is a 
isomorphism $\eta:\mathcal{P}_\bullet \mathcal{E} \rightarrow \mathcal{P}_\bullet \mathcal{E}'$ of 
$\mathcal{P}_\bullet \mathcal{E}$ and $\mathcal{P}_\bullet \mathcal{E}'$ as filtered vector bundles such that $\varphi'= \eta^{-1} \circ \varphi \circ \eta$.

\subsection{Quantization of $\CM_H$}\label{sec: EqLoc}

One of the major goals of this paper is to study the quantization of wild Hitchin moduli spaces,
\be
\left(\CM_H(\Sigma,G),k\omega_I\right) \quad \leadsto \quad \CH(\Sigma,G,k).
\ee
The quantization problem takes as input the symplectic manifold $(\CM_H(\Sigma,G),k\omega_I)$ --- the ``phase space,'' and aims to produce a space of quantum states --- the ``Hilbert space.'' In this particular case, the resulting space $\CH(\Sigma,G,k)$ can be interpreted as the Hilbert space of complex Chern-Simons theory at real level $k$ on $\Sigma$, with the complex connection developing singularities near the punctures.

Using the standard machinery of geometric quantization of K\"ahler manifolds, one can identify the Hilbert space with holomorphic sections of a ``prequantum line bundle''
\be
\CH(\CM_H(\Sigma,G),k\omega_I)=H^0(\CM_H,\CL^{\otimes k}). 
\ee
Here $\CL$ denotes the determinant line bundle over $\CM_H$ whose curvature is cohomologous to $\omega_I$,
\be
c_1(\CL)=[\omega_I].
\ee

For all quantization problems, a very interesting question is to find the dimension of the resulting Hilbert space. In the present case, the dimension of $\CH$ can be formally written as an integral over $\CM_H$,\footnote{We use integrals for pedagogical reasons. $\CM_H$ generically is not a manifold, and should be viewed as a stack.}
\be\label{InfiniteH}
\dim\, H^0(\CM_H, \CL^{\otimes k}) =\chi(\CM_H, \CL^{\otimes k})
= \int_{\CM_H} e^{k \omega_I} \wedge \mathrm{Td} (\CM_H).
\ee
In the above expression, we used the vanishing of higher cohomology groups\footnote{The vanishing theorem for unramified and tamely ramified cases was proved in \cite{2016arXiv160801754H} and \cite{Andersen:2016hoj}, and the vanishing is expected to hold also in the wild case --- morally, because of the Kodaira vanishing along the fibers of the Hitchin map.} to rewrite the dimension as an Euler characteristic, and then used index theorem to express it as an integral over the moduli space.

Just like their unramified or tamely ramified cousins, the wild Hitchin moduli spaces are also non-compact and would give rise to infinite-dimensional Hilbert spaces after quantization. This is seen quite clearly from the integral in \eqref{InfiniteH}, which diverges due to the non-compactness of $\CM_H$. 

However, as the $U(1)$ Hitchin action is Hamitonian (in particular it preserves $\omega_I$), it also acts on the Hilbert space $\CH$. Then the dimension of $\CH$ can be refined to the \emph{graded} dimension, defined as the character of the $U(1)$ action,
\be
\dim_{\ft} \CH = \sum_n \dim \CH_n \ft^n.
\label{graded}
\ee
Here $\ft$ is the fundamental character of $U(1)$, and $\CH_n$ is the subspace of $\CH$ where $U(1)$ acts with eigenvalue $n$. In \cite{Gukov:2015sna}, this Hitchin character was computed in the unramified or tamely ramified case, and was found to be given by a Verlinde-like formula, known as the ``equivariant Verlinde formula.'' The word ``equivariant'' comes from the fact that the Hitchin character can also be written as an integral, similar to \eqref{InfiniteH}, but now in the $U(1)$-equivariant cohomology of $\CM_H$,
\be
\dim_{\ft} \CH(\Sigma, G,k) = \chi_{U(1)}(\CM_H, \CL^{\otimes k}) = \int_{\CM_H} e^{c_1(\CL^{\otimes k},\, \beta)} \wedge \mathrm{Td} (\CM_H, \beta).
\label{equivInd}
\ee
Here, the second quantity is the equivariant Euler characteristic of $\CL^{\otimes k}$ which is then expressed as an integral over $\CM_H$ via the equivariant index theorem. This integral will actually converge, but we will need to first briefly review the basics of equivariant cohomology and introduce necessary notation. We will be very concise and readers  unfamiliar with this subject may refer to \cite{alekseev2000notes} for a more pedagogical account. 

Let $V$ be the vector field on $\CM_H$ generated by the $U(1)$ action; we pick $\beta$ to be the degree-$2$ generator of the equivariant cohomology of $H^\bullet_{U(1)}(pt)$ and is related to $\ft$ by $\ft=e^{-\beta}$. Using the Cartan model for equivariant cohomology, we define the equivariant exterior derivative as
\be
\hat{\delta} = \delta + \beta \iota_V
\ee
with ${\hat \delta}^{\,2} = 0$ over equivariant differential forms. One can then define the equivariant cohomology as
\be
H^\bullet_G(\CM_H)=\mathrm{ker}\,{\hat \delta}/\mathrm{im}\,{\hat \delta}.
\ee
For an equivariant vector bundle, one can also define the equivariant characteristic classes. For example, the equivariant first Chern class of $\CL$ is now
\be
c_1(\CL,\, \beta)={\tilde \omega}_I := \omega_I - \beta \mu.
\ee
And one can verify that it is equivariantly closed
\be
\hat{\delta}\, {\tilde \omega}_I = 0.
\ee
Similarly, one can define the equivariant Todd class $\mathrm{Td} (\CM_H, \beta)$ of the tangent bundle of $\CM_H$. 

Now we can see that the integral in \eqref{equivInd} has a very good chance of being convergent as $e^{c_1(\CL,\, \beta)}$ contains a factor $e^{-\beta\mu}$ which suppresses the contribution from large Higgs fields. Further, one can use the Atiyah-Bott localization formula to write \eqref{equivInd} as a summation over fixed points of the Hitchin action,
\be
\int_{\CM_H} e^{c_1(\CL^{\otimes k},\, \beta)} \wedge \mathrm{Td} (\CM_H, \beta) = \sum_{F_d} e^{-\beta k \mu(F_d)} \int_{F_d} \frac{{\rm Td}(F_d) \wedge e^{k {\omega}_I}}{\prod_i^{\text{codim}_{\mathbb{C}}F_d} (1-e^{-x_i - \beta n_i})}
\ee
where $F_d$ is a component of the fixed points, and $x_i+\beta n_i$ are the equivariant Chern roots of the normal bundle of $F_d$ with $n_i$ being the eigenvalues under the $U(1)$ action. For a Hitchin moduli space, there is finitely many $F_d$'s and each of them is compact, so the localization formula provides a way to compute the Hitchin character. To use the above expression, one must understand the fixed points and their ambient geometry --- something that is typically challenging. This makes the relation \eqref{CBIQuant} very useful, since it suggests that the Hitchin character, along with all the non-trivial geometric information about $\CM_H$ that it encodes, can be obtained in a completely different (and in many senses simpler) way from the Coulomb index of the 4d SCFT $T[\Sigma,G]$! This is precisely the approach taken in \cite{Gukov:2016lki} for tamely ramified $\Sigma$. We now proceed to study the Coulomb branch index of the general Argyres-Douglas theories to uncover the wild Hitchin characters.

We end this section with two remarks. The first is about the large-$k$ limit of the Hitchin character. In this limit, it is related to another interesting invariant of $\CM_H$ called the ``equivariant volume'' studied in \cite{moore2000integrating}
\be
{\rm Vol}_{\beta}(\CM_H) = \int_{\CM_H} \exp(k{\tilde \omega}_I) = \sum_{F_d} e^{-\beta \mu(F_d)} \int_{F_d}\frac{e^{ {\omega}_I}}{{\rm eu}_{\beta}(F_d)}
\label{equivVol}
\ee
where ${\rm eu}_{\beta}(F_d)$ is the equivariant Euler class of the normal bundle of $F_d$,
\be
{\rm eu}_{\beta}(F_d) = \prod_{i=1}^{\text{codim}_{\mathbb{C}}F_d} (x_i + \beta n_i).
\ee

The second remark is about the quantization of the monodromy parameter $\alpha$ (and also the $\alpha_1$ and $\alpha_2$). In the definition of the moduli space $\CM_H$, this parameter can take arbitrary values inside the Weyl alcove ${\rm Lie}(\mathbb{T})/W_{\text{aff}}$ subject to no restrictions. However, only for discrete values of the monodromy parameter, $\CM_H$ is quantizable. The allowed values are given by the characters of $G$ modulo $W_{\text{aff}}$ action (or equivalently integrable representations of $G$ at level $k$.)
\be
k\alpha\in \Lambda_{\text{char}}(G)/W_{\text{aff}}=\mathrm{Hom}(G,U(1))/W_{\text{aff}},
\ee 
which ensures the prequantum line bundle $\CL^{\otimes k}$ has integral periods over $\CM_H$ (see \cite{Gukov:2016lki} for completely parallel discussion of this phenomenon in the tame case.) For $G=SU(2)$, we often use the integral parameter
\be
\lambda=2k\alpha \in \{0,1,\ldots,k\}.
\label{MonodQuant}
\ee

The discretization of $\alpha$ can also be understood from the SCFT side. For a quantum field theory with flavor symmetry $^LG$ on $M_3\times \R$, one can deform the system --- and also its Coulomb branch --- by turning on a flavor holonomy in $\mathrm{Hom}(\pi_1M_3,^L\! G)/^LG$. When $M_3=L(k,1)$, the homomorphism $\pi_1=\Z_k\rightarrow ^L\!\! G$ up to conjugation is precisely classified by elements in 
\be
\Lambda_{\text{cochar}}(^LG)=\Lambda_{\text{char}}(G)
\ee
modulo affine Weyl symmetry.\footnote{In \cite{Gukov:2016lki}, the importance of distinguishing between $G$ and $^LG$ was emphasized. However, for the wild Hitchin moduli space that we study, the difference is not as prominent, because $\Sigma$ is now restricted to be $\mathbb{C}\mathbf{P}^1$, making the Hitchin character insensitive to global structure of the gauge group. In fact, the wild Hitchin characters we will consider are complete determined by the Lie algebra $\frak{g}$, provided that we analytically continuate $k\alpha$ to be a weight of $\frak{g}$. Because of this, we will use the simply-connected group --- $SU(2)$ in the rank-2 case --- for both the gauge group of the SCFT and the moduli space.}

\section{The Coulomb branch index of AD theories from $\CN = 1$ Lagrangian}\label{Sec 3: CBI}

Now our task is to compute the Coulomb branch index of Argyres-Douglas theories on the lens space $L(k,1)$. This is, however, a rather nontrivial problem, since these theories are generically strongly-interacting, non-Lagrangian SCFTs. Their original construction using singular loci of the Coulomb branch of ${\cal N}=2$ super Yang-Mills theory is not of much use: the IR R-symmetries are emergent, the Seiberg-Witten curves are derived from a subtle scaling limits (see \eg~\cite{Gaiotto:2010jf} for discussion of this issue), and the Higgs branches are intrinsic to the superconformal point itself \cite{Argyres:2012fu}. Also, no known dualities can relate them to Lagrangian theories. For example, in \cite{Gukov:2016lki} the generalized Argyres-Seiberg duality is very powerful for study of Coulomb index of class $\CS$ theories, but its analogue for Argyres-Douglas theories is not good enough to enable the computation of superconformal indices, since the two S-duality frames in general both consist of non-Lagrangian theories \cite{Buican:2014hfa, Xie:2016uqq, Cecotti:2015hca,Xie:2017vaf}. 

Recently, the author of \cite{Maruyoshi:2016tqk, Maruyoshi:2016aim, Agarwal:2016pjo} discovered that a certain class of four-dimensional ${\cal N}=1$ \textit{Lagrangian} theories exhibit supersymmetry enhancement under RG flow. In particular, some of them flow to ${\cal N}=2$ Argyres-Douglas theories. The $\CN=1$ description allows one to  track down the flow of R-charges and identify the flavor symmetry from the UV, making the computation of the full superconformal index possible. 

In this section we will use their prescription to calculate the Coulomb branch index of Argyres-Douglas theories on $S^1 \times L(k,1)$. Investigation of their properties, which is somewhat independent of the main subject of the paper, is presented in Appendix~\ref{sec:CBIProperty}, which consists of two subsections. The TQFT properties of the Coulomb branch indices make up Appendix~\ref{sec:TQFT}. When there is only tame ramifications, the lens space Coulomb branch index of $T[\Sigma]$ gives rise to a very interesting 2D TQFT on $\Sigma$ \cite{Gukov:2016lki}. In the presence of irregular singularities, the geometry of $\Sigma$ is highly constrained, and only a remnant of the TQFT cutting-and-gluing rules is present, which tells us how to close the regular puncture on the south pole to go from the $(A_1,D_{K+1})$ theory to $(A_1,A_{K-2})$.

In Appendix~\ref{sec: 3d red}, we consider the dimensional reduction of Argyres-Douglas theories, which will be relevant later when we discuss the large-$k$ behavior of the Hitchin character. The main motivation is to resolve an apparent puzzle: any fractional $U(1)_r$ charges in four dimensions should disappear upon dimensional reduction, since it is impossible to have fractional R-charges in the resulting three-dimensional ${\cal N}=4$ theory, whose R-symmetry is enhanced to $SU(2)_C \times SU(2)_H$. The solution lies in the mixing between the topological symmetry and the R-symmetry, similar to what was first discussed in \cite{Buican:2015hsa} using Schur index. Here we shall confirm the statement from Coulomb branch point of view directly.

In the following we begin with a brief review of the construction \cite{Maruyoshi:2016tqk, Maruyoshi:2016aim, Agarwal:2016pjo} and present an integral formula for the Coulomb branch index on lens spaces. 

\subsection{The construction}

In the flavor-current multiplet of a 4d ${\cal N}=2$ SCFT, the lowest component is known as the ``moment map operator'', which we will denote as ${\hat \mu}$. It is valued in $\frak{f}^*$, the dual of the Lie algebra of the flavor symmetry $F$, and transforms in the ${\bf 3}_0$ of the $SU(2)_R\times U(1)_r$ R-symmetry. In other words, if the Cartan generators of $SU(2)_R$ and $U(1)_r$ is $I_3$ and $r$, then 
\be
I_3({\hat \mu})=1, \quad \text{and}\quad r(\hat \mu)=0.
\ee

The idea of \cite{Maruyoshi:2016tqk, Maruyoshi:2016aim, Agarwal:2016pjo} is to couple the moment map operator ${\hat \mu}$ with an additional ${\cal N}=1$ ``meson'' chiral multiplet $M$ in the adjoint representation $\frak{f}$ of $F$ via the superpotential
\be
W = \langle\hat \mu, M\rangle
\label{N1superpotential}
\ee
and give $M$ a nilpotent vev $\langle M \rangle$. If the ${\cal N}=2$ theory we start with has a Lagrangian description (the case that we will be mainly interested in below), such deformation will give mass to some components of quarks, which would be integrated out during the RG flow.

The Jacobson-Morozov theorem states that a nilpotent vev $\langle M \rangle\in\frak{f}^+$ specifies a Lie algebra homomorphism $\rho: \mathfrak{su}(2) \rightarrow \mathfrak{f}$. The commutant of the image of $\rho$ is a Lie subalgebra $\frak{h}\subset\frak{f}$.  This subalgebra $\frak{h}$  is the Lie algebra of the residual flavor symmetry $H$. In the presence of the nilpotent vev, $\frak{f}$ (and similarly $\frak{f}^*$) can be decomposed into representations of $\frak{su}(2)\times \frak{h}$ as
\be
\frak{f} = \sum_j V_j \otimes R_j,
\label{adjDecomp}
\ee
where the summation runs over all possible spin-$j$ representations $V_j$ of $\mathfrak{su}(2)$, and $R_j$ carries a representation of $\frak{h}$. Both $M$ and $\hat\mu$ can be similarly decomposed
\be
M=\sum_{j,j_3}\tilde M_{j,j_3},\quad  \hat\mu=\sum_{j,j_3}\hat\mu_{j,j_3}
\ee
where $M_{j,j_3}$ also carries the $R_j$ representation of $\frak{h}$ that we omitted. Here $(j,j_3)$ is the quantum number for the $\frak{su}(2)$ representation $V_j$. Among them, $M_{1,1}$ will acquire a vev $v$, and we re-define $M$ to the fluctuation $M-\langle M\rangle$. Then, the superpotential \eqref{N1superpotential} decomposes as
\be
W =v \mu_{1,-1} + \sum_j \langle M_{j,-j}, {\hat \mu}_{j,j}\rangle.
\ee
Note that only the $-j$ component of the spin-$j$ representation of $\mathfrak{su}(2)$ for the $M$'s remains coupled in the theory, as the other components giving rise to irrelevant deformations \cite{Maruyoshi:2016aim}. 

Next, we examine the R-charge of the deformed theory. In the original theory, we denote $(J_+, J_-) = (2I_3, 2r)$ and a combination of them will be the genuine $U(1)_R$ charge of the ${\cal N} = 1$ theory, leaving the other as the flavor symmetry ${\cal F} = (J_+ - J_-)/2$. Upon RG flow to the infrared SCFT, the flavor symmetry would generally mix with the naive assignment of $U(1)_R$ charge:
\be
R = \frac{1}{2} (J_+ + J_-) + \frac{\epsilon}{2} (J_+ - J_-).
\label{Rflow}
\ee
The exact value of the mixing parameter $\epsilon$ can be determined via $a$-maximization \cite{Intriligator:2003jj} and its modification to accommodate decoupled free fields along the RG flow \cite{Kutasov:2003iy}. In the following, we summarize the ${\cal N}=1$ Lagrangian theory and the embedding $\rho$ found in \cite{Maruyoshi:2016aim, Agarwal:2016pjo} that are conjectured to give rise to Argyres-Douglas theories relevant for this paper.

\vspace{8pt} 
{\bf Lagrangian for $(A_1, A_{2N})$ theory.} The ${\cal N}=1$ Lagrangian is obtained by starting with ${\cal N}=2$ SQCD with $Sp(N)$ gauge group\footnote{We adopt the convention that $Sp(1) \simeq SU(2)$.} plus $2N+2$ flavors of hypermultiplets. The initial flavor symmetry is $F=SO(4N+4)$ and we pick the principal embedding, given by the partition $[4N+3, 1]$. The resulting ${\cal N}=1$ matter contents are listed in Table \ref{A1A2N_Lag}. Under the RG flow the Casimir operators $\Tr \phi^{2i}$ with $i = 1,2, \dots, N$ and $M_j$ with $j = 1,3, \dots, 2N+1$ and $M'_{2N+1}$ decouple. The mixing parameter in \eqref{Rflow} is
\be
\epsilon = \frac{7+6N}{9+6N}.
\ee

\begin{table}
\centering
    \begin{tabular}{ | x{6.0cm} | x{3.0cm} | x{3.0cm} |}
      \hline
      matter & $Sp(N)$ & $(J_+, J_-)$\\ \hline
        $q$   & $\Box$ & $(1,0)$ \\
        $q'$  & $\Box$ & $(1, -4N - 2)$ \\
        $\phi$ & adj & $(0,2)$ \\
        $M_j$, $j = 1,3, \dots 4N+1$ & {\bf 1} & $(0,2j+2)$ \\
        $M'_{2N+1}$ & {\bf 1} & $(0, 4N+4)$ \\ \hline
   \end{tabular}
\caption{The ${\cal N}=1$ matter content for the $Sp(N)$ gauge theory that flows to $(A_1, A_{2N})$ Argyres-Douglas theory. $\rho$ is given by the principal embedding, and $j$ takes values in the exponents of $\frak{f}$. For $\frak{f}=\frak{so}(4N+4)$, the exponents are $\{2N+1;1,3,\ldots,4N+1\}$.} \label{A1A2N_Lag}
\end{table}

\vspace{8pt}
{\bf Lagrangian for $(A_1, A_{2N-1})$ theory.} Similarly one starts with ${\cal N}=2$ SQCD with $SU(N)$ gauge group and $2N$ fundamental hypermultiplets with $SU(2N) \times U(1)_B$ flavor symmetry. We again take the principal embedding. The matter content is summarized in Table~\ref{A1A2N-1_Lag}. Using $a$-maximization we see that $M_j$ with $j = 1,2, \dots, N$, along with all Casimir operators, become free and decoupled. The mixing parameter in \eqref{Rflow} is
\be
\epsilon = \frac{3N+1}{3N+3}.
\ee
It is worthwhile to emphasize that the extra $U(1)_B$ symmetry would become the flavor symmetry of the Argyres-Douglas theory. In particular, when $N=2$, it is enhanced to $SU(2)_B$. This $U(1)_B$ symmetry is the physical origin of the gauge monodromy $\alpha$ in Section~\ref{sec: AD}. 

\begin{table}
\centering
    \begin{tabular}{ | x{6.0cm} | x{2.0cm} | x{2.0cm} | x{2.0cm} |}
      \hline
      matter & $SU(N)$ & $U(1)_B$ & $(J_+, J_-)$\\ \hline
        $q$   & $\Box$ & $1$ & $(1,-2N+1)$ \\
        $\tilde q$  & ${\overline \Box}$ & $-1$ & $(1, -2N + 1)$ \\
        $\phi$ & adj & $0$ & $(0,2)$ \\
        $M_j$, $j = 1,2, \dots 2N-1$ & {\bf 1} & $0$ & $(0,2j+2)$ \\ \hline
   \end{tabular}
\caption{The ${\cal N}=1$ matter content for the $SU(N)$ gauge theory that flows to $(A_1, A_{2N-1})$ Argyres-Douglas theory. $\rho$ is again the principal embedding, and $j$ ranges over the exponents of $su(2N)$.} \label{A1A2N-1_Lag}
\end{table}

\vspace{8pt}
{\bf Lagrangian for $(A_1, D_{2N+1})$ theory.}  Just as the $(A_1, A_{2N})$ theories, the starting point is the ${\cal N}=2$ SCFT with $Sp(N)$ gauge group and $2N+2$ fundamental hypermultiplets.  However, the nilpotent embedding $\rho$ is no longer the principal one; rather it is now given by the partition $[4N+1, 1^3]$, whose commutant subgroup is $SO(3)$ \cite{Agarwal:2016pjo}. The Lagrangian of the theory is given in Table~\ref{A1D2N+1_Lag}. Among mesons and Casimir operators $\Tr\,\phi^i$, only $M_j$ for $j = 2N+1, 2N+3, \dots, 4N-1$ remain interacting. The mixing parameter in \eqref{Rflow} is found to be
\be
\epsilon = \frac{6N+1}{6N+3}.
\ee
In this case, the UV $SO(3)$ residual flavor symmetry group is identified as the IR $SU(2)$ flavor symmetry coming from the simple puncture.

\begin{table}
\centering
    \begin{tabular}{ | x{6.0cm} | x{2.0cm} | x{2.0cm} | x{2.0cm} |}
      \hline
      matter & $Sp(N)$ & $SO(3)$ & $(J_+, J_-)$\\ \hline
        $q$   & $\Box$ & ${\bf 3}$ & $(1,0)$ \\
        $q'$  & ${\Box}$ & ${\bf 1}$ & $(1, -4N)$ \\
        $\phi$ & adj & ${\bf 1}$ & $(0,2)$ \\
        $M_j$, $j = 1,3, \dots 4N-1$ & {\bf 1} & ${\bf 1}$ & $(0,2j+2)$ \\ 
        $M'_{2N}$ & {\bf 1} & {\bf 3} & $(0, 4N+2)$ \\ 
        $M'_{0}$ & {\bf 1} & {\bf 3} & $(0, 2)$ \\ \hline
   \end{tabular}
\caption{The ${\cal N}=1$ matter content for the $Sp(N)$ gauge theory that flows to $(A_1, D_{2N+1})$ Argyres-Douglas theory.} \label{A1D2N+1_Lag}
\end{table}

\vspace{8pt}
{\bf Lagrangian for $(A_1, D_{2N})$ theory.} Similar to the $(A_1, A_{2N-1})$ case, we start with the $SU(N)$ gauge theory with $2N$ fundamental hypermultiplets, but choose $\rho$ to be the embedding given by the partition $[2N-1, 1]$. This leaves a $U(1)_a \times U(1)_b$ residual flavor symmetry, the first of which is the baryonic symmetry that we started with. The Lagrangian is summarized in Table~\ref{A1D2N_Lag}. Under RG flow, the decoupled gauge invariant operators are Casimir operators $\Tr \phi^i, i = 2, 3, \dots, N$, $M_j$ with $j = 0, 1, \dots, N-1$ and $(M, \tilde M)$. The $a$-maximization gives the mixing parameter
\be
\epsilon = 1 - \frac{2}{3N}.
\ee

\begin{table}
\centering
    \begin{tabular}{ | x{4.0cm} | x{2.0cm} | x{2.0cm} | x{2.0cm} | x{2.0cm} |}
      \hline
      matter & $SU(N)$ & $U(1)_a$ & $U(1)_b$ & $(J_+, J_-)$\\ \hline
        $q$   & $\Box$ & $1$ & $2N-1$ & $(1,0)$ \\
        $\tilde q$  & ${\overline \Box}$ & $-1$ & $-2N+1$ & $(1, 0)$ \\
        $q'$   & $\Box$ & $1$ & $-1$ & $(1,2-2N)$ \\
        ${\tilde q}'$  & ${\overline \Box}$ & $-1$ & $+1$ & $(1, 2-2N)$ \\
        $\phi$ & adj & $0$ & $0$ & $(0,2)$ \\
        $M_j$, $j = 0,1, \dots 2N-2$ & {\bf 1} & $0$ & $0$ & $(0,2j+2)$ \\ 
        $M$ & {\bf 1} & $0$ & $2N$ & $(0,2N)$ \\ 
        $\tilde M$ & {\bf 1} & $0$ & $-2N$ & $(0,2N)$ \\ \hline
   \end{tabular}
\caption{The ${\cal N}=1$ matter content for the $SU(N)$ gauge theory that flows to $(A_1, D_{2N})$ Argyres-Douglas theory.} \label{A1D2N_Lag}
\end{table}
In the IR, one combination of $U(1)_a$ and $U(1)_b$ would become the Cartan of the enhanced $SU(2)$ flavor symmetry.

\subsection{Coulomb branch index on lens spaces}

The $\CN=1$ constructions of the generalized Argyres-Douglas theories enable one to compute their $\CN=2$ superconformal index by identifying the additional R-symmetry with a flavor symmetry of the $\CN=1$ theory. As the ordinary superconformal index on $S^1 \times S^3$, the $\CN=1$ lens space index can be defined in terms of the trace over Hilbert space on $L(k,1)$ \cite{Razamat:2013opa, Nieri:2015yia} 
\be
{\cal I}_{\cN = 1} (p, q) = \Tr (-1)^F p^{j_1+j_2 + R /2} q^{j_2-j_1 + R /2} \xi^{\cF} \prod_i a_i^{f_i} \exp(-\beta' \delta')
\label{N=1index}
\ee
where $j_{1,2}$ are the Cartans of the $SO(4)_{E} \simeq SU(2)_1 \times SU(2)_2$ rotation group, $R$ counts the superconformal $U(1)_R$ charge of the states. We also introduce the flavor fugacity $\xi$ for the symmetry $\CF = (J_+ - J_-)/2$ inherited from the $\cN = 2$ R-symmetry. Finally, $\delta'$ is the commutator of a particular supercharge $Q$ chosen in defining the index. It is given by
\be
\delta' = \{ Q, Q^{\dagger} \} = E - 2j_1 + \frac{3R}{2}
\ee
where $E$ the conformal dimension. Supersymmetry ensures that only states annihilated by $Q$ contribute in \eqref{N=1index}; hence the results are independent of $\beta'$ and one can restrict the trace to be taken over the space of BPS states.

One advantage of the lens space index comes from the non-trivial fundamental group of $L(k,1)$, making it sensitive to the global structure of the gauge group \cite{Razamat:2013opa}. Also, the gauge theory living on $L(k,1)$ has degenerate vacua labelled by holonomies around the Hopf fiber, so the Hilbert space will be decomposed into different holonomy sectors. All of these make the lens index a richer invariant than the ordinary superconformal index.

For a theory with a Lagrangian, the lens space index can be computed by first multiplying contributions from free matter multiplets after $\Z_k$-projection, then integrating over the (unbroken) gauge group determined by a given holonomy sector, and finally summing over all inequivalent sectors. We introduce the elliptic Gamma function
\be
\Gamma(z; p, q) = \prod_{j,k=0}^{+\infty} \frac{1 - z^{-1} p^{j+1} q^{k+1}}{1 - z p^j q^k}.
\ee
Then, for a chiral superfield with gauge or flavor fugacity/holonomy $(b, m)$ we have
\be
I_{\chi}(m, b) = I^{\chi}_0(m,b) \cdot\Gamma\pbra{ (p q)^{\frac{R}{2}} q^{k-m} b; q^k, p q} \Gamma\pbra{ (p q)^{\frac{R}{2}} p^{m} b; p^k, p q}
\ee
with the prefactor related to the Casimir energy
\be
I^{\chi}_0(m, b) = \pbra{(pq)^{\frac{1-R}{2}}b^{-1}}^{\frac{m(k-m)}{2k}} \pbra{\frac{p}{q}}^{\frac{m(k-m)(k-2m)}{12k}}.
\ee
For a vector multiplet the contribution is
\be
I_{V}(m, b) = \frac{I^V_0(m,b)}{\Gamma\pbra{ q^{m} b^{-1}; q^k, p q} \Gamma\pbra{ p^{k-m} b^{-1}; p^k, p q}}
\ee
with 
\be
I^V_0(m, b) = \pbra{(pq)^{\frac{1}{2}}b^{-1}}^{-\frac{m(k-m)}{2k}} \pbra{\frac{q}{p}}^{\frac{m(k-m)(k-2m)}{12k}}.
\ee
Notice that we will not turn on flavor holonomy for the $U(1)$ flavor symmetry $\CF$ along the Hopf fiber. This is because it is part of the $\CN=2$ R-symmetry; turning on background holonomy for it will break the $\CN=2$ supersymmetry. 

To connect \eqref{N=1index} with $\cN = 2$ lens space index, recall the definition of the latter is \cite{Benini:2011nc, Razamat:2013jxa}
\be
{\cal I}_{\cN = 2} (p, q, t) = \Tr (-1)^F p^{j_1+j_2 + r} q^{j_2-j_1 + r} t^{R - r} \prod_i a_i^{f_i} \exp(-\beta'' \delta'')
\label{N=2index}
\ee
where the index counts states with $SU(2)_R \times U(1)_r$ charge $(R,r)$ that are BPS with respect to $\delta'' = E - 2j_2 - 2R - r$. To recover the above $\CN=2$ index from \eqref{N=1index}, we make the substitution
\be
\xi \rightarrow \pbra{t(pq)^{-\frac{2}{3}}}^{\gamma}
\label{N1toN2g}
\ee
for some constant $\gamma$ depending on how $U(1)_{\cal F}$ is embedded inside $SU(2)_R \times U(1)_r$. 

Finally, we take the ``Coulomb branch limit" of the $\cN = 2$ lens space index,
\be
p, q, t \rightarrow 0, \ \ \ \ \frac{pq}{t} = \ft\ \ \text{fixed}.
\ee
The trace formula \eqref{N=2index} then reduces to 
\be\label{traceC}
{\cal I}_{\cN = 2}^C = \Tr_{\! C} (-1)^F \ft^{r - R} \prod_i a_i^{f_i},
\ee
where the trace is taken over BPS states annihilated by both $\tilde{Q}_{1\dot{-}}$ and $\tilde{Q}_{2\dot{+}}$ (\ie~satisfying $E-2j_2-2R+r=E+2j_2+2R+r=0$.) Notice that, in our convention, $L(k,1)$ is a quotient of $S^3$ by $\Z_k\subset U(1)_{\mathrm{Hopf}}\subset SU(2)_1$. Since both $\tilde{Q}_{1\dot{-}}$ and $\tilde{Q}_{2\dot{+}}$ transform trivially under $SU(2)_1$, they are preserved after the $\Z_k$ quotient. Hence the trace formula \eqref{traceC} is well-defined. 

For all known examples the Coulomb branch operators have $R = 0$, so the above limit effectively counts $U(1)_r$ charge. For a Lagrangian theory, when $k = 1$ this limit counts the short multiplet ${\bar {\cal E}}_{r, (0,0)}$ \cite{Gadde:2011uv}, whose lowest component parametrizes the Coulomb branch vacua of the SCFT.  

Below we will list the integral formulae for the Coulomb branch indices of Argyres-Douglas theories that we are interested in throughout this paper. In computing the lens space index we have removed contributions from the decoupled fields.

\vspace{8pt}
{\bf $(A_1, A_{2N})$ theories.} We have
\be
{\cal I}_{(A_1, A_{2N})} & = \prod_{i=1}^{N} \frac{1}{1-\ft^{\frac{2(N+i+1)}{2N+3}}} \prod_{i=1}^{N} \frac{1-\ft^{\frac{2i}{2N+3}}}{1-\ft^{\frac{1}{2N+3}}}\\[0.5em]
& \times \sum_{m_i} \prod_{\alpha > 0} \pbra{\ft^{\frac{2}{2N+3}}}^{-\frac{1}{2}(\dbra{\alpha(m)} - \frac{1}{k}\dbra{\alpha(m)}^2)} \prod_{i=1}^N \pbra{\ft^{\frac{4(N+1)}{2N+3}}}^{\frac{1}{2}(\dbra{m_i} - \frac{1}{k}\dbra{m_i}^2)}  \\[0.5em]
& \times \frac{1}{|{\cal W}_m| }\oint [d{\bf z}] \prod_{\dbra{\alpha(m)} = 0}\frac{1-{\bf z}^{\alpha}}{1- \ft^{\frac{1}{2N+3}} {\bf z}^{\alpha}}\label{indexA1A2N}
\ee
where the integral is taken over the unbroken subgroup of $Sp(N)$ with respect to a given set of holonomies $\{ m_i \}$. Here, $|{\cal W}_m|$ is  the order of Weyl group for the residual gauge symmetry. The constant $\gamma$ \eqref{N1toN2g} is $\gamma = 1/(2N+3)$. We use the notation $\dbra{x}$ to denote the remainder of $x$ modulo $k$.

\vspace{8pt}
{\bf $(A_1, A_{2N-1})$ theories.} After taking $\gamma = 1/(N+1)$ and the Coulomb branch limit, we have
\be
{\cal I}_{(A_1, A_{2N-1})} & = \prod_{i=1}^{N-1} \frac{1}{1-\ft^{\frac{2N+1-i}{N+1}}} \prod_{i=1}^{N-1} \frac{1-\ft^{\frac{i+1}{N+1}}}{1-\ft^{\frac{1}{N+1}}}\\[0.5em]
& \times \sum_{m_i} \prod_{\alpha > 0} \pbra{\ft^{\frac{2}{N+1}}}^{-\frac{1}{2}(\dbra{\alpha(m)} - \frac{1}{k}\dbra{\alpha(m)}^2)} \prod_{i=1}^N \pbra{\ft^{\frac{2N}{N+1}}}^{\frac{1}{2}(\dbra{m_i + n} - \frac{1}{k}\dbra{m_i + n}^2)} \\[0.5em]
& \times \frac{1}{|{\cal W}_m|} \oint [d{\bf z}] \prod_{\dbra{\alpha(m)} = 0}\frac{1-z_i / z_j}{1- \ft^{\frac{1}{N+1}} z_i / z_j}
\label{indexA1A2N-1}
\ee
where we have introduced $U(1)$ flavor holonomy $n$ and the integral is taken over the (unbroken subgroup of) $SU(N)$. Specifically, suppose the gauge holonomy breaks the gauge group $SU(N)$ as
\be
SU(N) \rightarrow SU(N_1) \times SU(N_2) \times \dots SU(N_l) \times U(1)^r
\ee
where $N -1 = (N_1 - 1)+ (N_2 - 1) + \dots + (N_l - 1) + r$ then we have
\be
\frac{1}{|{\cal W}_m|} \oint [d{\bf z}] \prod_{\dbra{\alpha(m)} = 0}\frac{1-z_i / z_j}{1- \ft^{\frac{1}{N+1}} z_i / z_j} = \prod_{i=1}^l \prod_{j=1}^{N_l - 1} \frac{1-\ft^{\frac{1}{N+1}}}{1-\ft^{\frac{j+1}{N+1}}}.
\label{SUbreak}
\ee
To derive the general formula, we assume the $U(1)$ flavor holonomy $n$ is an integer. In fact, we will see in Section~\ref{Sec 4: Verlinde} that $n$ is allowed to take value in $\mathbb{Z} / N$. In fact, $n$ is the quantization of the monodromy around irregular puncture. Its allowed values differ from $\lambda$ in \eqref{MonodQuant} since they are identified respectively in the UV and IR. Their relation is $\lambda = \dbra{N n} = 2 k \alpha$. The index takes the following form
\be
\ft^{\frac{1}{N+1}\pbra{\dbra{Nn} - \frac{1}{k} \dbra{Nn}^2}}(1+\ldots),
\label{A2N-1Norm}
\ee
where the ellipsis stands for terms with only \emph{positive} powers of $\ft$. 

\vspace{8pt}
{\bf $(A_1, D_{2N+1})$ theories.} We have
\be
{\cal I}_{(A_1, D_{2N+1})} = & \prod_{j=1}^{N}\frac{1}{1-\ft^{\frac{4N+2-2j}{2N+1}}}  \prod_{j=1}^{N}\frac{1-\ft^{\frac{2j}{2N+1}}}{1-\ft^{\frac{1}{2N+1}}}\\[0.5em]
& \times \sum_{m_i} \prod_{\alpha > 0} \pbra{\ft^{\frac{2}{2N+1}}}^{-\frac{\dbra{\alpha(m)}(k-\dbra{\alpha(m)})}{2k}} \prod_i \pbra{\ft^2}^{\frac{\dbra{m_i}(k-\dbra{m_i})}{2k}} \pbra{\ft^{\frac{1}{2N+1}}}^{\frac{\dbra{m_i \pm 2n}(k-\dbra{m_i \pm 2n})}{2k}} \\[0.5em]
& \times \frac{1}{|{\cal W}_m| }\oint [d{\bf z}] \prod_{\dbra{\alpha(m)} = 0}\frac{1-{\bf z}^{\alpha}}{1- \ft^{\frac{1}{2N+1}} {\bf z}^{\alpha}}
\label{indexA1D2N+1}
\ee
where $n$ is regarded as the holonomy for $SU(2)$ symmetry in the IR,\footnote{The factor of $2$ in front of $n$ is due to the fact that the quarks $q$ in the UV transform in the triplet ${\bf 3}$ of $SU(2)$.} which is related to the quantized monodromy around the regular puncture at the south pole by $\lambda = \dbra{2n} = 2k \alpha$. The constant $\gamma$ here is $1/(2N+1)$. As in $(A_1, A_{2N})$ case, the integral is taken over the unbroken subgroup of $Sp(N)$. Note that here we allow $n$ to a half-integer. This fact also plays an important role when we discuss TQFT structure in Appendix~\ref{sec:TQFT}. As before, the closed expression of the index contains a normalization factor
\be
\ft^{\frac{N}{2N+1}\pbra{ \dbra{2n} - \frac{1}{k} \dbra{2n}^2 }}.
\label{D2N+1Norm}
\ee

\vspace{8pt}
{\bf $(A_1, D_{2N})$ theories.} Similarly, the index formula is
\be
{\cal I}_{(A_1, D_{2N})} = & \prod_{j=N+1}^{2N-1}\frac{1}{1-\ft^{\frac{j}{N}}}  \prod_{j=1}^{N-1}\frac{1-\ft^{\frac{j+1}{N}}}{1-\ft^{\frac{1}{N}}}\\[0.5em]
&\times  \sum_{m_i} \prod_{\alpha > 0} \pbra{\ft^{\frac{2}{N}}}^{-\frac{\dbra{\alpha(m)}(k-\dbra{\alpha(m)})}{2k}} \prod_i \pbra{\ft^{\frac{1}{N}}}^{\frac{\dbra{m_i + n_1+(2N-1)n_2}(k-\dbra{m_i + n_1+(2N-1)n_2})}{2k}}\\[0.5em]
& \times \prod_i \pbra{\ft^{\frac{2N-1}{N}}}^{\frac{\dbra{m_i + n_1 - n_2}(k-\dbra{m_i + n_1 - n_2})}{2k}}\\[0.5em]
&\times \frac{1}{|{\cal W}_m|} \oint [d{\bf z}] \prod_{\dbra{\alpha(m)} = 0}\frac{1-z_i / z_j}{1- \ft^{\frac{1}{N}} z_i / z_j}
\label{indexA1D2N}
\ee
where we have introduced $(n_1, n_2)$ to represent the $(U(1)_a, U(1)_b)$ flavor holonomy repsectively. The constant $\gamma = 1/N$, and the integral is over the (unbroken subgroup of) $SU(N)$. Its precise value is given in \eqref{SUbreak} by substituting $\ft^{1/(N+1)}$ with $\ft^{1/N}$. In \eqref{indexA1D2N} the computation was done assuming $n_{1,2} \in \mathbb{Z}$ so that the gauge holonomies $m_i$ are all integers.  However, the allowed set of values are in fact larger. We will return to this issue in Section~\ref{Sec 4: Verlinde}. The relations to monodromies around wild and simple punctures are given by, respectively,
\be
\lambda_1 = \dbra{N n_1} = 2 k \alpha_1, \ \ \ \ \lambda_2 = \dbra{2N n_2} = 2 k \alpha_2.
\label{D2NMonod}
\ee
Again, the evaluation of \eqref{indexA1D2N} gives a normalization factor
\be
(u)^{{\frac{N-1}{2N}}\pbra{\dbra{2N n_2}-\frac{1}{k}\dbra{2N n_2}^2} +{\frac{1}{2N}}\pbra{\dbra{Nn_1+Nn_2}-\frac{1}{k}\dbra{Nn_1+Nn_2}^2}+{\frac{1}{2N}}\pbra{\dbra{Nn_1-Nn_2}-\frac{1}{k}\dbra{Nn_1-Nn_2}^2}}.
\label{D2NNorm}
\ee

\section{Wild Hitchin characters}\label{Sec 4: Verlinde}

Now that we have the integral expressions for the Coulomb branch indices of Argyres-Douglas theories \eqref{indexA1A2N}, \eqref{indexA1A2N-1}, \eqref{indexA1D2N+1} and \eqref{indexA1D2N}, we will evaluate them explicitly in this section. 

Before presenting the results, we remark that the Coulomb indices have several highly non-trivial properties. Anticipating the equality between the index and wild Hitchin characters, we can often understand these properties from geometry. 
\begin{enumerate}
	\item \textbf{Positivity.} The Coulomb branch index as a series in $\ft$ always has positive coefficients. This phenomenon is not obvious from the integral expression. 
From the geometric side, this is a simple corollary of the ``vanishing theorem'' for the wild Hitchin moduli space
	\be
	H^i(\CM_H,\CL^{\otimes k})=0 \quad \text{for $i>0$}.
	\ee
	This further implies that, on the physics side, all Coulomb BPS states on $L(k,1)$ are bosonic. This positivity phenomenon is the analogue of those observed in \cite{Gukov:2016gkn} and \cite{GPPV} with wild ramifications. 
	
	\item \textbf{Splitting.} The indices always turn out to be rational functions. Further, they split as a sum over fixed points --- a form predicted by the Atiyah-Bott localization formula from the geometry side \eqref{equivInd}. This will allow us to extract geometric data for moduli spaces directly. However, the interpretation of this decomposition is not clear at the level of the BPS Hilbert spaces $\CH_{\text{Coulomb}}$. It is not even clear that the $\CH_{\text{Coulomb}}$ can be decomposed in similar ways, as the individual contributions from some fixed points do not have positivity. 
	
	\item \textbf{Fractional dimensions.} One notable feature of Argyres-Douglas theories is the fractional scaling dimensions of their Coulomb branch operators. From the point of view of the Hitchin action, this comes from the fractional action on the $z$ coordinate. For example, the $U(1)$ action on $\CM_{2,2N+1}$ involves a rotation of the base curve $\mathbb{C}\mathbf{P}^1$ with coordinate $z$ by
	\be
	\rho_\theta:\quad z\mapsto e^{-i\frac{2}{2N+3}\theta} z.
	\ee
  Therefore only the $(2N+3)$-fold cover of the $U(1)$ defines a (genuine non-projective) group action, and the Hitchin character will be a power series in $\ft^{\frac{1}{2N+3}}$. In all four families of moduli spaces ($\CM_{2,K}$ versus $\tilde{\CM}_{2,K}$;  $K$ either even or odd) $K+2$ is always the number of Stokes rays centered at the irregular singularity, and the Hitchin character will be a power series in $\ft^{\frac{1}{K+2}}$. When $K$ is even, one can check that the $(K+2)/2$-fold cover of the $U(1)$ given by $\rho_\theta$ defines a group action, and the Hitchin character will contain integral powers of $\ft^{\frac{2}{K+2}}$ as a consequence. 
\end{enumerate}

We will start this section by giving formulae for the wild Hitchin characters in Section~\ref{sec:WHC}. In Section~\ref{sec: 4dto3d}, the large-$k$ limit of the wild Hitchin character is discussed. This limit effectively reduces the theory to three dimensions; by taking the mirror symmetry $\CM_H$ is realized as the Higgs branch of a 3d $\CN = 4$ quiver gauge theory. This is in accordance with the mathematical work \cite{boalch2008irregular}. By comparing 3d index and 4d index, we will see how good this approximation is on the nilpotent cone. As a byproduct, we give a physical interpretation of the fixed points from the 3d mirror point of view.

In Appendix~\ref{sec: fixed}, we will present mathematical calculations that directly confirm the physical prediction: the Coulomb branch index of Argyres-Douglas theory indeed computes the wild Hitchin character for $\CM_H(\Sigma, PSL(2,\mathbb{C})) := {^L}\CM_H$. 

As we have explained --- and we will soon offer another explanation from the physics perspective --- the Hitchin character is not sensitive to the difference between $\CM_H(\Sigma, SL(2,\mathbb{C}))$ and $\CM_H(\Sigma, PSL(2,\mathbb{C}))$ when $\Sigma$ is a sphere with at most two punctures. In fact, one can directly check that the fixed points are exactly the same with identical ambient geometry. As a consequence, the Hitchin character for $\CM_H(\Sigma, SL(2,\mathbb{C}))$ can be obtained via ``analytic continuation'' of $\lambda$, $\lambda_1$ and $\lambda_2$ by allowing them to take odd values. So we will not emphasize the difference between $\CM_H$ and $^L\CM_H$ in this section, unless specified.

\subsection{The wild Hitchin character as a fixed-point sum} \label{sec:WHC}

\subsubsection{The moduli space $\CM_{2, 2N+1}$}

A nice illustrative example to start is the $(A_1, A_2)$ theory with no flavor symmetry at all. The Coulomb branch index is
\be
{\cal I}_{(A_1, A_2)} = \frac{1}{(1-\ft^{\frac{2}{5}})(1-\ft^{\frac{3}{5}})} + \frac{\ft^{\frac{k}{5}}}{(1-\ft^{\frac{6}{5}})(1-\ft^{-\frac{1}{5}})}.
\label{A1A2V}
\ee
On the other hand, the moduli space $\CM_{2,3}$ has two complex dimensions, and we have the fixed points and the associated eigenvalues of the circle action on normal bundles obtained in Appendix~\ref{sec: fixed}: 
\be
\varphi^*_0 = \left( \begin{array}{cc} 0 & z\\ z^2 & 0 \end{array} \right) dz, \ \ \ \ \varphi^*_1 & = \left( \begin{array}{cc} 0 & 1\\ z^3 & 0 \end{array} \right) dz, 
\label{A2fixedpoint}
\ee
with moment maps $\mu = 1/40$ and $9/40$ respectively. After shifting the two moment maps simultaneously by $1/40$,\footnote{We normalize the Hitchin character such that the $\ft=0$ limit gives 1. The ambiguity of multiplying a monomial $\ft^{\Delta \mu}$ to the Hitchin character corresponds to redefining the $U(1)$ action such that it rotates the fiber of the line bundle $\CL$ as well.} we get $\mu_1 = 0$ and $\mu_2 = 1/5$. These are precisely the power entering the numerator of each term in \eqref{A1A2V}! Furthermore, from the denominator of each term, we are able to read off the weights of the circle action on the two-dimensional normal bundle of each fixed points --- they are respectively $(2/5, 3/5)$ and $(6/5, -1/5)$. This is directly checked in Appendix~\ref{sec: fixed} from geometry, providing strong evidence for our proposal \eqref{CBIQuant}. Also, notice the ubiquity of  number ``5'' --- the number of Stokes rays associated with the irregular singularity.

The formula \eqref{A1A2V} encodes various interesting information about the geometry and topology of the moduli space.  As in the tame case, the moment map (which agrees with \eqref{moment Mu} at fixed points) is expected to be a perfect Morse function on $\CM_H$. The fixed points are critical points of $\mu$, and the positive- (negative-)eigenvalue subspaces of the normal bundle correspond to the upward (downward) Morse flows. In particular, we know that the top fixed point in $\CM_{2,3}$ has Morse index $2$ and the downward flow from it coincides with the nilpotent cone --- the singular fiber of the Hitchin fibration with Kodaira type \Rnum{2} \cite{Argyres:2015ffa}. Then the Poincar\'{e} polynomial of $\CM_{2,3}$ is
\be
\CP(\CM_{2,3}) = 1 + \fr^2.
\ee
Another important quantity is the equivariant volume of $\CM_{2,3}$ as given in \eqref{equivVol}
\be
{\rm Vol}_{\beta}(\CM_{2,3}) = \frac{25}{6 \beta^2} (1-e^{-\frac{1}{5}\beta}).
\ee
Note that as $\beta \rightarrow +\infty$, the volume scale as $\beta^{-2}$, with the negative power of $\beta$ being the complex dimension of $\CM_{2,3}$. This is unlike the tame situation, where $\beta$ scales according to half the dimension of $\CM_H$. Intuitively, this is because, while Higgs field is responsible for half of the dimensions of $\CM_H$ in tame case, they are responsible for all dimensions in the wild Hitchin moduli space, as a $G$-bundle has no moduli over $\Sigma$ in the cases that we consider. 

We now give a general formula of the wild Hitchin character for $\CM_{2, 2N+1}$, predicted by the Coulomb index and proved in Appendix \ref{sec: fixed}. There are $N+1$ fixed points in the moduli space $P_0, P_1, \dots, P_{N}$. They have moment maps given by 
\be
\mu_i = \frac{i(i+1)}{2(2N+3)}, \ \ i =0, 1, 2, \dots, N
\ee
where we have already shifted a universal constant so that $P_0$ as moment map $0$. The weights are given in \eqref{M2N+1weights}, and the wild Hitchin character reads
\be
{\cal I}({\cal M}_{2, 2N+1}) = \sum_{i = 0}^N \frac{\ft^{\frac{i(i+1)}{2(2N+3)}k}}{\prod_{l=1}^{i} \pbra{1-\ft^{\frac{2(N+l+1)}{2N+3}}} \pbra{1-\ft^{-\frac{2l-1}{2N+3}}} \prod_{l=i+1}^{N} \pbra{1-\ft^{\frac{2l+1}{2N+3}}} \pbra{1-\ft^{\frac{2(N-l+1)}{2N+3}}} }.
\label{M2N+1V}
\ee
The Morse index of $P_i$ is $2i$, so the Poincar\'{e} polynomial of $\CM_{2, 2N+1}$ is 
\be
\CP({\cal M}_{2, 2N+1}) = 1+ \fr^2 + \fr^4 + \dots + \fr^{2N} = \frac{1-\fr^{2N+2}}{1-\fr^2}.
\ee

\subsubsection{The moduli space ${\widetilde \CM}_{2, 2N-1}$}

A closely related moduli space is ${\widetilde \CM}_{2, 2N-1}$, which has regular puncture at the south pole of $\Sigma$ in addition to the irregular puncture $I_{2, 2N-1}$ at the north pole. Then the gauge connection has monodromy $A \sim \alpha d\theta$ around the regular puncture, and $\lambda=2k\alpha=\{0,1,\ldots,k\}$ is quantized and are integrable weights of $\hat{\frak{su}}(2)_k$.\footnote{$\lambda$ starts life as a weight of $SO(3)$, since the physical set-up computes the Hitchin character of ${^L}{\widetilde \CM}_{SU(2)}={\widetilde \CM}_{SO(3)}$ according to \eqref{CBIQuant}. As we have explained, from the geometric side, the difference between ${\widetilde \CM}_{SU(2)}$ and ${\widetilde \CM}_{SO(3)}$ is almost negligible for the purpose of studying wild Hitchin characters --- one only needs to analytically continuate $\lambda$ to go from one moduli space to another. This phenomenon has a counterpart in the index computation as well. Being an $SU(2)$ flavor holonomy, a natural set of values for $\lambda$ without violating charge quantization condition is $0, 2, \dots, 2\lfloor k/2  \rfloor$ \cite{Gukov:2016lki}. However, in the expression \eqref{indexA1D2N+1}, there is no problem with simply allowing $\lambda=2n$ to take odd values. This can be understood from the perspective of the $\CN = 1$ Lagrangian theories listed in Table \ref{A1D2N+1_Lag}. There all the matter contents are assembled either in the trivial or the vector representation of the global $SO(3)$ symmetry, and these two representations cannot distinguish $SU(2)$ from $SO(3)$; as a consequence if we expand the full superconformal index and look at the BPS spectrum of Argyres-Douglas theory, only representations for $SO(3)$ will appear. This means odd $\lambda$ does not violate the charge quantization condition, and can be allowed. Furthermore, since the superconformal index of $(A_1, A_{2N})$ can be obtained from $(A_1, D_{2N+3})$ by closing the regular puncture through \eqref{OddCap}, one immediately concludes that the Hitchin characters for $\CM_{2, 2N+1}$ and for the Langlands dual ${^L}\CM_{2, 2N+1}$ are exactly the same. 
} Again we will absorb the normalization constant \eqref{D2N+1Norm} appearing in the superconformal index so that the index as a series in $\ft$ will start with 1.

Next we present the wild Hitchin character for the moduli space ${\widetilde \CM}_{2, 2N-1}$. We begin with the example ${\widetilde \CM}_{2, 1}$, or Argyres-Douglas theory of type $(A_1, D_3)$. Denote $\lambda := 2k\alpha=2n$ valued in $\{0, 1, \dots, k\}$. Then, we have
\be
{\cal I}_{(A_1, D_3)} = \frac{1}{(1-\ft^{\frac{1}{3}})(1-\ft^{\frac{2}{3}})} + \frac{\ft^{\frac{\lambda}{3}} + \ft^{\frac{k - \lambda}{3}}}{(1-\ft^{-\frac{1}{3}})(1-\ft^{\frac{4}{3}})}.
\ee
This formula tells us that ${\widetilde \CM}_{2, 1}$ has three fixed points under the Hitchin action. One of them has the lowest moment map $0$ with weights on the normal bundle $(1/2, 2/3)$, while the other two have moment maps $\mu_1^{(1)} = 2\alpha / 3$ and $\mu_1^{(2)} = (1-2\alpha) / 3$. 
These results are also confirmed by mathematical calculations in Appendix \ref{sec: fixed}. Using Morse theory, we get the Poincar\'{e} polynomial of ${\widetilde \CM}_{2, 1}$
\be
\CP({\widetilde \CM}_{2, 1}) = 1 + 2 \fr^2.
\ee
And the equivariant volume is given by
\be
{\rm Vol}_{\beta}({\widetilde \CM}_{2, 1}) = \frac{9}{4\beta^2} (2 - e^{-\frac{2\alpha}{3}\beta} - e^{-\frac{1 - 2\alpha}{3}\beta}).
\ee

\begin{figure*}[htbp]
\begin{adjustwidth}{-0.4cm}{}
      \begin{subfigure}[t]{.4\textwidth}
        \includegraphics[width=7cm]{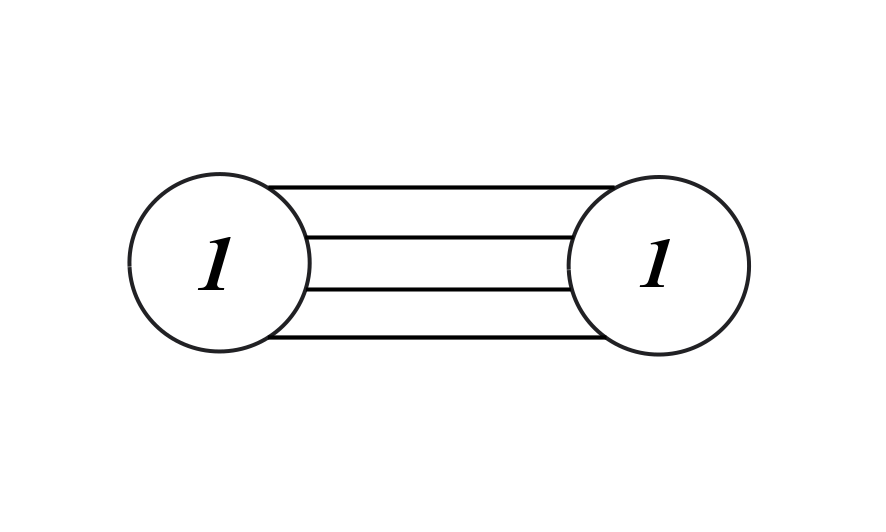}
        \caption*{}
      \end{subfigure}
      \hspace{1.5cm}
      \begin{subfigure}[t]{.4\textwidth}
        \includegraphics[width=8cm]{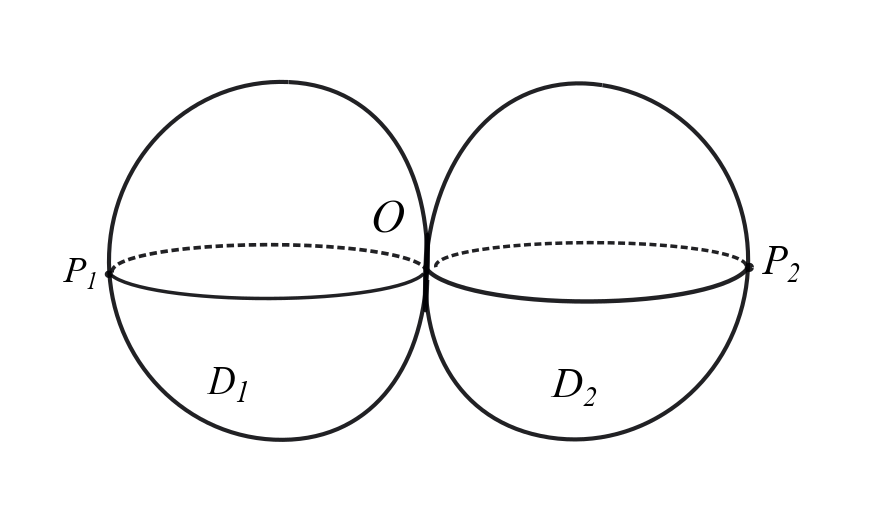}
        \caption*{}
      \end{subfigure}
  \end{adjustwidth}
 \caption[The affine $A_1$ Dynkin diagram and the nilpotent cone of Kodaira type \Rnum{3}]{Left: the affine $A_1$ Dynkin diagram. Right: the nilpotent cone of Hitchin fibration for ${\widetilde \CM}_{2, 1}$, consisting of two $\mathbb{C}\mathbf{P}^1$ intersecting at $O$ with intersection number $2$. Together with $P_1$, $P_2$, they comprise the three fixed points of the Hitchin moduli space ${\widetilde \CM}_{2, 1}$.}
\label{Kodaira3}
\end{figure*}

As ${\widetilde \CM}_{2, 1}$ has hyper-K\"{a}hler dimension one, it is an elliptic surface in complex structure $I$. The only singular fiber is the nilpotent cone with Kodaira type \Rnum{3} \cite{Argyres:2015ffa} (\ie~labeled by the affine $A_1$ Dynkin diagram, see Figure \ref{Kodaira3}). It consists of two $\mathbb{C}\mathbf{P}^1$ with the intersection matrix given by
\be
\left( \begin{array}{cc} -2 & 2 \\ 2 & -2 \end{array} \right).
\ee
The null vector of the intersection matrix should be identified with the homology class of the Hitchin fiber,
\be
\left[ {\bf F} \right] = 2\left[ D_1 \right] + 2\left[ D_2 \right]. 
\ee
This relation translates into (see \cite{Gukov:2010sw} and \cite{Gukov:2016lki} for review of this relation as well as examples with tame ramifications)
\be
{\rm Vol}({\bf F}) = 2{\rm Vol}(D_1) + 2{\rm Vol}(D_2)
\label{D3Vol}
\ee
which is indeed visible from the Hitchin character. It is not hard to see that for each $\mathbb{C}\mathbf{P}^1$, the volumes are ${\rm Vol}(D_1) = 3 \mu_1^{(1)} = 2\alpha$ and ${\rm Vol}(D_2) = 3 \mu_1^{(2)} = 1-2\alpha$ respectively. (The factor ``$3$'' is due to the weights $-1/3$ that corresponds to the downward Morse flow.) Consequently, we see \eqref{D3Vol} is exactly true, with ${\rm Vol}({\bf F}) = 2$ in our normalization.

We now give a general statement for the wild moduli space ${\widetilde \CM}_{2, 2N-1}$. There are $2N+1$ fixed points, divided into $N+1$ groups. We label them as $P_{i}^{(1,2)}$, $i = 0, 1, \dots, N$. The $i$-th group contains two fixed points for $i > 0$ and one fixed points for $i = 0$. The $U(1)$ weights on the $2N$-dimensional normal bundle to $P_i$ is given by 
\be
& \epsilon_l = -\frac{2l-1}{2N+1},   \ \ \ \ {\tilde \epsilon}_l = \frac{2N+2l}{2N+1},\ \ \ \ l = 1,2,\dots, i\\[0.5em]
& \epsilon_l = \frac{2l-1}{2N+1}, \ \ \ \ {\tilde \epsilon}_l = \frac{2N+2-2l}{2N+1}, \ \ \ \ l = i+1, i+2, \dots N.
\ee
The normal bundle can be decomposed into the tangent space to the nilpotent cone plus its orthogonal complement, and  $\epsilon_l$ and ${\tilde \epsilon}_l$ correspond respectively to the former and the latter. 

For the $0$-th fixed point the moment map is $0$, while for the $i$-th group with $i>0$, the two moment map values are
\be
\mu_i^{(1)} = \frac{i(i+1)}{2(2N+1)} - \frac{i}{2N+1}(2\alpha), \ \ \ \ \mu_i^{(2)} = \frac{(i-1)i}{2(2N+1)} + \frac{i}{2N+1}(2\alpha)
\label{tM2N-1moment}
\ee
where $\alpha$ is again the monodromy around the simple puncture. Then the wild Hitchin character is
\be
{\cal I}\pbra{{\widetilde {\cal M}}_{2, 2N-1}} & = \frac{1}{\prod_{l=1}^N \pbra{1-\ft^{ \frac{2l-1}{2N+1}}}\pbra{1-\ft^{\frac{2N+2-2l}{2N+1}}}}\\[0.5em]
& + \sum_{i=1}^N  \frac{\ft^{k \mu_i^{(1)}} + \ft^{k \mu_i^{(2)}}}{\prod_{l=1}^{i} \pbra{1-\ft^{\frac{2N+2l}{2N+1}}}\pbra{1-\ft^{ -\frac{2l-1}{2N+1}}}\prod_{l=i+1}^{N} \pbra{1-\ft^{\frac{2l-1}{2N+1}}}\pbra{1-\ft^{ \frac{2N+2-2l}{2N+1}}}}
\label{tM2N-1V}
\ee
which precisely agrees with the mathematical calculation in Appendix~\ref{sec: fixed}. The Morse index of $P_i$ is again $2i$, giving the Poincar\'{e} polynomial of the moduli space
\be
\CP({\widetilde {\cal M}}_{2, 2N-1}) = 1 + 2 \fr^2 + 2 \fr^4 + \dots 2 \fr^{2N}.
\ee

\subsubsection{The moduli space $\CM_{2, 2N}$}

Compared to its cousin $\CM_{2, 2N+1}$, the moduli space $\CM_{2, 2N}$ depends on an additional parameter $\alpha$ giving the formal monodromy of the gauge field around the irregular singularity, again subject to the quantization condition $2k\alpha=0,1,\ldots,k$. On the physics side, it is identified with the holonomy of the $U(1)_B$ flavor symmetry of the $(A_1, A_{2N-1})$ theory. 

From this point forward, the level of difficulty in finding fixed points via geometry increases significantly; on the contrary, the physical computation is still tractable, yielding many predictions for the moduli space.

When $N=1$ the physical theory is a single hypermultiplet, and the index is just a multiplicative factor \eqref{A2N-1Norm}. When $N = 2$ the moduli space is isomorphic to ${\widetilde {\cal M}}_{2, 1}$; and two Argyes-Douglas theories $(A_1, A_3)$ and $(A_1, D_3)$ are identical \cite{Xie:2012hs}. Hence in this section we begin with the next simplest example $\CM_{2, 6}$. After absorbing the normalization constant \eqref{A2N-1Norm} similar to previous examples, we arrive at the expression
\be
{\cal I}_{(A_1, A_5)} & = \frac{\ft^{\frac{k - \lambda}{2}} + \ft^{\frac{\lambda}{2}} + \ft^{\frac{k}{2}}}{(1-\ft^{\frac{6}{4}})(1-\ft^{\frac{5}{4}})(1-\ft^{-\frac{2}{4}})(1-\ft^{-\frac{1}{4}})} +  \frac{\ft^{\frac{k - \lambda}{4}} + \ft^{\frac{\lambda}{4}}}{(1-\ft^{\frac{3}{4}})(1-\ft^{\frac{5}{4}})(1-\ft^{\frac{1}{4}})(1-\ft^{-\frac{1}{4}})} \\[0.5em]
& + \frac{1}{(1-\ft^{\frac{3}{4}})(1-\ft^{\frac{2}{4}})(1-\ft^{\frac{2}{4}})(1-\ft^{\frac{1}{4}})}.
\ee

The index formula predicts that there are six fixed points under the Hitchin action, with their weights on the normal bundle manifest in the denominators. The Poincar\'{e} polynomial is then
\be
\CP(\CM_{2, 6}) = 1 + 2\fr^2 + 3 \fr^4.
\ee
And the equivariant volume is
\be
{\rm Vol}_{\beta}({\CM}_{2, 6}) = \frac{64}{15 \beta^4} \pbra{e^{-\frac{1-2\alpha}{2}\beta} + e^{-\frac{2\alpha}{2}\beta} + e^{-\frac{1}{2}\beta} - 4 e^{-\frac{1-2\alpha}{4}\beta} - 4e^{-\frac{2\alpha}{4}\beta} + 5}.
\ee

We now write down the general formula for the Hitchin character of $\CM_{2, 2N}$. The moduli space has $N$ groups of fixed points. We label the group by $i = 0, 1, \dots, N-1$ with increasing Morse indices. The $i$-th group contains $i+1$ isolated fixed points $P_{i}^{(j)}$ with $j = 0, 1, \dots, i$. The weights on the normal bundle for each group are as follows:
\be
\epsilon_l & = \frac{N+1+l}{N+1}, \ \ {\tilde \epsilon}_l = -\frac{l}{N+1}, \ \ \ \ l = 1, 2, \dots , i\\[0.5em]
\epsilon_l & = \frac{N-l}{N+1}, \ \ {\tilde \epsilon}_l = \frac{l+1}{N+1}, \ \ \ \ l = i+1, i+2, \dots , N-1.
\ee
Within the group the moment maps are organized in a specific pattern:
\be
&j\ \text{odd:}\ \ \ \   \mu_i^{(j)} = \frac{(2i - j + 1)(j+1)}{4(N+1)} - \frac{i-j+1}{N+1} (2\alpha)\\[0.5em]
&j\ \text{even:}\ \ \ \   \mu_i^{(j)} = \frac{(2i - j + 2)j}{4(N+1)} + \frac{i-j}{N+1}(2\alpha).\\[0.5em]
\label{M2Nmoment}
\ee
Then the wild Hitchin character is
\be
{\cal I}({\cal M}_{2, 2N}) = \sum_{i= 0}^{N-1} \frac{\sum_{j=0}^i \ft^{k \mu_i^{(j)}}}{\prod_{l=1}^i \pbra{1-\ft^{\frac{N+1+l}{N+1}}}\pbra{1-\ft^{-\frac{l}{N+1}}} \prod_{l=i+1}^{N-1} \pbra{1-\ft^{ \frac{N-l}{N+1}}}\pbra{1-\ft^{ \frac{l+1}{N+1}}}}
\label{M2NV}
\ee
and from it we can write down immediately the Poincar\'{e} polynomial 
\be
\CP({\cal M}_{2, 2N}) = 1 + 2 \fr^2 + 3 \fr^4 + 4 \fr^6 + \dots + N \fr^{2(N-1)}.
\ee

In the large-$k$ limit, some of the moment maps $\mu_i^{(j)}$ in the numerator of \eqref{M2NV} will stay at $O(1)$ and become large after multiplied by $k$, even when $\lambda=2k\alpha$ is fixed, and the contribution from the corresponding fixed points will be exponentially suppressed. We see that for each group in \eqref{M2Nmoment} only one fixed point survives, namely the one with $j = 0$. These fixed points are the only ones visible in the three-dimensional reduction of Argyres-Douglas theories. We will revisit this problem in Section~\ref{sec: 4dto3d}.

\subsubsection{The moduli space ${\widetilde \CM}_{2, 2N-2}$} 

We now turn to the last of the four families of wild Hitchin moduli spaces, ${\widetilde \CM}_{2, 2N-2}$, which is arguably also the most complicated. It is the moduli space associated with Riemann sphere with one irregular singularity $I_{2, 2N-2}$ and one regular singularity, with monodromy parameters $\alpha_1$ and $\alpha_2$. The corresponding Argyres-Douglas theory $(A_1, D_{2N})$ generically has $U(1) \times SU(2)$ flavor symmetry, and $\lambda_1=2k\alpha_1$ and $\lambda_2=2k\alpha_2$ in \eqref{D2NMonod} label their holonomies along the Hopf fiber of $L(k,1)$. 

Let us again start from the simplest example: ${\widetilde \CM}_{2, 2}$ or $(A_1, D_4)$ Argyres-Douglas theory. The hyper-K\"{a}hler dimension of this moduli space is again one; we thus expect to understand the geometric picture more concretely. Modulo the normalization constant, \eqref{D2NNorm}, we have
\be
{\cal I}_{(A_1, D_4)} = \frac{\ft^{k \mu^{(0)}_1} + \ft^{k \mu^{(1)}_1} + \ft^{k \mu^{(2)}_1}}{(1-\ft^{\frac{3}{2}})(1-\ft^{-\frac{1}{2}})} + \frac{1}{(1-\ft^{\frac{1}{2}})(1-\ft^{\frac{1}{2}})}.
\label{A1D4V}
\ee
The moment map values are
\be
\mu_1^{(0)} & = \frac{1}{2} - \frac{1}{2k} \max  \pbra{\dbra{\lambda_1 + \frac{\lambda_2}{2}}, \lambda_2}\\[0.5em]
\mu_1^{(1)} & =  \frac{1}{2k}\min\pbra{\dbra{\lambda_1 + \frac{\lambda_2}{2}}, \lambda_2}\\[0.5em]
\mu_1^{(2)} & = \frac{1}{2k} \max  \pbra{\dbra{\lambda_1 + \frac{\lambda_2}{2}}, \lambda_2} - \frac{1}{2k} \min  \pbra{\dbra{\lambda_1 + \frac{\lambda_2}{2}}, \lambda_2}.
\label{momentD4}
\ee
Here, when $(\lambda_1+\lambda_2/2) \notin \mathbb{Z}$, the character formula \eqref{A1D4V} shall be set to zero. 

\begin{figure*}[htbp]
\begin{adjustwidth}{0.4cm}{}
      \begin{subfigure}[t]{.4\textwidth}
        \includegraphics[width=5cm]{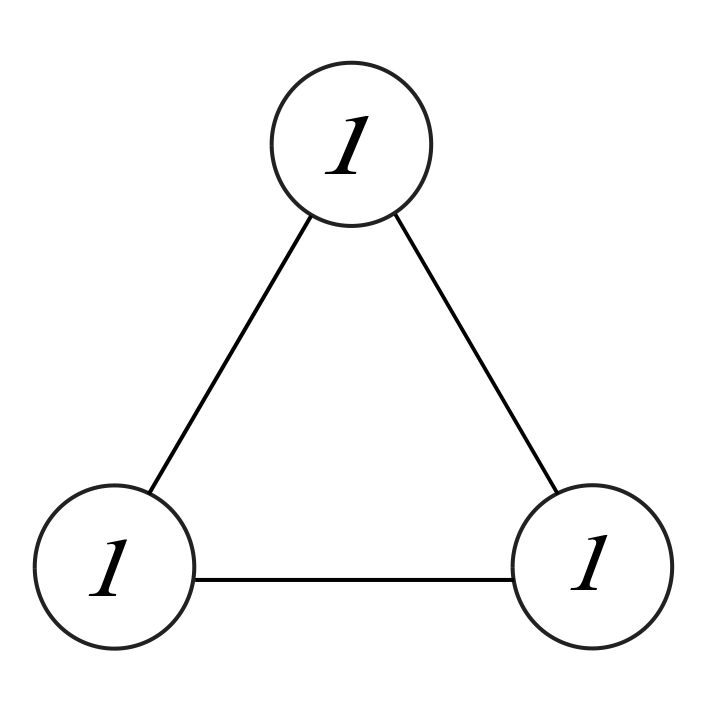}
        \caption*{}
      \end{subfigure}
      \hspace{1.5cm}
      \begin{subfigure}[t]{.4\textwidth}
        \includegraphics[width=6cm]{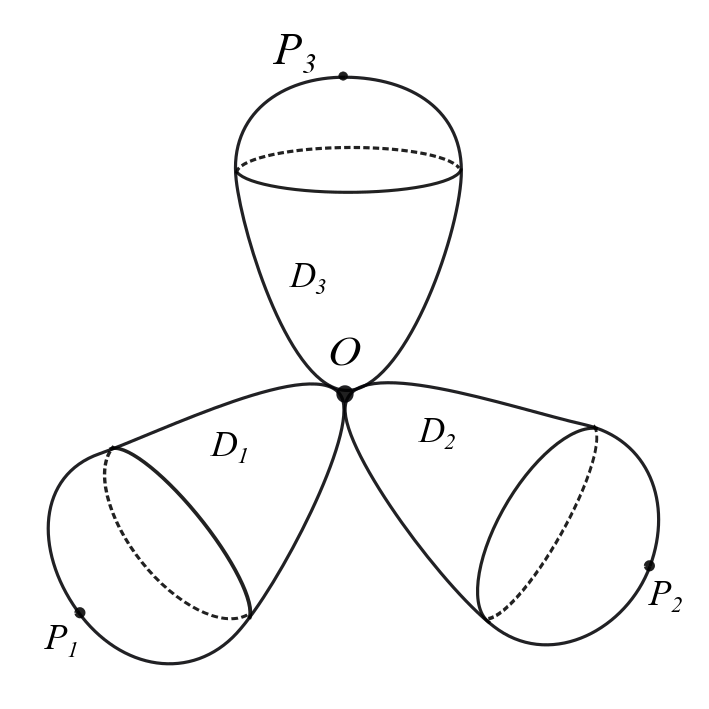}
        \caption*{}
      \end{subfigure}
  \end{adjustwidth}
 \caption[The affine $A_2$ Dynkin diagram and the nilpotent cone of Kodaira type \Rnum{4}]{Left: the affine $A_2$ Dynkin diagram, with Dynkin label indicated at each node. Right: the nilpotent cone of singular fibration, consisting of three $\mathbb{C}\mathbf{P}^1$ intersecting at $O$. The spheres are distorted a little to accommodate the common intersection. Together with $P_1$, $P_2$ and $P_3$, they comprise the four fixed points of the Hitchin moduli space ${\widetilde \CM}_{2, 2}$.}
\label{Kodaira4}
\end{figure*}

From the wild Hitchin character \eqref{A1D4V}, we know the Poincar\'{e} polynomial is
\be
\CP({\widetilde \CM}_{2, 2}) = 1 + 3 \fr^2.
\ee
${\widetilde \CM}_{2, 2}$ is another elliptic surface, and the nilpotent cone is of Kodaira type \Rnum{4} \cite{Argyres:2015ffa}, labeled by the affine $A_2$ Dynkin diagram. It contains three $\mathbb{C}\mathbf{P}^1$'s, which we denote as $D_{1,2,3}$, and the intersection matrix is given by
\be
\left( \begin{array}{ccc} -2 & 1 & 1 \\ 1 & -2 & 1 \\ 1 & 1 & -2 \end{array} \right).
\ee
$D_{1,2,3}$ each contains one of the three fixed points with Morse index $2$, see Figure \ref{Kodaira4} for illustration. The null vector of the intersection matrix gives the homology class of the Hitchin fiber,
\be
\left[ {\bf F} \right] = 2\left[ D_1 \right] + 2\left[ D_2 \right] + 2\left[ D_3 \right] 
\ee
which can be translated into a relation about the volumes
\be
{\rm Vol}({\bf F}) = 2{\rm Vol}(D_1) + 2{\rm Vol}(D_2) + 2{\rm Vol}(D_3).
\label{D4Vol}
\ee
Indeed, the three moment map values \eqref{momentD4} satisfy 
\be
4\mu_1^{(0)} + 4\mu_1^{(1)} + 4\mu_1^{(2)} = 2={\rm Vol}({\bf F}).
\ee

We now write down the general wild Hitchin character for the moduli space ${\widetilde \CM}_{2, 2N-2}$. There are $N$ groups of fixed points, we label them as $i = 0, 1, \dots, N-1$. The $i$-th group contains $2i+1$ fixed points with Morse index $i$. The expression looks like
\be
{\cal I}({\widetilde \CM}_{2, 2N-2}) = \sum_{i=0}^{N-1} \frac{\sum_{j=0}^{2i} \ft^{k \mu_i^{(j)}}}{\prod_{l=1}^i \pbra{1-\ft^{\frac{l+N}{N}}}\pbra{1-\ft^{-\frac{l}{N}}}\prod_{l=i+1}^{N-1} \pbra{1-\ft^{\frac{l}{N}}}\pbra{1-\ft^{\frac{N-l}{N}}}}
\label{tM2NV}
\ee
Explicit formulae for the moment map $\mu_i^{(j)}$ when $\lambda_1$ and $\lambda_2$ are zero are given after \eqref{D2Nh}. In general, they are functions of $\dbra{\lambda_1 + \lambda_2 / 2}$ and $\lambda_2$, 
with the quantization condition of  $(\lambda_1+\lambda_2/2)$ being an integer. Moreover, for the $i$-th group of fixed points, the sum of the moment map values,
\be
\sum_{j=0}^{2i}\mu_i^{(j)} = \frac{1}{6N} i (i+1)(2i+1),
\ee
is independent of the monodromy parameters.

We can similarly obtain the Poincar\'{e} polynomial for this moduli space,
\be
\CP({\widetilde \CM}_{2, 2N-2}) = 1 + 3 \fr^2 + 5\fr^4 + \dots + (2N - 1) \fr^{2N-2}.
\ee

\subsection{Fixed points from the three-dimensional mirror theory}\label{sec: 4dto3d}

One interesting limit of the superconformal index on $S^1 \times L(k,1)$ is the large-$k$ limit, where the Hopf fiber shrinks and the spacetime geometry effectively becomes $S^1 \times S^2$.  In this limit, the 4d ${\cal N}=2$ theory becomes a three-dimensional ${\cal N}=4$ theory $T_{\text{3d}}[\Sigma, G]$. Its 3d mirror $T_{\text{3d}}^{\text{mir.}}[\Sigma, G]$ sometimes admits a Lagrangian description \cite{Benini:2010uu, Nanopoulos:2010bv}. The original Coulomb branch vacua of $T_{\text{3d}}[\Sigma, G]$ becomes the Higgs branch vacua in the mirror frame. What is the relation between the Hitchin moduli space $\CM_H$ and the Coulomb branch $\CM^*$ of $T_{\text{3d}}[\Sigma, G]$? Intuitively, we expect that the latter is an ``approximation'' of the former because some degrees of freedom become massive and integrated out. More precisely, under the RG flow to the IR, we zoom in onto a small neighborhood of the origin of the Coulomb branch. As a consequence, the Coulomb branch $\CM^*$ of $T_{\text{3d}}[\Sigma]$ is a linearized version of $\CM_H$, given by a finite-dimensional hyper-K\"ahler quotient of vector spaces --- in other word, $\CM^*$ is a quiver variety consisting of holomorphically trivial $G_\C$-bundle over $\Sigma$.

This precisely agrees with the discovery of \cite{boalch2008irregular}: there it was proved mathematically that the wild Hitchin moduli space $\CM_H$ contains the quiver variety $\CM^*$ as an open dense subset, parametrizing irregular connections on a trivial bundle on $\mathbb{C}\mathbf{P}^1$. Furthermore, $\CM^*$ contains a subset of the $U(1)$ fixed points in $\CM_H$. These fixed points can be identified with massive vacua of $T_{\text{3d}}^{\text{mir.}}[\Sigma, G]$ on the Higgs branch, giving much easier access to them compared with the rest.\footnote{Note that no analogue exists in four dimensions, simply because Coulomb branch cannot be lifted without breaking supersymmetry.} To recap, we have the following relations:
\begin{equation}
\boxed{
\begin{aligned}
\text{Hitchin moduli space}\ \CM_H  & \leadsto \ \text{quiver variety}\ \CM^* \\[0.5em]
\text{Coulomb branch of}\ T[\Sigma] \ \text{on}\ S^1 & \leadsto \ \text{Higgs branch of $T_{\text{3d}}^{\text{mir.}}[\Sigma]$} \\[0.5em]
\text{``lowest'' fixed points on } \CM_H & \leadsto \ \text{massive Higgs branch vacua}
\end{aligned}
}\ .
\end{equation}

These relations also suggest that there is  a relation between the Hitchin character and the Higgs branch index of $T_{\text{3d}}^{\text{mir.}}[\Sigma]$, as we will show below. Recall that the 3d $\cN = 4$ index is given by \cite{Razamat:2014pta} 
\be
{\cal I}^{3\text{d}}_{\CN = 4} = {\rm Tr}_{\cal H} (-1)^F \fq^{j_2 + \frac{1}{2}(R_H + R_C)} \fv^{R_H - R_C} e^{-2\beta({\tilde E} -R_H - R_C - j_2)},
\label{3dN4Ind}
\ee
where $j_2$ is the angular momentum with respect to the Cartan of the $SO(3)$ Lorentz group and $R_{C,H}$ are respectively the Cartans of $SU(2)_C \times SU(2)_H$ R-symmetry. There are two interesting limits:
\be
& \text{Coulomb limit}: \ \ \ \ \fq, \fv \rightarrow 0, \ \ \ \ \frac{\fq^{\frac{1}{2}}}{\fv} = \ft \ \ {\rm fixed}, \\[0.5em]
& \text{Higgs limit}: \ \ \ \ \fq, \fv^{-1} \rightarrow 0, \ \ \ \ \fq^{\frac{1}{2}} \fv = \ft' \ \ {\rm fixed}.
\label{3dLimit}
\ee
As we will work with $T_{\text{3d}}^{\text{mir.}}[\Sigma]$ in the mirror frame, the Higgs branch limit is that one that interests us.

\subsubsection*{3d mirror of $(A_1,A_{2N-1})$ theory}
To begin with, let us first turn to $(A_1, A_{2N-1})$ theory whose three-dimensional mirror is $\CN = 4$ SQED with $N$ fundamental hypermutiplets. The Higgs branch has an $SU(N)$ flavor symmetry while the Coulomb branch has $U(1)_J$ topological symmetry that can be identified with the flavor symmetry of the initial $(A_1, A_{2N-1})$ theory. Let $(z_i, m_i)$ be the fugacities and monopole numbers for the $SU(N)$ flavor symmetry and let $(b, n)$ be the fugacity and monopole number for the $U(1)_J$ topological symmetry.  The fugacities $z_i$ are subject to the constraint $\prod_i z_i = 1$, while $m_i$ will all be zero. The Higgs branch index is given by
\be
{\cal I}^{3\text{d}}_H & = (1-\ft') \prod_{i = 1}^N \delta_{m_i, 0}  \oint \frac{dw}{2\pi i w} w^{Nn} \prod_{i = 1}^N \frac{1}{(1-{\ft'}^{\frac{1}{2}} w z_i)(1-{\ft'}^{\frac{1}{2}} w^{-1} z_i^{-1})}\\[0.5em]
& = \pbra{\prod_{i = 1}^N \delta_{m_i, 0} } \sum_{i = 1}^N {\ft'}^{\frac{|Nn|}{2}} z_i^{-|Nn|}\prod_{j \neq i} \frac{1}{1-\ft' z_j / z_i} \frac{1}{1 - z_i / z_j}.
\label{Higgs3d}
\ee
To recover the $k \rightarrow +\infty$ limit of the $(A_1, A_{2N-1})$ Coulomb branch index \eqref{M2NV}, we make the following substitution:
\be
z_i \rightarrow {\ft'}^{(N+1-2i)/(2N+2)}, \ \ \ \ i = 1,2,\dots, N.
\label{topMix}
\ee
This substitution \eqref{topMix} can be interpreted as the mixing between topological symmetry and $SU(2)_C$ symmetry on the Coulomb branch of $T_{\text{3d}}[\Sigma]$, which is further examined in Appendix~\ref{sec: 3d red}. After the substitution, the index can be written as
\be
{\cal I}^{3\text{d}}_H & = \ft'^{\frac{1}{N+1} |Nn|} \sum_{i = 1}^N \frac{\ft'^{\frac{i-1}{N+1} |Nn|}}{\prod_{j \neq i} \pbra{1-\ft'^{\frac{N+1+i-j}{N+1}}} \pbra{1-\ft'^{\frac{j-i}{N+1}}} },
\label{A1A2N-13d}
\ee
where each term in the summation is the residue at a massive vacuum. Comparing to the Hitchin character \eqref{M2NV}, one finds that only a subset of fixed points in $\CM_H$ contribute to ${\cal I}^{3\text{d}}_H$.  Namely, these are fixed points that live in $\CM^*\subset\CM_H$.  

For pedagogy, we describe these massive supersymmetric vacua explicitly. Our description is again in the mirror frame and one can easily interpret them in the original frame. First we turn on the real FI parameter $t_{\mathbb{R}}$, and the Higgs branch (which is a hyper-K\"{a}hler cone) gets resolved to be $T^*\mathbb{C}\mathbf{P}^{N-1}$. The $SU(N)$ flavor symmetry and $SU(2)_H$ acts on $T^*\mathbb{C}\mathbf{P}^{N-1}$, and the $U(1)$ Hitchin action is embedded into the Cartan of $SU(N)\times SU(2)_H$, with the embedding given by \eqref{topMix}. Then, one can study the fixed points under this $U(1)$ subgroup. It turns out that there are $N$ of them, computed in Appendix~\ref{app: Massive}. As the equivariant parameters of the $SU(N)$ flavor symmetry are the masses of hypermultiplets, these fixed points can be interpreted as massive vacua of the theory when mass parameters are turned on according to the mixing \eqref{topMix}. 

On the other hand, from the perspective of $\CM_H$, the contributing fixed points are also straightforward to identify: they are precisely the ones whose moment map values multiplied by $k$ remain finite in the large-$k$ limit, and there are precisely $N$ of them. Summing up their contributions gives back \eqref{A1A2N-13d}. 

\subsubsection*{3d mirror of $(A_1,D_{2N})$ theory}
\begin{figure}[htbp]
\centering
\includegraphics[width=7cm]{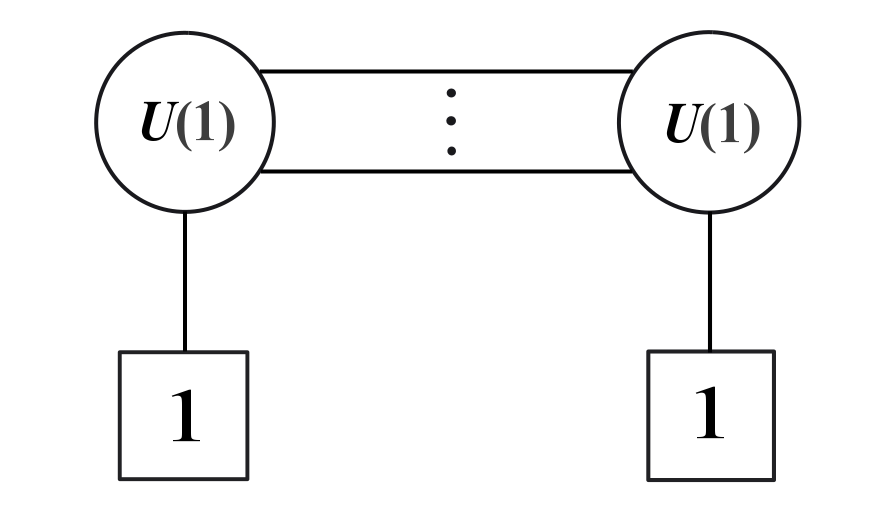}
\caption[Quiver diagram for $(A_1, D_{2N})$ 3d mirror]{The 3d mirror of $(A_1, D_{2N})$ theories. There are $N-1$ hypermultiplet between two $U(1)$ gauge nodes, and there are additional one hypermultiplet charged under each node.}
\label{D2Nquiver}
\end{figure}

Now we turn to Argyres-Douglas theories of type $(A_1, D_{2N})$, which are also known to have three-dimensional mirrors with Lagrangian descriptions \cite{Xie:2012hs}. The mirror theory of $(A_1, D_{2N})$ is given by a quiver $U(1) \times U(1)$ gauge theory, with $N-1$ charged hypermultiplets between two gauge nodes. These hypermultiplets enjoy an $SU(N-1)$ flavor symmetry. Moreover, there is one hypermultiplet only charged under the first $U(1)$ gauge group while another hypermultiplet is charged only under the second $U(1)$ gauge group. There is also an additional $U(1)$ flavor symmetry that rotates $N+1$ hypermultiplets together with charge $1/2$. See the quiver diagram in Figure~\ref{D2Nquiver}. 

The index computation is similar. We will use $N (n_2 - n_1)$ and $N(n_1 + n_2)$ to denote monopole numbers for the $U(1)\times U(1)$ topological symmetry on the Coulomb branch. They come from the combination of flavor holonomies of the parent Argyres-Douglas theory. Besides the fugacity $z$ for $U(1)$ flavor symmetry, we also include $a_i$, $i = 1, \dots, N-1$ as the fugacities for the extra $SU(N-1)$ flavor symmetry, subject to the constraint $\prod a_i = 1$. The associated background flavor monopole numbers all vanish, similar to the previous case. Then we have the index formula:
\be
{\cal I}^{3\text{d}, D_{2N}}_H = (1-\ft')^2 & \oint \frac{dw_1}{2\pi i w_1}\frac{dw_2}{2\pi i w_2} w_1^{N (n_2 - n_1)} w_2^{N(n_1 + n_2)}\\[0.5em]
& \times \frac{1}{1-{\ft'}^{\frac{1}{2}} (w_1 z^{\frac{1}{2}})^{\pm}} \frac{1}{1-{\ft'}^{\frac{1}{2}} (w_2 z^{\frac{1}{2}})^{\pm}} \prod_{i=1}^{N-1} \frac{1}{1-{\ft'}^{\frac{1}{2}} (w_1 w_2^{-1} a_i z^{\frac{1}{2}})^{\pm}}.
\ee
In the computation we have set $z = 1$ as it will not mix with the R-symmetry (see Appendix~\ref{sec: 3d red} for more details). To evaluate the integral, we can assume without loss of generality that $n_2 > n_1 > 0$. Then summing over residues gives
\be
{\cal I}^{3\text{d}, D_{2N}}_H & = {\ft'}^{Nn_2} \prod_{i=1}^{N-1} \frac{1}{(1-{\ft'}^{\frac{1}{2}} a_i)(1-{\ft'}^{\frac{1}{2}} a_i^{-1})}\\[0.5em]
& + \sum_{j=1}^{N-1} \frac{(\ft' a_j)^{N(n_1+n_2)} {\ft'}^{\frac{N}{2}(n_2-n_1)}}{(1-{\ft'}^{\frac{3}{2}} a_j)(1-{\ft'}^{-\frac{1}{2}} a_j^{-1})} \prod_{i \neq j} \frac{1}{(1-\ft' a_j / a_i)(1-a_i / a_j)}\\[0.5em]
& + \sum_{j=1}^{N-1} \frac{{\ft'}^{Nn_2} \pbra{{\ft'}^{\frac{1}{2}} a_j^{-1}}^{N(n_2 - n_1)}}{(1-{\ft'}^{\frac{3}{2}} a^{-1}_j)(1-{\ft'}^{-\frac{1}{2}} a_j)} \prod_{i \neq j} \frac{1}{(1-\ft' a_i / a_j)(1-a_j / a_i)}.
\label{D2N+13d}
\ee
It is not hard to see the following substitution would recover the parent Hitchin character \eqref{tM2NV} at $k \rightarrow +\infty$:
\be
a_j \rightarrow {\ft'}^{\frac{j}{N} - \frac{1}{2}}.
\label{topMix2}
\ee

Similarly, the residue sums in \eqref{D2N+13d} are in one to one correspondence with massive vacua of the 3d mirror theory, which are also identified with the fixed points under the $U(1) \subset SU(N-1)\times SU(2)_H$ action on the Higgs branch. Explicit calculations done in Appendix \ref{app: Massive} show that there are precisely $2N+1$ fixed points, which, from Hitchin moduli space point of view, are exactly those with vanishing moment map in the large-$k$ limit.

In summary, considering the three-dimensional mirror theory gives physical interpretation to the fixed points in $\CM^*$ as discrete vacua of the mass-deformed theory. The fixed-point sum can be thought of as a sum of residues in the Higgs branch localization \cite{Peelaers:2014ima}. 

\section{Chiral algebras} \label{Sec: CA}

In previous sections, we have given a very strong test of the proposed isomorphism \eqref{CBIQuant} for Argyres-Douglas theories. In this section, we enrich this correspondence to the triangle \eqref{Triangle} by introducing another player into the story --- chiral algebras. 

\subsection{Chiral algebra from geometric Langlands correspondence}

One motivation for incorporating chiral algebras is the celebrated geometric Langlands correspondence (see \cite{Frenkel:2005pa} and \cite{Frenkel:2009ra} for pedagogical reviews on this subject), which conjectures the equivalence of two derived categories,
\be\label{GeomLang}
\boxed{\text{$\CD$-modules on $\mathrm{Bun}_{G_\C}$}} \quad {=} \quad \boxed{\text{coherent sheaves on $\mathrm{Loc}_{^LG_{\mathbb{C}}}$}}\ .
\ee
The gauge theory approach to the geometric Langlands program, started by \cite{Kapustin:2006pk}, suggests that the above relation naturally fits inside a triangle,
\be\label{LangTriangle}
\begin{array}{rcl}
\text{A-branes in $(\CM_H,\omega_K)$}&\overset{\textcircled{\raisebox{-0.9pt}{1}}}{\longleftrightarrow} &\text{B-branes in $(^L\CM_H,J)$}\\
\\
\textcircled{\raisebox{-0.9pt}{2}}$\rotatebox[origin=c]{-45}{$\longleftrightarrow$}$   &  & $\rotatebox[origin=c]{45}{$\longleftrightarrow$} $\textcircled{\raisebox{-0.9pt}{3}} \\
\\
& \text{$\CD$-modules on $\mathrm{Bun}_{G_{\mathbb{C}}}$ }&.
\end{array}
\ee
The geometric Langlands correspondence  \eqref{GeomLang} now becomes the arrow \textcircled{\raisebox{-0.9pt}{3}} on the bottom-right of \eqref{LangTriangle}, as the B-brane category of $^L\CM_H$ is closely related to the derived category of coherent sheaves on $\mathrm{Loc}_{^LG_{\mathbb{C}}}$. The arrow \textcircled{\raisebox{-0.9pt}{1}} on the top is the homological mirror symmetry (or S-duality from the 4d guage theory viewpoint). The arrow \textcircled{\raisebox{-0.9pt}{2}},  a new relation, was proposed in Section~11 of \cite{Kapustin:2006pk} and is related to the ``brane quantization'' of $\mathrm{Bun}_{G_{\mathbb{C}}}$ \cite{Gukov:2008ve} (see also \cite{ Witten:2015dta} for more examples and \cite{2006math......4379N,2006math.....12399N} for an alternative way to establish the equivalence).  

Now let us return to the diagram
\be
\begin{array}{rcl}
\text{Coulomb index of $\CT$}&\longleftrightarrow &\text{quantization of ${^L}{\CM}_{\CT}$}\\
\\
$\rotatebox[origin=c]{-45}{$\longleftrightarrow$}$   &  & $\rotatebox[origin=c]{45}{$\longleftrightarrow$} $ \\
\\
& \text{chiral algebra $\chi_{\CT}$}&
\end{array}.
\label{triangle2}
\ee
The top arrow for class $\CS$ theories explained in \cite{Gukov:2016lki} is in fact the result of \textcircled{\raisebox{-0.9pt}{1}} in \eqref{LangTriangle} as we review below. Then one expects there is a chiral algebra that fits into the diagram, giving rise to $\CD$-modules via the conformal block construction (see \eg~part \Rnum{3} of \cite{Frenkel:2005pa}). 

To understand the top arrow from homological mirror symmetry, one first rewrites the Coulomb BPS states on $L(k,1)$, view as $T^2$ fibered over an interval,\footnote{As observed in \cite{Gukov:2016lki} and \cite{Gukov:2015sna}, the Coulomb index is the same as a topologically twisted partition function. This enables us to treat the physical theory as if it is a TQFT and freely deform the metric on $L(k,1)$. } in the categorical language
\be\label{AHom}
\CH_{\text{Coulomb}}=\mathrm{Hom}_{\CC_A}(A_0,ST^kS\cdot A_0).
\ee  
Here $\CC_A$ is the category of boundary conditions on $T^2$  (or ``A-branes'' in $\CM_H$) of the Argyres-Douglas theory, and $A_0\in \CC_A$ is the boundary condition given by the solid torus $D^2\times S^1$, and $ST^kS$ is an element of $SL(2,\Z)$ that acts on $\CC_A$ via the modular group action on $T^2$. Suppressing one $S^1$ circle and the time direction, the geometry near the endpoint of the interval is given by the tip of a cigar, and the brane $A_0$ associated with this geometry is conjectured to be the ``oper brane.'' The generator $S\in SL(2,\Z)$ acts as homological mirror symmetry, transforming $\CC_A$ into $\CC_B$ --- the category of B-branes in $^L\CM_H$, and the mirror of $A_0$ is expected to be $S\cdot A_0=B_0=\CO$, the structure sheaf of $^L\CM_H$. Then acting on \eqref{AHom} by $S$ gives
\be\label{A=BHom}
\mathrm{Hom}_{\CC_A}(A_0,ST^kS\cdot A_0)=\mathrm{Hom}_{\CC_B}(B_0,T^k\cdot B_0).
\ee
As $T\in SL(2,\Z)$ acts on objects in $\CC_B$ by tensoring with the determinant line bundle $\CL$, the right-hand side is precisely the geometric quantization of $^L\CM_H$,
\be\label{BHom}
\CH(\Sigma,^L\!\! G,k)=H^\bullet\left(^L\!\CM_H,\CL^{\otimes k}\right)=\mathrm{Hom}_{\CC_B}(B_0,T^k\cdot B_0).
\ee
If a chiral algebra fits into the triangle \eqref{triangle2} via the correspondence between A-branes and $\CD$-modules, there should be a modular tensor category $\CC_\chi$ of representations of the chiral algebra, and there is a similar vector space
\be\label{CHom}
\mathrm{Hom}_{\CC_\chi}(\chi_0,ST^kS\cdot \chi_0).
\ee
The module $\chi_0$ corresponding to the oper brane $A_0$ is expected to be the vacuum module, and $ST^kS$ acts by modular transform. The ``geometric Langlands triangle'' \eqref{LangTriangle} states that all the above three vector spaces are isomorphic, which implies, at the level of dimensions,
\be\label{ChiralMod}
\dim \CH(\CM_H) = \CI_{\text{Coulomb}}=(ST^kS)_{0,0}.
\ee  
As the first two quantities can be refined by $\ft$, one expects the $S$- and $T$-matrices for the chiral algebra should also be refined. However, for the chiral algebras that will appear (such as Virasoro minimal models), the refinement is not known, and we will only check the relation \eqref{ChiralMod} at a root of unity $\ft=e^{2\pi i}$.\footnote{As the wild Hitchin character involves fractional powers of $\ft$, such limit is different from $\ft \rightarrow 1$ and is in fact associated with a non-trivial root of unity. Also, the ambiguity of normalizing the Hitchin character by a monomial in $\ft$ now becomes the ambiguity of a phase factor in matching \eqref{ChiralMod}. }

\subsubsection*{With flavor holonomy}

Moreover, with flavor symmetry $G$ from the singularities of the Riemann surface, we also consider the Coulomb index on $L(k,1)$ in the presence of a flavor holonomy along the Hopf fiber labeled by $\lambda\in \Lambda_{\text{cochar}}(G)/k\Lambda_{\text{cochar}}(G)$. This is equivalent to inserting a surface defect at the core of a solid torus in the decomposition of $L(k,1)$, carrying a monodromy determined by $\lambda$. It will change \eqref{AHom} into   
\be\label{AHom'}
\CH_{\text{Coulomb}}(\lambda)=\mathrm{Hom}_{\CC_A}(A_0, ST^k S\cdot  A_\lambda),
\ee  
where
\be
A_\lambda=L_\lambda A_0
\ee
with $L_\lambda$ representing the action of the surface defect on boundary conditions. These defects are analogous to the  't Hooft line operators --- in fact, they are constantly referred to as ``'t Hooft-like operators'' in \cite{Gukov:2006jk} --- and change the parabolic weights at the singularities on $\Sigma$. Then, the relation between A-branes and $\CD$-modules predicts that there exists a corresponding operator (which we again denote as $L_\lambda$) in the category $\CC_\chi$. Now, the chiral algebra has $\hat{\frak{g}}$ affine Kac-Moody symmetry, whose modules are labeled by the weights $\lambda$ of $\hat{\frak{g}}$, and one expects the action of $L_{\lambda}$ on the vacuum module is given by 
\be
L_\lambda\cdot \chi_0=\chi_{-\lambda}.
\ee
Then, in the presence of flavor holonomies, one expects the following relation
\be\label{ChiralMod2}
\dim \CH(\CM_H,\lambda) = \CI_{\text{Coulomb},\lambda}=(ST^kS)_{0,-\lambda}.
\ee

At this stage we do not know \textit{a priori} what is the right chiral algebra when $\CM_H$ is a wild Hitchin moduli space, but we conjecture that it is given by the chiral algebra under the ``SCFT/chiral algebra correspondence'' discovered in \cite{Beem:2013sza, Beem:2014rza, Lemos:2014lua, Buican:2015ina, Cordova:2015nma, Xie:2016evu}. Indeed, for theories of class $\CS$, this correspondence gives, for each maximal tame puncture, an affine Kac-Moody symmetry at the critical level --- the one that gives rise to a specific type of $\CD$-modules central to the geometric Langlands program known as Hecke eigensheaves. In the rest of this section, we will review this correspondence and check that the above relations \eqref{ChiralMod} and \eqref{ChiralMod2} hold for wild Hitchin moduli spaces. It will be an interesting problem to explain why this construction gives the correct $\CD$-modules relevant for this particular problem.

Moreover, as shown in \cite{Cordova:2016uwk}, general characters of certain $2d$ chiral algebras can be expressed by the Schur indices with line operator insertion of corresponding $4d$ theory. Our results can be interpreted as a relation between the Coulomb branch indices and the modular transformation of Schur indices with line operator insertion of AD theories. The modular properties of indices without any operator insertion of $4d$ theories are studied in \cite{Spiridonov:2012ww, Razamat:2012uv} and their modular properties are related to the 't Hooft anomalies of the theory. It is interesting to further study the $4d$ interpretation of modular $S$ transformations on indices with line operator insertion and their relation with Coulomb branch indices.

\subsection{2d chiral algebras from 4d SCFTs}\label{sec: 4dto2d CA}

As was first discovered in \cite{Beem:2013sza}, every four-dimensional $\CN = 2$ superconformal theory contains a protected subsector of BPS operators, given by the cohomology of certain nilpotent supercharge $\mathbbmtt{Q}$, when these operators lie on a complex plane inside $\R^4$. These BPS operators are precisely the ones that enter into the Schur limit of the 4d $\CN=2$ superconformal index \cite{Gadde:2011uv}. Moreover, the operator product expansion (OPE) of these operators are meromorphic, and they can be assembled into a two-dimensional \textit{chiral algebra}. The central charges of the 4d SCFT and the 2d chiral algebra are related by
\be
c_{2d} = -12 c_{4d}
\label{2d4dC}
\ee
which implies that all chiral algebras obtained in this way are necessarily non-unitary. If the parent four-dimensional theory enjoys a global symmetry given by a Lie group, then it will be enhanced to an affine Lie symmetry on the chiral algebra side. The relation between the flavor central charge and the level for the affine symmetry is given by
\be
k_{2d} = -\frac{1}{2} k_{4d}.
\ee

Examples of these chiral algebras are identified on a case-by-case basis \cite{Beem:2014rza, Lemos:2014lua, Buican:2015ina, Cordova:2015nma, Xie:2016evu}. We listed some examples of Argyres-Douglas theories in Table~\ref{table:ADexamples}. For the case of $(A_1, A_{2N-1})$ and $(A_1, D_{2N})$, the chiral algebras are identified very recently in \cite{CreutzigVOA}.

\begin{table}
\centering
\begin{tabular}{c|c}
AD theory & chiral algebra \\
\hline
$(A_1, A_{2N})$ & $(2, 2N+3)$ minimal model\\
$(A_1, A_{2N-1})$ & $\CB_{N+1}$ algebra \\
$(A_1, D_{2N+1})$ & ${\widehat{\mathfrak{sl}}}(2)_k$ at level $k = -\frac{4N}{2N+1}$ \\
$(A_1, D_{2N})$ & $\CW_N$ algebra
\end{tabular}
\caption{\label{table:ADexamples}Examples of Argyres-Douglas theories and corresponding chiral algebras. To be more precise, in the $(A_1, A_{2N-1})$ case, it is the subregular quantum Hamiltonian reduction of ${\widehat{\mathfrak{sl}}}(N)_k$ at level $k=-N^2/(N+1)$ \cite{Creutzig:2013pda, CreutzigVOA}. In the $(A_1, D_{2N})$ case, it is the non-regular quantum Hamiltonian reduction of ${\widehat{\mathfrak{sl}}}(N+1)_k$ with $k=-(N-1)^2/N$  \cite{CreutzigVOA}. For details about quantum Hamiltonian reduction, see \cite{kac2004quantum}.}
\label{ADandCA}
\end{table}

As was mentioned, the chiral algebra has a very close relationship with the Schur operators. In particular, the Schur limit of the superconformal index is equal to the vacuum character of the chiral algebra.\footnote{On the other hand, the Schur index that incorporates line defects maybe used to probe non-vacuum modules, see \cite{Cordova:2016uwk}.} In contrast, Coulomb branch operators do not enter into the $\mathbbmtt{Q}$-cohomology and are not counted by the Schur index. However, it turns out that the Coulomb branch index is related to the chiral algebra in a quite surprising manner --- the \textit{modular transformation} property of the latter is captured by the Coulomb branch index, as we have motivated using the geometric Langlands correspondence in \eqref{ChiralMod} and \eqref{ChiralMod2}. 

To check these relations explicitly, we need to identify the relevant representation categories $\CC_\chi$ of the chiral algebras listed in Table~\ref{ADandCA} that are closed under modular transforms. For the $(A_1,A_{2N})$ series, the answer is clear --- the $(2,2N+3)$ minimal model specifies a category of highest-weight modules of the Virasoro algebra. For the rest, we will also give the relevant category later in this section. But what about a more general theory $\CT$? Once we obtain the chiral algebra $\chi_\CT$, how is  the category $\CC_{\chi_\CT}$ that is relevant for the Coulomb index of $\CT$ constructed?

An obvious candidate would be the category of all representations of $\chi_\CT$, but it cannot be the right answer as it is too large and there are many non-highest-weight modules whose conformal dimensions are not bounded from below nor above. Nonetheless, there is a natural procedure, called ``semi-simplification" \cite{Creutzig:2016fms}, that gives precisely the category we are interested in. Specifically, one forms a new quotient category, denoted as $\CO^{\rm s}_{\chi_\CT}$, by modding out the negligible morphisms \cite{haahr1995fusion, bakalov2001lectures} and keeping only simple objects with non-zero categorical dimensions.  This category is believed to be a modular tensor category \cite{Creutzig:2016fms}, and in each class of modules there is at least one module with bounded conformal dimensions (the ``highest-weight" module). And we conjecture 
\be
\CO^{\rm s}_{\chi_\CT}=\CC_{\chi_\CT}
\ee
is the category fitting in the triangle \eqref{Triangle}.

This conjecture will be verified in the four series of Argyres-Douglas theories that we study in this paper. In the following we show that the wild Hitchin character (or Coulomb branch index) at $\ft \rightarrow e^{2\pi i}$ is indeed given by a matrix element of the modular transformation $ST^k S$ in $\CC_{\chi}$. In fact, in order for the relation \eqref{ChiralMod} to be correct for all $k$, it is necessary to have a one-to-one correspondence between fixed points in $\CM_H$ and modules in the category $\CC_\chi$.

\subsection{Chiral algebras of Argyres-Douglas theories}

\subsubsection{$(A_1, A_{2N})$ theories and Virasoro minimal models}

The observation of \cite{Cordova:2015nma}, by comparing the central charge \eqref{2d4dC}, indicates that the associated chiral algebra for $(A_1, A_{2N})$ Argyres-Douglas theory is the $(2, 2N+3)$ Virasoro minimal model. (Recall that $2N+3$ is also the number of Stokes rays centered at the irregular singularity.) The minimal model contains a finite number of highest-weight representations labeled by the conformal dimension $h_{r,s}$, where $s = 0$ and $1 \leq r + 1 \leq 2N+2$.\footnote{Unlike the usual convention in the literature here we shift $r$ and $s$ by $1$ so that the vacuum corresponds to $(r,s) = (0,0)$.} Among these representations, there are $N+1$ independent ones given by $r = 0, 1, \dots, N$ --- exactly the same as the number of fixed points in the wild Hitchin moduli space $\CM_{2, 2N+1}$!

In \cite{Fredrickson-Neitzke}, the one-to-one correspondence between the fixed points in $\CM_{2,2N+1}$ and representations in the Virasoro minimal model is spelled out.
Namely, if one defines the effective central charge
\be
c_{{\rm eff}} = c - 24 h_{r,s},
\ee
then there is a simple relation between $c_{{\rm eff}}$ and the moment map $\mu$
\be
\boxed{\mu = \frac{1}{24}(1- c_{{\rm eff}})}\ .
\label{A2NtoMinimal}
\ee
Here the moment map values are calculated around 
\eqref{muM2N+1},  \textit{without} the further shift we made in the last section. Later, we extend this observation to all the other types of wild rank-two Hitchin moduli spaces, with emphasis on the perspective of modular transformations, where this correspondence finds its natural home. 

To see the relation between the wild Hitchin character \eqref{M2N+1V} of $\CM_{2,2N+1}$ and the modular transformation of $(2,2N+3)$ minimal model, recall that characters of these $N+1$ modules form an $N+1$-dimensional representation of $SL(2, \mathbb{Z})$, with the $S$- and $T$-matrices given by
\be
\label{eq:STminimalModel}
\mathcal{S}_{r,\rho}&=\frac{2}{\sqrt{2N+3}}(-1)^{N+r+\rho} \sin \left(\frac{2\pi (r+1) (\rho+1)}{2N+3} \right),\\[0.5em]
\mathcal{T}_{r,\rho}&=\delta_{r \rho}e^{2\pi i(h_{r,\rho}-c/24)},
\ee
where $r$ and $\rho$ run from $0$ to $N$. With the help of \eqref{eq:STminimalModel} one can show that\footnote{We thank T.~Creutzig and D.~Gaiotto for discussion.}
\be
{\cal I}(\CM_{2, 2N+1}) = \ft^{\frac{k}{8(2N+3)}} \CI_{(A_1, A_{2N})} |_{\ft \rightarrow e^{2\pi i}} =
e^{\frac{ \pi i k}{12} } \left(\CS\CT^k\CS\right)_{0,0}.
\ee

\subsubsection{$(A_1, D_{2N+1})$ theories and Kac-Moody algebras}

It was conjectured in \cite{Cordova:2015nma, Xie:2016evu} that the corresponding chiral algebra is the affine Kac-Moody algebra $\widehat{\mathfrak{su}}(2)_{k_F}$ for which
\be
k_F = -2 + \frac{2}{2N+1}.
\ee
which is a boundary admissible level \cite{kac1988modular}. Notice that $-2$ is the critical level for $\widehat{\mathfrak{su}}(2)$, while $2N+1$ is again the number of Stokes rays on $\Sigma$. There is a notion of ``admissible representations"  for the Kac-Moody algebra, which is the analogue of integrable representations for Kac-Moody algebra at positive integer level (see \eg~\cite[Sec.~18]{francesco2012conformal}). These representations are highest-weight modules, and are objects in the quotient category $\CO^{\rm s}_\chi$. Their fusion rules and representation theory remained controversial for years, and were completely solved and understood (in the case of $N = 1$ for instance) recently in \cite{Creutzig:2012sd, Creutzig:2013yca} (see also the reference therein).

Let ${\hat \omega}_0$ and ${\hat \omega}_1$ be the fundamental weights of $\widehat{\mathfrak{su}}(2)$. A highest-weight representation for $\widehat{\mathfrak{su}}(2)_{\kappa}$ is called \textit{admissible}, if the highest weight ${\hat \lambda} = \left[ \lambda_0,  \lambda_1 \right]:= \lambda_0 {\hat \omega}_0 + \lambda_1 {\hat \omega}_1$, can be decomposed as
\be
{\hat \lambda} = {\hat \lambda}^I - (\kappa+2) {\hat \lambda}^F.
\ee
Here, if we write $\kappa = t / u$ with $t \in \mathbb{Z}\backslash \{ 0 \}$, then $u \in \mathbb{Z}^+$ and $(t, u) = 1$. In our case $t=-4N$ and $u=2N+1$. ${\hat \lambda}^I$ and ${\hat \lambda}^F$ are integrable representations for $\widehat{\mathfrak{su}}(2)$ at level $k^I = u(\kappa+2) - 2$ and $k^F = u - 1$ respectively. Specializing to our case, we see that ${\hat \lambda}^I = 0$ and ${\hat\lambda}^F = 2N$, so the admissible representations are in one-to-one correspondence with the $2N+1$ integrable representations of $\widehat{\mathfrak{su}}(2)_{2N}$. This is again the same number as the fixed points of the moduli space ${\widetilde \CM}_{2, 2N-1}$! Let us see if there is a similar relation for the moment maps.

For each admissible module, the conformal dimension is given by
\be
h_{{\hat \lambda}} = \frac{\lambda_1 (\lambda_1 + 2)}{4(\kappa+2)}.
\ee
If we denote the highest weight of the $i$-th integrable representation of $\widehat{\mathfrak{su}}(2)_{2N}$ as $\bbra{2N- i, i}$ for $i = 0, 1, \dots, 2N$, then we have
\be
\lambda^i_1 = -\frac{2i}{2N+1}, \ \ \ \ h^i_{{\hat \lambda}} = - \frac{i(2N+1 - i)}{2(2N+1)}.
\label{admis}
\ee
In order to see the relation between \eqref{admis} with the values of the moment map in \eqref{tM2N-1moment}, we relabel the indices. Additionally, to get rid of overall phase factors, we shift the moment map
\be
\mu \rightarrow \mu + \frac{1}{8(2N+1)} + \frac{2N}{2N+1} \alpha.
\ee
Such shift is not as \textit{ad hoc} as it appears --- the second term is the minimal moment map value computed in \eqref{D2N+1minMoment} for $\alpha = 0$, while the third term comes from the linear piece of the normalization factor \eqref{D2N+1Norm}. Then, we have the following correspondence
\be
\boxed{\mu = \pbra{h_{{\hat \lambda}} - \frac{c}{24} + \frac{1}{8}} - \lambda_1 \alpha}\ .
\ee
Hence the moment maps in \eqref{tM2N-1V} are in one-to-one correspondence with admissible representations of the Kac-Moody algebra $\widehat{\mathfrak{su}}(2)_{k_F}$. This also explains why the fixed points are assembled into groups --- the two fixed points in each group are precisely the ones that are related by an outer-automorphism of the Kac-Moody algebra (recall that the outer-automorphism group is $\Z_2$, the same as the center of $SU(2)$).

The characters of admissible modules of $\widehat{\mathfrak{su}}(2)_{\kappa}$ also form a representation of the modular group and the $S$- and $T$-matrices are
\begin{equation}
\label{eq:STsu2}
\begin{split}
\CS_{\hat{\lambda},\hat{\mu}}
&=\sqrt{\frac{2}{u^2(\kappa+2)}}(-1)^{\mu^F_1(\lambda^I_1+1)+\lambda^F_1(\mu^I_1+1)}\\[0.5em]
&\hspace{1cm}\times e^{-i\pi\mu^F_1\lambda^F_1(\kappa+2)}
\sin\left[\frac{\pi(\lambda^I_1+1)(\mu^I_1+1)}{\kappa+2}\right],\\[0.5em]
\CT_{\hat{\lambda},\hat{\mu}}
&=\delta_{\hat{\lambda}\hat{\mu}}
e^{2\pi i (h_{\hat\lambda}-c/24)},
\end{split}
\end{equation}
with $\kappa=t/u$ being the level of the affine $\widehat{\mathfrak{su}}(2)$. Using \eqref{eq:STsu2} we have
\begin{equation}
{\cal I}(\tilde{\CM}_{2,2N-1}) (\ft \rightarrow e^{2\pi i}, \lambda = 0) = e^{\frac{k\pi i}{4}}\left(\CS\CT^k\CS\right)_{0,0}.
\end{equation}
When the monodromy is non-zero, the moment map changes accordingly. In fact we have
\begin{equation}
{\cal I}(\tilde{\CM}_{2,2N-1})(\ft \rightarrow e^{2\pi i}, \lambda) = e^{\frac{k\pi i}{4}}\left(\CS\CT^k\CS\right)_{0,(2N+1-\lambda)}.
\end{equation}

\subsubsection{$(A_1, D_{2N})$ theories and $\CW_{N}$ algebra}\label{sec: D2N CA}

As we have seen in Table \ref{ADandCA}, the chiral algebra in this case is given by the $\CW_N$ algebra, which is a non-regular quantum Hamiltonian reduction of affine Kac-Moody algebra $\widehat{\mathfrak{sl}}(N+1)_k$ at level $k = -(N-1)^2 / N$. The set of modules generated by spectral flow are considered in \cite{CreutzigVOA}. For a given chiral algebra $\chi$, in general there are two types of modules: the ``local" modules and the ``twisted" modules.  A local module \cite{pareigis1995braiding} in the braided category $\CO_{\chi}^{\rm s}$ (\textit{cf.}~Section \ref{sec: 4dto2d CA}) is a module $M$ of $\chi$ with no non-trivial monodromy. A twisted module is attached to the automorphism of $\chi$ \cite{dong1994twisted}, similar to the twisted sectors in string theory on orbifolds. For our $\CW_N$ algebra, the precise details of the modules depend on whether $N$ is even or odd. For simplicity, we will focus in this section on the even case where all local modules are closed under modular transformations \cite{CreutzigVOA}. They are parametrized by the set 
\be
(s, s') \in \{ -N \leq s \leq N-1, 0 \leq s' \leq N-1, s + s' \in 2 \mathbb{Z} \}.
\ee
It is not hard to see that the number of local modules is $N^2$ --- exactly the same as the number of fixed points on Hitchin moduli space ${\widetilde \CM}_{2, 2N-2}$.

By picking suitable representatives of local modules, their conformal dimensions are bounded from below and given by
\be
h_{(s, s')} = \frac{s^2 - {s'}^2}{4N} - \frac{|s|}{2} + \begin{cases} & 0, \ \ \ \ \text{for}\ \ |s+s'| \leq N \ \text{and}\ |s - s'| \leq N, \\ 
                                                                                                  & (s + s') / 2 - N / 2, \ \ \ \ \text{for}\ s+s' > N, \\
                                                                                                  & (s' - s) / 2 - N/2, \ \ \ \ \text{for}\ s-s' < -N.
                                                                            \end{cases}
\label{D2Nh}
\ee
Then, we find that for vanishing flavor holonomies, there is the relation
\be
\boxed{\mu(\lambda_1 = \lambda_2 = 0) = h - \frac{c}{24} + \frac{1}{6}}
\ee
where the central charge of $\CW_N$ algebra is given by $c = 4 - 6N$. We also have the modular transformation data among those $N^2$ modules \cite{CreutzigVOA},
\be
{\CT}_{(\ell, \ell'), (s, s')} & = \delta_{\ell, s} \delta_{\ell', s'} \exp\bbra{2\pi i \pbra{h_{(s,s')} - \frac{c}{24}}}, \\[0.5em]
{\CS}_{(\ell, \ell'), (s, s')} & = \frac{1}{N}\exp \bbra{-\frac{\pi i}{N}(s \ell - s' \ell')}.
\ee
It can be verified that
\be
{\cal I}({\tilde \CM}_{2, 2N-2})(\ft \rightarrow e^{2 \pi i}, \lambda_1 = \lambda_2 = 0) = e^{\frac{k \pi i}{3}} \pbra{\CS \CT^k \CS}_{00}.
\ee

We note that the above matching becomes subtle when $N$ is odd, where modular transformation turns local modules into twisted modules. Moreover, the vacuum module (which is local), is half-integer graded and thus have ``wrong statistics'' \cite{TensorVOA}. On the contrary, our index formula for the Hitchin moduli space ${\widetilde{\CM}}_{2, 2N-2}$ does not exhibit drastic difference between odd and even $N$. It will be interesting to understand the precise relation here.

\subsubsection{$(A_1, A_{2N-3})$ theories and $\CB_N$ algebra}
 
 Finally, we remark on the last type of Argyres-Douglas theory. We will be very brief here. As the $(A_1, A_{2N-3})$ theory is related to the $(A_1, D_{2N})$ theory via Higgsing, the chiral algebra $\CB_N$ in Table~\ref{ADandCA} can be similarly constructed via quantum Hamiltonian reduction of the $\CW_N$ algebra introduced above. As in previous case, the representation theory of the chiral algebra again depends on the parity of $N$. For $N$ odd, local modules are preserved under modular transformation \cite{CreutzigVOA, AugerVOA}. By carefully picking a set of basis, it is clear that the modules are in one-to-one correspondence with fixed points (the total number is $N(N-1)/2$), and the moment map values match with effective central charges. When $N$ is even, much less is known about the relevant categorical property. It will be interesting to understand this situation further.

\subsection{Other examples}

In fact, the correspondence between fixed manifolds on the Hitchin moduli space under the circle action and modules in $\CO^{\rm s}_\chi$ of chiral algebras is much more general. To supplement our previous discussion focused on Argyres-Douglas theories, here we list such correspondence for other $T[\Sigma]$'s where the chiral algebras are known. For a tame puncture decorated by a parabolic subgroup of $G_\C$ (usually in the $A_{N-1}$ series), we will use $\left[ s_1, s_2, \dots, s_l \right]$ to denote the associated Young tableau with each column of heights $s_1, \dots, s_l$. If for a given Young tableau there is $n_s$ columns with height $s$, then the flavor symmetry associated with the puncture is $S(\prod_s U(n_s))$. In this notation the maximal puncture is $[1,1, \dots, 1]$. 

\begin{enumerate}

\item[$\bullet$] $(A_1, D_4)$ Argyres-Douglas theory. The chiral algebra is ${\widehat{\mathfrak{su}}}(3)_{-\frac{3}{2}}$. The Hitchin moduli space ${\tilde \CM}_{2,2}$ has four fixed points, corresponding to the four admissible modules of the affine Kac-Moody algebra. The relation between effective central charge and moment maps are checked in Section \ref{sec: D2N CA}, but one can also check directly using results from the Kac-Moody algebra. One again sees that $\mu(\lambda_1 = \lambda_2 = 0) = -c_{\rm eff}/24 + 1/6$.

\item[$\bullet$] $SU(2)$ gauge theory with four hypermultiplets. The Hitchin moduli space has $SU(2)$ gauge group, defined on $S^2$ with four tame punctures. There are five fixed manifolds --- one $\mathbb{C}\mathbf{P}^1$ plus four points, and they all lie on the nilpotent cone of Kodaira type ${\text{\Rnum{1}}}_0^*$. When the holonomies are set to zero, the moment map values are $\{ 0,0,0,0,1 \}$. The chiral algebra is $\widehat{\mathfrak{so}}(8)_{-2}$. There are five highest-weight modules belonging to the category $\CO^{s}_\chi$, which for Kac-Moody algebras always coincide with Bernstein-Gelfand-Gelfand's category $\CO$ \cite{arakawa2015joseph}. The corresponding highest weights are $\{ -2\omega_1, -2\omega_3, -2\omega_4, -\omega_2 ,0 \}$ with conformal dimensions $\{ -1, -1, -1, -1, 0 \}$ \cite{perse2012note}. Then we see that $\mu(\lambda_{1,2,3,4} = 0) = -c_{\rm eff}/24 + 5/12$.

\item[$\bullet$] $T_3$ theory \cite{Minahan:1996fg}. The Hitchin moduli space is associated with $S^2$ with three maximal tame punctures, with gauge group $SU(3)$. The moduli space has seven fixed manifolds: one $\mathbb{C}\mathbf{P}^1$ plus six fixed points lying on the nilpotent cone of Kodaira type ${\rm \Rnum{4}}^*$ \cite{Gukov:2016lki}. The associated chiral algebra is the affine Kac-Moody algebra $\widehat{\mathfrak{e}_6}$ at level $-3$ \cite{Beem:2014rza, Lemos:2014lua}. There are exactly seven highest-weight modules in the category $\CO$ \cite{arakawa2015joseph}. The highest weights are, respectively, $\{ 0, -\omega_4,  - 2\omega_2 + \omega_3 - \omega_4, \omega_2 - 2\omega_3, -2\omega_1 + \omega_2 - 2\omega_3 + \omega_4, -2 \omega_5 + \omega_6, -3\omega_6\}$ with conformal dimension $\{0, -2, -2, -2, -2, -2, -2\}$. It is not hard to check from the results of \cite{Gukov:2016lki} that the relation between moment maps and effective central charges with zero holonomy is  given by $\mu = -c_{\rm eff}/24+11/12$.

\item[$\bullet$] $E_7$ SCFT \cite{Minahan:1996cj}. The associated Hitchin system has $G=SU(4)$, and $\Sigma$ is a sphere with three tame punctures. Two of them are maximal punctures, while the third one is a next-to-minimal puncture $[2,2]$ \cite{Chacaltana:2010ks}. By comparing the central charges, it is not hard to see that the chiral algebra should be the affine Kac-Moody algebra $\widehat{\mathfrak{e}_7}$ at level $-4$. Although \cite{Gukov:2016lki} did not present the calculation of Hitchin character in this case, the steps of calculation were outlined using generalized Argyres-Seiberg duality. The fixed manifolds consist of one $\mathbb{C}\mathbf{P}^1$ plus seven points, all of which stay on the nilpotent cone of Kodaira type ${\rm{\Rnum{3}}}^*$. Again there are in total eight highest-weight modules of the chiral algebra \cite{arakawa2015joseph}.

\item[$\bullet$] $E_8$ SCFT \cite{Minahan:1996cj}. Now $G=SU(6)$ and $\Sigma$ is a three-punctured sphere, with one maximal puncture, one $[2, 2, 2]$ puncture and one $[3, 3]$ puncture. The moduli space contains nine fixed manifolds --- one $\mathbb{C}\mathbf{P}^1$ and eight fixed points all lying on the nilpotent cone of Kodaira type ${\rm{\Rnum{2}}}^*$. One finds the chiral algebra is the  affine Kac-Moody algebra $\widehat{\mathfrak{e}_8}$ at level $-6$, which has exactly nine highest-weight modules in the category $\CO$ \cite{arakawa2015joseph}.

\end{enumerate}

It is also quite curious to note that in all cases, the vacuum module corresponds to the \textit{top} fixed point with largest moment map. This is in line with the relation between the vacuum module and the oper brane --- the support of the latter is on the Hitchin section, which intersects the nilpotent cone at the top. 

Based on the above observations, we formulate the general conjecture that relates the Coulomb branch vacua and the representation of chiral algebra as follows.

{\bf Conjecture.} Given a four-dimensional $\CN = 2$ SCFT $\CT$, the fixed points on the Coulomb branch $\CM_{\CT}$ on $S^1\times \R^3$ under the $U(1)_r$ action are in one-to-one correspondence with the highest-weight modules of the chiral algebra $\chi_\CT$ associated with $\CT$, in the modular tensor category $\CO^{\rm s}_{\chi_\CT}$ obtained from semi-simplification,
\be\label{FinalConj}
\boxed{\text{$U(1)_r$ fixed points in $\CM_\CT$}}\; \longleftrightarrow\; \boxed{\text{objects in $\CO^{\rm s}_{\chi_\CT}$}}\ .
\ee

One may also wish to formulate the correspondence on the categorical level, not just on the level of objects. For this one needs to find the replacement on the left-hand side, and a natural candidate is the following. Consider the theory $\CT$ on $\R_{\text{time}}\times D^2 \times S^1$, then weakly gauging $U(1)_{r-R}$ (a subgroup of the R-symmetry group $SU(2)_R\times U(1)_r$ generated by $j_r-j_{3,R}$) will break half of the supersymmetries. The resulting theory $\CT'$ will have vacua given by connected components of $U(1)$ fixed points in $\CM_\CT$. Then we have the category of boundary conditions at the spacial infinity $\partial(D^2\times S^1)=T^2$, which we denote as $\CT'(T^2)$. This is a modular tensor category, on which the modular group acts via the mapping class group action of the spacial boundary $T^2$. Then the above conjecture may be formulated as the equivalence between two modular tensor categories --- the ``categorical SCFT/chiral algebra correspondence'' --- as 
\be
\CT'\left(T^2\right)=\CO^{\rm s}_{\chi_\CT}.
\ee




\acknowledgments{We thank Jørgen E.~Andersen, Tomoyuki Arakawa, Chris Beem, Matthew Buican, Clay C\'ordova, Mykola Dedushenko, Davide Gaiotto, Tam\'as Hausel, Lotte Hollands, Victor Kac, Can Koz\c{c}az, Conan Leung, Leonardo Rastelli, Shamil Shakirov, Shu-Heng Shao, Jaewon Song, Richard Wentworth, Dan Xie, Shing-Tung Yau and Matthew Young for helpful discussion. We would especially like to thank Philip Boalch, Sergei Gukov, Andrew Neitzke and Steve Rayan for reading our draft and offering their helpful comments, and Thomas Creutzig for sharing his notes as well as extensive communication and discussion. We are grateful for the organizers of the Simons Summer Workshop 2016, ``RTG Workshop on the Geometry and Physics of Higgs bundles" held in University of Illinois at Chicago, and ``exact operator algebras for superconformal field theories workshop" held at the Perimeter Institute for Theoretical Physics for the hospitality. DP and KY are supported by the DOE Grant DE-SC0011632, the Walter Burke Institute for Theoretical Physics. Additionally, the work of DP is also supported by the center of excellence grant ``Center for Quantum Geometry of Moduli Space" from the Danish National Research Foundation (DNRF95). WY is supported by the Center for Mathematical Sciences and Applications at Harvard University.}

\appendix

\section{Properties of the Coulomb branch index}\label{sec:CBIProperty}
   \subsection{TQFT structure}\label{sec:TQFT}
As the $\CN = 2$ superconformal index of the class $\CS$ theories $T[\Sigma_{g,s}; G]$ does not depend on complex moduli of $\Sigma$, it has a TQFT structure \cite{Gadde:2011uv}. This further implies that the index can be computed by cutting and gluing the Riemann surface. As all Riemann surfaces can be reduced to cylinders and pairs of pants, one should be able to recast the superconformal index into the form
\be
{\cal I}(T[\Sigma_{g,s}; G]; {\bf a}_1, \dots, {\bf a}_s) = \sum_{\alpha} (C_{\alpha \alpha \alpha})^{2g-2+s} \prod_{i=1}^s \psi^{\alpha}({\bf a}_i)
\label{TQFT}
\ee
by choosing a basis in the TQFT Hilbert space to make the ``fusion coefficients'' $C_{\alpha\beta\gamma}$ associated with a pair of pants diagonal, and the ``metric'' $\eta_{\alpha \beta}$ associated with a cylinder proportional to the identity matrix $\delta_{\alpha \beta}$. Here $C_{\alpha \alpha \alpha}$ is also known as the ``structure constant,'' $\psi^{\alpha}({\bf a}_i)$ is called the ``wave function" with flavor fugacity ${\bf a}_i$ at the puncture.\footnote{The diagonalizability of the TQFT structure constant is not a guaranteed property when the TQFT Hilbert space is infinite-dimensional (\eg, for Schur limit of lens space index, it seems that one could not simultaneously diagonalize flavor fugacity variable and flavor holonomy variable \cite{Alday:2013rs}). But the cutting and gluing rules still apply.}

Now let us specialize to the Coulomb branch index for class $\CS$ theories on $S^1 \times L(k,1)$ and recall the TQFT structure studied in \cite{Gukov:2016lki}. Unlike the usual lens space index where the holonomies take integral values, in \cite{Gukov:2016lki} the authors defined the ``full index" by summing over 't Hooft fluxes, allowing fractional holonomies as long as charge quantization condition is satisfied. In the case of theories of type $\mathfrak{g} = \mathfrak{su}(2)$, this means that the holonomy $m_i$ at each puncture takes value in $\{0, 1/2, 1, \dots, k/2\}$. These holonomies form the Hilbert space of the TQFT, and are essentially the set of integrable representations of ${\widehat{\mathfrak{su}}(2)}_k$. After appropriate normalization of the states, \eqref{TQFT} has the following form \cite{Gukov:2016lki, Gukov:2015sna}:
\be
{\cal I}(T[\Sigma_{g,s}; {\widehat{su}(2)}]; m_1, \dots, m_s) = \sum_{l = 0}^k C_l^{2g-2+s} \prod_{i=1}^{s} \psi^l (m_i)
\ee
where
\be
C_l = \frac{L_l^{-1}}{\sqrt{1-\ft} \sin \theta_l |1 - \ft\, e^{2i\theta_l} |^2}
\label{VerlindeStrConst}
\ee
and
\be
\psi^l (m) = \sqrt{1-\ft}\ L_l \times \begin{cases}  (1+ \ft) \sin \theta_l , \ \ \ \ m = 0, \\ \sin 2\theta_l, \ \ \ \ m = 1/2, \\ \sin 3\theta_l - \ft \sin \theta_l, \ \ \ \ m = 1, \\ \sin 4\theta_l - \ft \sin 2\theta_l, \ \ \ \ m = 3/2, \\ \vdots  \\ \sin k \theta_l - \ft \sin (k-2)\theta_l, \ \ \ \ m = (k-1)/2, \\ \sin (k+1) \theta_l - \ft \sin (k-1)\theta_l, \ \ \ \ m = k/2. \end{cases} 
\label{VerlindeWave}
\ee
Here the normalization constant is 
\be
L_l^{-2} = \frac{k+2}{2} |1 - \ft\, e^{2i \theta_l} |^2 + 2\ft\cos 2\theta_l -2 \ft^2
\ee
and those $\theta_l$'s are the $k+1$ solutions in $(0,\pi)$ to the Bethe ansatz equation,
\be
e^{2 i k \theta} \pbra{\frac{e^{i\theta} - \ft\, e^{-i\theta}}{\ft\, e^{i\theta} - e^{-i\theta}}}^2=1.
\ee
Moreover the metric in this basis is given by $ \eta^{\lambda \lambda} = ( 1-\ft^2, 1-\ft, \dots, 1-\ft, 1-\ft^2)$.

What happens when irregular punctures are present? It may not even make sense to talk about TQFT structure, because for a Riemann surface $\Sigma_{g, \ell, \{ n_{\alpha} \} }$ with arbitrary genus $g$ plus $\ell$ regular punctures and an arbitrary number of irregular ones labeled by $\{ n_{\alpha} \}$, the $U(1)_r$ symmetry is broken and the resulting theory is generically asymptotically free \cite{Nanopoulos:2010zb, Bonelli:2011aa} instead of superconformal. For instance,  consider gauging the diagonal $SU(2)$ group of $(A_1, D_{K})$ and $(A_1, D_M)$ theory by an $SU(2)$ vector multiplet. Each side has a flavor central charge $k_{SU(2)} = 4(K-1)/K$ and $k'_{SU(2)} = 4(M-1)/M$; the gauging would contribute to the one-loop running of gauge coupling as
\be
b_0 = 2\pbra{\frac{1}{K} + \frac{1}{M}} > 0.
\ee
If one tries to extend the superconformal index of Argyres-Douglas theory to an arbitrary Riemann surface $\Sigma_{g, \ell, \{ n_{\alpha} \} }$ by cutting and gluing,  the interpretation of the ``index" obtained at the end it is not obvious. In the case of the Schur index and the Macdonald index, it turns out that the cutting-and-gluing procedure computes the index of the UV fixed point, consisting of free multiplets with canonical choice of scaling dimensions \cite{Song:2015wta}.

Let us now examine the Coulomb branch limit. In order to define a viable TQFT structure as \eqref{TQFT}, a necessary condition is that one has to be able to consistently close the regular puncture. This means we should be able to reduce $(A_1, D_{K+1})$ to $(A_1, A_{K-2})$ theory since the Riemann sphere associated with the two theories differ only by an extra regular puncture. On the field theory side, one observes the Coulomb branch scaling dimensions of $(A_1, D_{K+1})$ and $(A_1, A_{K-2})$ theories are very similar, giving further evidence that these two theories are related.

In the language of TQFT, there is a natural ``cap state" that tells us how to close a regular puncture. Let us begin with $(A_1, D_{2N+1})$ and $(A_1, A_{2N-2})$ theories. Recall the lens space index \eqref{indexA1D2N+1} of $(A_1, D_{2N+1})$ contains a normalization factor \eqref{D2N+1Norm} which can be absorbed in the redefinition of the states (labeled by the holonomy $n$) inserted in the regular puncture. Then it is not hard to check that if we define
\be
\langle \phi' | = \langle 0' | - \ft^{\frac{2N}{2N+1}} \langle 1' |
\label{OddCap}
\ee
then this is precisely the cap that reduces the index of $(A_1, D_{2N+1})$ theories into $(A_1, A_{2N-2})$ theories. Recall that in the equivariant Verlinde TQFT, the cap state is decomposed as
\be\label{OldCap}
\langle \phi | = \langle 0| - \ft \langle 1 |.
\ee 
The only difference is the $\ft$ here is replaced with $\ft^{\frac{2N}{2N+1}}$ in \eqref{OddCap}. This is due to the fact that, in the presence of an irregular singularity, the $U(1)$ Hitchin action will also rotate the $\Sigma$, and the neighborhood of south pole (at $z=0$) is also rotated,
\be
\rho_\theta:\quad z\mapsto e^{-i\frac{2}{2N+1}\theta}z.
\ee 
So the state $\langle \phi' | $ is no longer associated with the ordinary cap, but with the ``rotating cap'', and similarly for $ \langle 0' |$ and $\langle 1' |$. 

From the cap states \eqref{OddCap}, it is not hard to argue that the structure constants and wavefunctions associated with regular puncture cannot remain simultaneously the same as those in \eqref{VerlindeStrConst} and \eqref{VerlindeWave}. This is simply because the cap state is given by $\sum_l {C}_l^{-1} \eta_{nn} \psi^l(n)$ which should depend on $N$.

Let us now turn to the $(A_1, D_{2N+2})$ and $(A_1, A_{2N-1})$ case. Unlike the previous situation, the latter theory contains an additional $U(1)$ flavor symmetry so that the existence of the cap state $\langle \phi' |$ is more non-trivial. Similarly, there is a normalization constant for each theory that needs to be absorbed. For the $(A_1, A_{2N-1})$ theory, the normalization constant is \eqref{A2N-1Norm} which shall be absorbed in the definition of irregular puncture wavefunction ${\hat \psi}^l_{2N}$; while for $(A_1, D_{2N})$ theory, the quantity is \eqref{D2NNorm}. Note that there is ``entanglement" between the two factors of the $U(1) \times SU(2)$ flavor symmetry, and one cannot split it into a product of two functions that depend on $n_1$ and $n_2$ separately. 

In order to go from $(A_1, D_{2N+2})$ to $(A_1, A_{2N-1})$, we should properly identify the residual $U(1)$ symmetry and which combination of $n_1$ and $n_2$ is enhanced to $SU(2)$ in the IR. In fact, \cite{ Agarwal:2016pjo} shows that the mixing to $SU(2)$ is given by $(1/2N+2) U(1)_b$. Therefore, we  identify $(N+1) n_2$ as the $SU(2)$ holonomy, while the residual symmetry is identified as
\be
n \sim \frac{N+1}{N} n_1.
\ee
Then it is a straightforward computation to see that the cap state for the regular puncture of $(A_1, D_{2N+2})$  can be defined as:
\be
\langle \phi' | = \langle 0' | - \left \langle \pbra{\frac{1}{N+1}}' \right | \times \begin{cases} \ft, \ \ \ \ \text{for}\ \ n_1 = 0 \\[0.5em]
                                                                                                              \ft^{\frac{N}{N+1}}, \ \ \ \ \text{for}\ \ n_1 > 0 
                                                                                                              \end{cases}
\label{EvenCap}                                                                                                       
\ee
Here, the value inside the bra is for $n_2$. Note the following peculiar behavior: when $n_1$ (the holonomy for $U(1)$ symmetry carried by the irregular puncture) is zero, then the cap state becomes the ordinary one in the tame case \cite{Gukov:2016lki, Gukov:2015sna}, while for non-zero $n_1$ the irregular puncture starts to affect in a non-local way the regular puncture on the other side. Similar to the previous case, one can argue that the structure constants and the wave function for the regular puncture cannot be made identical to the tame case \eqref{VerlindeStrConst} and \eqref{VerlindeWave} simultaneously.

We do not yet know what this quantity computes for arbitrary $\Sigma_{g, \ell, \{ n_{\alpha} \} }$ wild quiver gauge theories via cutting and gluing. What we have found above is a consistent way to define the TQFT structure \eqref{TQFT} solely for Argyres-Douglas theories. A clear picture may be achieved once the irregular states in TQFT are better understood, as was studied in CFT \cite{Gaiotto:2009ma, Gaiotto:2012sf, Gaiotto:2013rk}. 

\subsection{Symmetry mixing on the Coulomb branch}\label{sec: 3d red}

In Section \ref{sec: 4dto3d},  we mentioned that \eqref{topMix} and \eqref{topMix2} can be interpreted as the mixing between $U(1)_r$ symmetry and topological symmetry on the Coulomb branch. We now explain why this is so. We focus on the $T_{\text{3d}}[\Sigma]$ side instead of its mirror $T_{\text{3d}}^{\text{mir.}}[\Sigma]$, and the fugacities assigned on the Higgs branch of $T_{\text{3d}}^{\text{mir.}}[\Sigma]$ become those for the topological symmetry on the Coulomb branch of $T_{\text{3d}}[\Sigma]$. The trace formula \eqref{3dN4Ind} in the Coulomb limit becomes
\be
{\cal I}^{3\text{d}}_C = \Tr_{\CH_C} \ft^{R_C - R_H} {\bf z}^{{\bf f}_J}
\label{3dCoulomb}
\ee
with the BPS Hilbert space $\CH_C$ containing those states satisfying ${\tilde E} = R_C$ and $R_H = - j_2$. Here ${\bf f}_J$ is the charge under topological symmetry. To further simplify \eqref{3dCoulomb}, we claim $R_H = 0$. To see this, let us go back to 4d $\CN = 2$ index and ask what type of short multiplets are counted by Coulomb branch limit. In general, two types will enter \cite{Gadde:2011uv}: they are of type ${\bar{\CE}}_{r, (j_1,0)}$ and ${\bar{\CD}}_{0, (j_1,0)}$. It was shown in \cite{Buican:2014qla} that for Argyres-Douglas theories considered in this paper, no short multiplet of above two types with $j_1 > 0$ occur. Since ${\bar{\CD}}_{0, (0,0)}$ is a subclass of ${\bar{\CE}}_{r, (0,0)}$ it suffices to say that the Coulomb branch index only counts the ${\bar{\CE}}_{r, (0,0)}$ multiplet for Argyres-Douglas theories. After dimensional reduction, it  becomes clear that $R_H = 0$ in \eqref{3dCoulomb} since ${\bar{\CE}}_{r, (0,0)}$ carries the trivial representation of $SU(2)_R$.

Therefore, the substitution we have made in \eqref{topMix} and \eqref{topMix2} only mixes topological symmetry with $SU(2)_C$ symmetry. Under mirror symmetry, $SU(2)_C$ and $SU(2)_H$ are exchanged, and the topological symmetry becomes the flavor symmetry in the mirror frame. To see explicitly the operator mapping, consider $(A_1, A_{2N-1})$ theories with a rank-$(N-1)$ Coulomb branch, for which the mixing is given by \eqref{topMix} and \eqref{A1A2N-13d}. After comparing with \eqref{M2NV}, we see that the 4d $\CN = 2$ Coulomb branch operators come from the $\ft' z_j / z_i$ term with $i = N$ and $j = 1,2,\dots, N-1$. They are precisely the Higgs branch operators $X^j Y_1$, where $(X^i, Y_i)$ are two $\CN = 2$ chiral fields in the $i$-th hypermultiplet.\footnote{The results here differ slightly from that of \cite{Buican:2015hsa} due to a different choice of matrix representations of Cartan element. The two conventions can be mapped to each other. We thank Matthew Buican for discussion and clarification.}

We now turn to the $(A_1, D_{2N})$ Argyres-Douglas theory, whose three-dimensional mirror is given in Figure~\ref{D2Nquiver} \cite{Xie:2012hs}. The Higgs branch index is given by \eqref{D2N+13d} and the substitution made there is \eqref{topMix2}. Note that we set the $U(1)$ fugacity to be $1$, implying that this symmetry does not mix with the R-symmetry. In particular, when $N=2$, the non-abelian part of the topological symmetry is trivial, so we have no mixing at all! This is actually quite reasonable, because the $U(1)_r$ charge ($1/2)$ of the Coulomb branch operator of $(A_1, D_4)$ theory automatically satisfies the $SU(2)_C$ quantization condition.

For general $(A_1, D_{2N})$ theories with $N > 2$ the Coulomb branch operators no longer have half-integral scaling dimensions, so the symmetry mixing \eqref{topMix2} should be non-trivial. It is not hard to single out the term in the denominator of \eqref{D2N+13d} that gives rise to those Coulomb branch operators.

Unfortunately, it is not known in the current literature what is the three-dimensional mirror of $(A_1, A_{2N})$ and $(A_1, D_{2N+1})$ Argyres-Douglas theories. The absence of Higgs branch in the $(A_1, A_{2N})$ theories indicates that their 3d mirror cannot be given by quiver theory. The computation of Coulomb branch index and $k \rightarrow +\infty$ limit shows that the $T_{\text{3d}}[\Sigma]$ must have topological symmetry. 

\section{Massive vacua of three-dimensional quiver theory}\label{app: Massive}

In this appendix we give explicit steps in solving the massive vacua for certain three-dimensional $\CN = 4$ quiver gauge theories. These are the mirrors of three-dimensional reduction of Argyres-Douglas theories. As mentioned in Section \ref{sec: 4dto3d}, the problem of finding the $U(1)$ fixed points is equivalent to the problem of finding the massive vacua with masses turned on according to the embedding $U(1)\subset G_{\text{R-sym}}\times G_{\text{flavor}}$. More precisely, this embedding will specify a one-dimensional subspace of the Lie algebra of $\frak{g}_{\text{R-sym}}\oplus \frak{g}_{\text{flavor}}$ and its dual, where mass parameters lives.\footnote{Turning on mass parameters associated with R-symmetry will in general break supersymmetry. For us, it will break 3d $\CN=4$ to 3d $\CN=2$.} However, as the number of massive vacua are the same for a generic embedding and $U(1)_{\text{Hitchin}}$ is generic (in the sense that fixed points are isolated), we will work with a generic choice of mass parameters to simplify the notation, which will still lead to the right number of vacua.

\subsection{$(A_1, A_{2N-1})$ Argyres-Douglas theory}

The three dimensional mirror is $\CN = 4$ SQED with $N$ flavors of hypermultiplets. Let us denote $(X_i, Y_i)$ where $i = 1, 2, \dots, N$ as the chiral component for the $N$ hypermultiplets, and $\Phi$ ($\sigma$) as the complex (real) scalar in the $U(1)$ vector multiplet. We turn on complex masses $m_{\mathbb{C}}^i$ and real FI parameter $t_{\mathbb{R}} < 0$, and denote the induced action $(\mathbb{C}^*)_m$. The BPS equations are:
\be
& X \cdot Y = 0, \ \ \ \ \ \ \ \ \ |X|^2 - |Y|^2 + t_{\mathbb{R}} = 0,\\[0.5em]
& (\Phi + m_{\mathbb{C}}) \cdot X = 0, \ \ \sigma \cdot X = 0,\\[0.5em]
& (\Phi + m_{\mathbb{C}}) \cdot Y = 0, \ \ \sigma \cdot Y = 0.
\ee
The solution is easy to describe, given by
\be
\sigma = 0, \ \ \ \ \Phi = -m_{\mathbb{C}}^i, \ \ \ \ Y = 0, \ \ \ \ X = (0, \dots, 0, \sqrt{-t_{\mathbb{R}}}, 0, \dots, 0),
\label{A2N-1Vacua}
\ee
for $i = 1, 2, \dots, N$. So there are $N$ fixed points under $(\mathbb{C}^*)_m$ action.

\subsection{$(A_1, D_{2N})$ Argyres-Douglas theory}

The three dimensional mirror is a $U(1) \times U(1)$ quiver gauge theory with $N-1$ hypermultiplets $(X_i, Y_i)$ stretching between two gauge nodes, one single hypermultiplet $(A_1, B_1)$ only charged under the first $U(1)$, and another single hypermultiplet only charged under the second $U(1)$. The superpotential of the theory is
\be
W = \sum_{i=1}^{N-1} (\Phi_1 - \Phi_2 + m^i_{\mathbb{C}}) X_i Y_i + (\Phi_1 + M_1) A_1 B_1 + (\Phi_2 + M_2) A_2 B_2
\ee
where $m^i_{\mathbb{C}}, M_{1,2}$ are the complex masses. We have the following constraints on the space of allowed vacua:
\be
& (\Phi_1 - \Phi_2 + m^i_{\mathbb{C}}) X_i = 0, \ \ \ \  (\Phi_1 - \Phi_2 + m^i_{\mathbb{C}}) Y_i = 0,\\[0.5em]
&  (\Phi_1 + M_1) A_1 = 0, \ \ \ \ (\Phi_1 + M_1) B_1 = 0,\\[0.5em]
&  (\Phi_2 + M_2) A_2 = 0, \ \ \ \ (\Phi_2 + M_2) B_2 = 0,\\[0.5em]
& \sum_{i=1}^{N-1} X_i Y_i + A_1 B_1 = 0, \ \ \ \ -\sum_{i=1}^{N-1} X_i Y_i + A_2 B_2 = 0,
\ee
where $\Phi_{1,2}$ are the complex scalar in the gauge group. Since we have set the real mass to be zero, the vevs of real scalars $\sigma_{1,2}$ in the vector multiplet will automatically be zero. We also must impose the D-term equation, 
\be
& \sum_{i=1}^{N-1} (|X_i|^2 - |Y_i|^2) + |A_1|^2 - |B_1|^2 = t^1_{\mathbb{R}},\\[0.5em]
&  \sum_{i=1}^{N-1} (|X_i|^2 - |Y_i|^2) + |A_2|^2 - |B_2|^2 = t^2_{\mathbb{R}}.
\ee

For simplicity and without loss of generality, we will assume that the real FI parameters $t^{1,2}_{\mathbb{R}} > 0$. Let us try to solve the above equations.

\begin{enumerate}

\item[(a)] Suppose $\Phi_1 - \Phi_2 + m^i_{\mathbb{C}} \neq 0$ for all $i$.

This means that $X_i = Y_i = 0$ for all $i$. Then we get $A_1 B_1 = A_2 B_2 = 0$. But they cannot be simultaneously zero, otherwise the D-term condition would be violated. Therefore we see that only $B_1 = B_2 = 0$, and $|A_1| = \sqrt{t^{1}_{\mathbb{R}}}$, $|A_2| = \sqrt{t^{2}_{\mathbb{R}}}$. This fixes $\Phi_1 =- M_1$ and $\Phi_2 =- M_2$. This gives one solution.

\item[(b)] There exists one $i$ such that $\Phi_1 - \Phi_2 + m^i_{\mathbb{C}} = 0$. 

This implies that $X_j = Y_j = 0$ whenever $ j \neq i$ since the $m^i_{\mathbb{C}}$'s are kept generic. Now if we assume neither $\Phi_1 + M_1$ and $\Phi_2 + M_2$ is zero, then we should have $A_1 = A_2 = B_1 = B_2 = 0$. Then we see that $|X_i|^2 - |Y_i|^2$ equals both to $t^{1}_{\mathbb{R}}$ and $t^{2}_{\mathbb{R}}$, which is impossible since the real FI parameters are also generic.

We conclude that $\Phi_1 = -M_1$ or $\Phi_2 = - M_2$ (they cannot simultaneously hold). If the former is true, then $A_2 = B_2 = 0$, and $X_i Y_i = A_1 B_1 = 0$. We then see that $Y_i = 0$ and $|X_i| = \sqrt{t^{2}_{\mathbb{R}}}$, and $|A_1|^2 - |B_1|^2 = t^{1}_{\mathbb{R}} - t^{2}_{\mathbb{R}}$. Depending on whether $t^{1}_{\mathbb{R}} > t^{2}_{\mathbb{R}}$ or $t^{1}_{\mathbb{R}} < t^{2}_{\mathbb{R}}$ we can solve for $A_1$ and $B_1$. In this way we get $N-1$ solutions.

Similarly, if the latter is true, we also get $N-1$ solutions. So in total, we have $2N-1$ solutions, which is exactly what we want.

\end{enumerate}

\section{Fixed points under $U(1)$ Hitchin action}\label{sec: fixed}

In this appendix we give the explicit form of fixed points by solving the Hitchin equations. We only consider moduli spaces $\CM_{2, 2N+1}$ and ${\tilde \CM}_{2, 2N-1}$. In the case $\CM_{2,2N+1}$, the fixed points and corresponding values of $\mu$ are described
in \cite{Fredrickson-Neitzke}.  We check in detail the weights on the normal bundle for each fixed point and argue that they agree precisely with physical interpretations.  In the case ${\tilde \CM}_{2, 2N-1}$, we generalize the methods in \cite{Fredrickson-Neitzke} to describe the fixed points, and then check the weights.
Throughout this section, we adopt the convention specified around \eqref{Hitchin2}.

\subsection{Fixed points on $\CM_{2, 2N+1}$}
 
For given $N$, the $U(1)$ fixed points are labeled by an integer $\ell=0, 1, \dots, N$ up to gauge equivalence. In terms of the triple $({\bar \partial}_E, h, \varphi)$, they are given by
\be\label{U1fixedpoint}
& \delbar_E = \delbar, \\[0.5em]
& \varphi^*_{\ell} = \left( \begin{array}{cc} 0 & z^{N-\ell}\\ z^{N+1+\ell} & 0 \end{array} \right) dz, \\[0.5em]
& h = \left( \begin{array}{cc} |z|^{\frac{1+2\ell}{2}}e^{U} \\ & |z|^{-\frac{1+2\ell}{2}} e^{-U} \end{array} \right),
\ee
where $U = U(|z|)$ is the unique solution of the ordinary differential equation \cite{McCoyTracyWu}
\be
\left(\frac{d^2}{d|z|^2} + \frac{1}{|z|} \frac{d}{d|z|} \right) U = 8|z|^{2N+1}\sinh(2U) 
\ee
satisfying the following boundary conditions:
\be
& U(|z|) \sim - \frac{1+2\ell}{2} \ln |z| + \dots \ \ \ \ |z| \rightarrow 0,\\[0.5em]
& U(|z|) \sim 0, \ \ \ \ |z| \rightarrow \infty.
\ee
The boundary condition at $|z| = 0$ guarantees that the Hermitian metric $h$ is smooth there; therefore the Chern connection $D = \partial + \delbar + h^{-1} \del h$ has trivial monodromy. The gauge transformation $g_\theta$ which undoes the $U(1)$ action \eqref{circleAction'} on \eqref{U1fixedpoint} is
\be\label{gthetaval}
g_\theta = \begin{pmatrix} e^{\frac{1+2\ell}{2(2N+3)} i \theta} & \\ & e^{-\frac{1+2\ell}{2(2N+3)} i \theta} \end{pmatrix}.
\ee

The moment map \eqref{moment Mu} 
can be interpreted as a regularized $L^2$-norm of the Higgs field. Consequently, at the $U(1)$ fixed point labeled by the integer $\ell$, we have from \eqref{moment Mu}:
\be
\mu_{\ell} & = \frac{i}{\pi} \int |z|^{2N+1} \pbra{\cosh 2U - 1} dz \wedge d{\bar z}\\[0.5em]
& = \frac{(1+2\ell)^2}{8(2N+3)}. 
\label{muM2N+1}
\ee

The $U(1)$ action also acts on the tangent space $T_{(\delbar, \varphi, h)} \mathcal{M}_{2,2N+1}$ to each fixed point. Let $\dot{\varphi} \in \Omega^{(1,0)}(\mathbb{C}\mathbf{P}^{1}; {\rm End} E)$ be the variation of the Higgs field. We say that the $U(1)$ action acts on $\dot{\varphi}$ with weight $\varpi$ if  
\be
e^{i \theta} \rho_\theta^*\dot{\varphi}=e^{i \varpi \theta} g_\theta^{-1} \dot{\varphi}\, g_\theta
\label{weights}
\ee
where $g_\theta$ is given in \eqref{gthetaval}.

As in \cite{hitchin1987self, Gothenrank3}, one can define the complex symplectic form on the tangent space $(\dot A, \dot \varphi)$ as
\be
\omega'(({\dot A}_1, {\dot \varphi}_1), ({\dot A}_2, {\dot \varphi}_2)) = \int \Tr ({\dot \varphi}_2 \wedge \Psi_1 - {\dot \varphi}_1 \wedge \Psi_2)
\label{complexSymForm}
\ee
where $\Psi$ is the image of the identification from $\Omega^1(\mathbb{C}\mathbf{P}^1, {\rm ad} (P))$ to $\Omega^{(0,1)}(\mathbb{C}\mathbf{P}^1, {\rm ad} (P) \otimes \mathbb{C})$. Then it is immediate that the complex symplectic form $\omega'$ has charge $1$ under the circle action. The existence of such form implies that the weights are paired on the tangent space: if there is a weight $\varpi$ on the tangent space, there is also a weight $1 - \varpi$. This statement will be confirmed in examples shortly.

Our strategy in determining these weights relies heavily on permissible deformations of Higgs field and \eqref{weights}. By the word ``permissible" we mean that, (\rnum{1}) its spectral curve must be that of \eqref{AN-1SW} with $K = 2N+1$ with vanishing coupling constants; (\rnum{2}) it does not originate from infinitesimal meromorphic gauge transformation ${\dot \varphi} = \left[ \varphi, \varkappa \right]$ for $\varkappa \in \mathfrak{sl}(2, \mathbb{C})$, and (\rnum{3}) it does not introduce extra singularities; (\rnum{4}) it does not alter leading nilpotent coefficient matrix. The goal is then to enumerate these inequivalent permissible deformations. Moreover, it suffices to consider the deformation to the linear order and ignore all higher order terms. 

Let us begin with the case $\CM_{2,3}$, pick a small parameter $\upsilon$ and focus on the first fixed points
\be
\varphi^*_1 & = \left( \begin{array}{cc} 0 & 1\\ z^3 & 0 \end{array} \right) dz.
\ee
To preserve the spectral curve \eqref{AN-1SW}, there are two simple linear deformations one could write down:
\be
{\dot \varphi}_1 & = \left( \begin{array}{cc} 0 & 0 \\ \upsilon & 0 \end{array} \right)dz, \ \ \ \  \left( \begin{array}{cc} \upsilon & 0 \\ 0 & -\upsilon \end{array} \right) dz.
\ee
However, the second deformation is a gauge artifact, while the first one is legitimate with the weight being $6/5$. We then conclude that the other paired weight must be $-1/5$. Indeed one could find the corresponding deformation as
\be
{\dot \varphi}_1 = \left( \begin{array}{cc} \upsilon z^2 & 0 \\ 0 & -\upsilon z^2 \end{array} \right)dz + o(\upsilon).
\ee
The determinant of $\varphi_1^* + {\dot \varphi}_1$ equals to $-z^3 dz^2$ up to quadratic terms in $\upsilon$, so such deformation stays on the nilpotent cone.

On the other hand, we have another fixed point
\be
\varphi^*_0 & = \left( \begin{array}{cc} 0 & z\\ z^2 & 0 \end{array} \right) dz.
\ee
We see that the diagonal deformation is allowed at this time, since gauge transformation with essential singularity is forbidden. This deformation has weight $3/5$, whose paired weight is $2/5$. The associated deformation for the latter weight is then
\be
{\dot \varphi}_0 = \left( \begin{array}{cc} 0 & -\upsilon \\  \upsilon z & 0 \end{array} \right) dz+ o(\upsilon).
\ee

Now we generalize the above procedure to wild Hitchin moduli space $\CM_{2, 2N+1}$ with $N > 1$. Let us consider the $\ell$-th fixed point in \eqref{U1fixedpoint}. For $j=0, \cdots, \ell-1$, the following family of deformations come from infinitesimal deformations $\dot{\varphi}$ of the lower-left entry of the Higgs field:
\be\label{Deform}
 {\dot \varphi}^{(j)}_{\ell} = \left( \begin{array}{cc} 0 & 0 \\ \upsilon z^j & 0 \end{array} \right) dz.
\ee
The associated determinant that enters spectral curve is
\begin{equation}
    -\det (\varphi_{\ell}^{(j)}) = (z^{2N+1} + \upsilon z^{N-\ell+j})dz^2. 
\end{equation}
So \eqref{Deform} is a permissible deformation. The associated series of weights are
\be \label{series1}
 \varpi_j = \frac{2(N+\ell+1-j)}{2N+3} > 1, \qquad j=0, \dots, \ell-1.
\ee
The moment map is largest at the fixed point $\ell=N$. There are $N$ such deformations,
and this family of deformations at $\ell = N$ should be thought of as (the analogue of) the Hitchin section.

Because of the complex symplectic form $\omega'$ in \eqref{complexSymForm}, there are weights that are paired with those in \eqref{series1}: 
\be \label{series2}
  \varpi_j = \frac{-1-2j}{2N+3} < 0,  \qquad j=0, \dots, \ell-1
\ee
where we have relabeled the indices. They are downward Morse flows, so must stay on the nilpotent cone. In other words, the corresponding family of deformations $\varphi^{(j)}_{\ell}$ preserves the spectral curve $-\det(\varphi^{(j)}_{\ell}) = z^{2N+1} dz^2$:
\begin{equation}
 {\dot \varphi}^{(j)}_{\ell} = \left( \begin{array}{cc} \upsilon z^{N+j+1} & 0 \\ 0 & -  \upsilon z^{N+j+1} \end{array} \right) dz + o(\upsilon).
 \label{defFR}
\end{equation}
This particular type of deformation, \eqref{defFR} also appears in \cite{FredricksonRayan}.

The remaining $2(N-\ell)$ weights are between $0$ and $1$.  Let us consider one family of deformations labeled by $j = 0, \dots, N - \ell - 1$, which is the diagonal deformation:
\be
{\dot \varphi}^{(j)}_{\ell} = 
  \left( \begin{array}{cc} {\upsilon} z^j & 0 \\ 0 &  -{\upsilon} z^j \end{array} \right)d z,
\ee
and the determinant is $ -\det (\varphi^{(j)}_{\ell}) = z^{2N+1} dz^2$, meaning such deformation stays on the nilpotent cone. The associated series of weights are
\begin{equation} \label{eq:series3}
 \varpi_j =\frac{2N+1-2j}{2N+3}, \qquad j=0, \dots, N - \ell - 1.
\end{equation}

The rest weights correspond to deformations $\dot{\varphi}$ which involve both the upper-right and lower-left entries. They can be written as:
\begin{equation}
{\dot \varphi}^{(j)}_{\ell} = \left( \begin{array}{cc} 0 & - \upsilon z^j  \\  \upsilon z^{1+2\ell + j}
  & 0 \end{array} \right) dz + o(\upsilon),
\end{equation}
whose determinant can be verified to lie in the Hitchin base $\CB$. The associated weights are
\be\label{eq:series4}
\varpi_j =\frac{2(N-\ell-j)}{2N+3}, \qquad j=0, \dots, N-\ell-1.
\ee
These weights, after a reordering of indices, pair with the weights in \eqref{eq:series3}. In summary, we have the following weights for the $\ell$-th fixed points on the tangent space:
\begin{subequations}
\begin{align}
& \varpi_j = \frac{2(N+1+j)}{2N+3}, \ \ \ \ j = 1, 2, \dots, \ell,\\[0.5em]
& \varpi_j = -\frac{2j-1}{2N+3}, \ \ \ \ \ \ \ \ j = 1, 2, \dots, \ell,\\[0.5em]
& \varpi_j = \frac{2j+1}{2N+3}, \ \ \ \ \ \ \ \ \ \ j = \ell+1, \ell+2, \dots, N,\\[0.5em]
&\varpi_j = \frac{2(N - j+1)}{2N+3}, \ \ \ \ j = \ell+1, \ell+2, \dots, N.
\end{align}
\label{M2N+1weights}
\end{subequations}
These weights are precisely matched with the wild Hitchin character for $\CM_{2,2N+1}$ in Section \ref{Sec 4: Verlinde}.

\subsection{Fixed points on ${\tilde \CM}_{2, 2N-1}$}

The fixed points on ${\tilde \CM}_{2, 2N-1}$ are quite straightforward to obtain: one merely allows a regular singularity at $z = 0$, whose monodromy for gauge connection is denoted as $\alpha$.  Expressed in terms of a triple $({\bar \partial}_E, h, \varphi)$ these fixed points are
\be
\delbar_E & = \delbar,\\[0.5em]
\varphi & =  \begin{pmatrix} 0 & z^{\ell}\\
z^{2N - 1 - \ell} & 0 \end{pmatrix} dz, \\[0.5em]
h & = \begin{pmatrix}
|z|^{\frac{2N - 1 -2 \ell}{2}} e^{U} & 0 \\
0 & |z|^{-\frac{2N-1-2\ell}{2}} e^{-U} \end{pmatrix}.
\label{tM2N-1fixed}
\ee
where the index $\ell$ is an integer such that $-1 < \ell + 2 \alpha< 2N$ \cite{McCoyTracyWu}. The function $U(|z|)$ is the unique solution of
\be \label{eq:PainleveIII'}
\left(\frac{d^2 }{d|z|^2} + \frac{1}{|z|} \frac{d}{d|z|}  \right) U= 8 |z|^{2N-1} \sinh(2U)
\ee
satisfying the following boundary conditions
\be
& U(|z|) \sim \left(- \frac{2N-1-2\ell}{2}  + 2 \alpha \right) \ln |z| + \dots \ \ \ \ |z| \rightarrow 0,\\[0.5em]
& U(|z|) \sim 0, \ \ \ \ |z| \rightarrow \infty.
\ee

The asymptotics of $U(|z|)$ guarantees that near $z \sim 0$, the harmonic metrics all satisfy
\be
 h \sim \begin{pmatrix} |z|^{2\alpha} & 0 \\ 0 & |z|^{-2\alpha} \end{pmatrix}
\ee
so that the gauge connection indeed has monodromy $A \sim \alpha d \theta$. Computing the regularized value of the moment map \eqref{moment Mu} at each of these $U(1)$ fixed points, we get
\be 
\mu'(\ell) =\frac{1}{2(2N+1)} \left(- \frac{2N-1-2\ell}{2}  + 2 \alpha \right)^2.
\ee

In our case, $2\alpha \in (0, 1)$, these $2N+1$ fixed points are unique up to gauge transformation and are labeled by $\ell=-1, \cdots, 2N-1$. As in previous case, to match the physical predication we usually need to subtract the lowest moment map value. The minimal value, $\mu'_{\min}$ occurs at
$\ell= N-1$:
\be
\mu'_{\min}= \frac{1}{2(2N+1)} \left( -\frac{1}{2} + 2\alpha \right)^2.
\label{D2N+1minMoment}
\ee
Letting 
\be
\mu=\mu'- \mu'_{\min},
\ee
the values of $\mu$ are
\be
\mu = \frac{i(i+1)}{2(2N+1)} - \frac{i}{2N+1} (2 \alpha), \ \ \ \ \ \ \ i = N, N -1, \dots, -N+1, -N,
\ee
where we have relabeled the indices by setting $i = N - \ell - 1$. Note that these are precisely the values of the moment map appearing in \eqref{tM2N-1moment}.

Now we turn to the weights on the normal bundle of these fixed points. Notice that we do not have to compute everything from scratch, because the fixed points in \eqref{tM2N-1fixed}, except $\ell = -1$, are automatically fixed points for the moduli space $\CM_{2, 2N-1}$, \textit{cf.}~\eqref{U1fixedpoint}. However, we are missing two weights since
\be
\dim_\C \widetilde{\mathcal{M}}_{2,2N-1}= \dim_\C \mathcal{M}_{2,2N-1}+2.
\ee
These two additional weights are very easy to obtain, since the associated deformations of the Higgs fields involve $z^{-1}$. We then have:
\be
\epsilon_{N} = \frac{2N-1}{2N+1}, \qquad \tilde{\epsilon}_{N} = \frac{2}{2N+1}.
\ee
The weights for $\ell = -1$ are new, but they are computed in a similar way and we omit the details.

\newpage

\bibliographystyle{JHEP_TD}
\bibliography{draft}

\end{document}